\documentclass[aps,prX,nofootinbib,longbibliography,10pt,twocolumn,amsmath,amssymb,floatfix,showkeys]{revtex4-1}

\usepackage[utf8]{inputenc}
\usepackage[T1]{fontenc}

\usepackage{graphicx} 
\usepackage{enumitem}
\usepackage{bbm}
\usepackage{xcolor}
\usepackage{siunitx}

\extrafloats{10}

\definecolor{linkColor}{RGB}{0,50,150}

\usepackage[
	allcolors=blue,
	colorlinks=true, linkcolor = linkColor, citecolor = linkColor,
	pdfborder={0 0 0},
	pdfencoding = auto,
]{hyperref}

\usepackage[normalem]{ulem}

\usepackage{examplep}

\usepackage{enumitem}

\newlist{romanlist}{enumerate}{1}
\setlist[romanlist]{label=(\textit{\roman*\kern0.15em})}

\newcommand\dt{\partial_t}

\newcommand{\gradx}{\partial_x}

\newcommand{\integral}[3]{\int_{#1}^{#2} \kern-0.2em \mathrm{d}#3 \,}

\newcommand\uvec{\mathbf{u}}
\newcommand\ustat{\widetilde{\mathbf{u}}}
\newcommand\meq{m^{*}}
\newcommand\ceq{c^{*}}
\newcommand\nstat{\widetilde{n}}
\newcommand\mstat{\widetilde{m}}
\newcommand\cstat{\widetilde{c}\kern0.075em} 

\newcommand\qmax{q_\mathrm{max}}

\def\nLat{n_\text{lat}}
\def\Lint{\ell_\text{int}}
\def\etaInfty{\eta_0^\infty}
\def\nInf{n_\text{inf}}

\def\snc{s_\text{nc}}

\AtBeginDocument{%
    \newwrite\bibnotes
    \def\bibnotesext{Notes.bib}
    \immediate\openout\bibnotes=\jobname\bibnotesext
    \immediate\write\bibnotes{@CONTROL{REVTEX41Control}}
    \immediate\write\bibnotes{@CONTROL{%
    apsrev41Control,author="08",editor="1",pages="1",title="0",year="1"}}
     \if@filesw
     \immediate\write\@auxout{\string\citation{apsrev41Control}}%
    \fi
}%

\begin{document}


\title{Phase-space geometry of mass-conserving reaction--diffusion dynamics}

\author{Fridtjof Brauns}
\author{Jacob Halatek}
\author{Erwin Frey} 
\affiliation{%
 Arnold Sommerfeld Center for Theoretical Physics and Center for NanoScience, Department of Physics, Ludwig-Maximilians-Universti\"at M\"unchen, Theresienstra\ss e 37, D-80333 M\"unchen, Germany
}

\preprint{}

\date{\today}

\begin{abstract}

Experimental studies of protein-pattern formation have stimulated new interest in the dynamics of reaction--diffusion systems. However, a comprehensive theoretical understanding of the dynamics of such highly nonlinear, spatially extended systems is still missing. Here we show how a description in phase space, which has proven invaluable in shaping our intuition about the dynamics of nonlinear ordinary differential equations, can be generalized to mass-conserving reaction--diffusion (MCRD) systems. We present a comprehensive analysis of two-component MCRD systems, which serve as paradigmatic minimal systems that encapsulate the core principles and concepts of the \emph{local equilibria theory} introduced in the paper. The key insight underlying this theory is that shifting local (reactive) equilibria---controlled by the local total density---give rise to concentration gradients that drive diffusive redistribution of total density. We show how this dynamic interplay can be embedded in the phase plane of the reaction kinetics in terms of simple geometric objects: the reactive nullcline (line of reactive equilibria) and the diffusive flux-balance subspace. On this phase-space level, physical insight can be gained from geometric criteria and graphical constructions. The effects of nonlinearities on the global dynamics are simply encoded in the curved shape of the reactive nullcline. In particular, we show that the pattern-forming `Turing instability' in MCRD systems is a mass-redistribution instability, and that the features and bifurcations of patterns can be characterized based on regional dispersion relations, associated to distinct spatial regions (plateaus and interfaces) of the patterns.
In an extensive outlook section, we detail concrete approaches to generalize local equilibria theory in several directions, including systems with more than two-components, weakly-broken mass conservation, and active matter systems.

\end{abstract}
\maketitle


\section{Introduction}

\subsection{Motivation and background}

Nonlinear systems are as prevalent in nature as they are difficult to deal with conceptually and mathematically~\cite{Strogatz:Book, Izhikevich:Book, Mikhailov:1994, Murray:Book, Jackson:Book_Vol_1, CrossGreenside:Book, Ott:Book}. 
Cases in which the equations describing such systems can be solved in closed analytical form are rare, making nonlinear problems appear inaccessible to mathematical analysis at first sight.
A key insight, going back to the work of Poincar\'e \cite{Poincare:Book}, was that geometric structures in the phase space of a system can provide qualitative information about the \emph{global} dynamics (trajectories in phase space) without an explicit solution of the differential equations.
The essence of this geometric reasoning can be understood by considering simple systems with only two independent variables; see e.g.\ Refs.~\cite{Izhikevich:Book,Strogatz:Book}. 
In this case, the key geometric objects are nullclines, defined as curves in phase space along which one of the system's two variables is in equilibrium. The points at which nullclines intersect mark equilibria (fixed points) of the system. 
These geometric objects organize phase-space flow, and thereby allow one to infer the qualitative dynamics from the shapes and intersections of the nullclines.
Key concepts like linear stability, excitability, multi-stability, and limit cycles can be understood in the context of such a geometric analysis~\cite{Izhikevich:Book, Strogatz:Book}. Transitions (bifurcations) between qualitatively different regimes are revealed by structural changes of the flow in phase space as the control parameters are varied. 
One key advantage of such a geometric approach to nonlinear dynamical systems is that it yields systematic physical insights into the processes driving dynamics without requiring the explicit solution of the full set of equations. 

Generalizing these methods, developed for ordinary differential equations (ODEs), to phenomena that explicitly require a description on a spatially extended domain---and therefore involve partial differential equations (PDEs)---poses a huge, ongoing challenge.
Instances where this has been successfully achieved are rare.
One classical approach for nonlinear systems in one spatial dimension is the construction of steady-state patterns (including both stationary patterns and traveling-wave solutions in a co-moving frame). 
Mathematically, the steady state in this case is described by a set of ODEs, which can be analyzed based on their phase-space geometry (see e.g.\ Refs.~\cite{Jones:1994a, Kerner.Osipov1994}). 
An elementary example of this is the phase-plane analysis of traveling waves of the Fisher-KPP equation~\cite{Fisher:1937a, Kolmogorov:1937a} as described in Ref.~\cite{Jones:Book}.
Here, we go beyond this approach, and gain physical insight into the \emph{global dynamics} of spatially extended systems from the analysis of geometric objects in a low-dimensional phase space.
Crucially, such a theory should be able to explain both, the dynamic process of pattern formation---initiated, for instance, by a lateral (Turing) instability---as well as the final stationary patterns in terms of the same concepts and principles.

In its full generality, this is likely a futile task.
Here, we restrict ourselves to mass-conserving reaction--diffusion (MCRD) dynamics.
A broad class of systems that can be described by MCRD dynamics are models for intracellular protein-pattern formation \cite{Halatek:2018b}, which is essential for the spatiotemporal organization of many cellular processes, including cell division, motility and differentiation. 
Moreover, as we discuss in the \hyperref[sec:outlook]{Outlook}, many pattern-forming systems are governed by a combination of mass-conserving dynamics and source terms that break mass conservation.
Studying such systems in the \emph{nearly} mass-conserving limit may help to tackle long-standing questions like pattern selection (wavelength selection) in the highly nonlinear regime \cite{Brauns:2020a}. 

Recent results have indicated ways of making progress towards a general theory rooted in mass-conservation~\cite{Halatek:2018a}.
Based on numerical simulations, this study suggests a new way of thinking about pattern formation, namely in terms of mass redistribution that gives rise to \emph{moving local equilibria}:
A dissection of space into (notional) local compartments allows the spatiotemporal dynamics to be characterized on the basis of the ODE phase space of local reactions.
As (\emph{globally} conserved) masses are spatially redistributed, the \emph{local} masses in the compartments act as parameters for the reactive phase-space flow within them. The properties (position and stability) of the \emph{local reactive equilibria} in the compartments are shown to depend on local masses and thus act as proxies for the local phase-space flow \cite{Note:Onsager}.\phantom{\cite{Onsager:1931a}}

Diffusion acts to redistribute the conserved quantities between neighboring compartments and thereby induces changes in the local phase-space structure. 
This level of description proved to be very powerful in explaining chemical turbulence and transitions from chemical turbulence to long range order (standing and traveling waves) far from onset of the (subcritical) lateral instability. 
The prediction of chemical turbulence at onset, based on numerical simulations in Ref.~\cite{Halatek:2018a}, was recently confirmed experimentally \cite{Denk:2018a}. Hence, the advances in this ``proof-of-principle'' study \cite{Halatek:2018a} suggest that a comprehensive theory of pattern formation in reaction--diffusion systems with conserved total densities (masses) can be developed based on the concept of mass-redistribution. 

Here, we put this overarching idea on a general theoretical foundation. To this end, we develop a number of new theoretical concepts, exemplified by two-component mass-conserving reaction-diffusion (2C-MCRD) systems and based on simple geometric structures in the phase space of the reaction kinetics. 
From these concepts, general geometric criteria for lateral (Turing) instability and stimulus-induced pattern formation emerge and allow us to obtain the features and bifurcations of patterns from graphical constructions.
Moreover, these advances reveal connections to other pattern forming phenomena like liquid-liquid phase separation and shear banding in complex fluids.

In our work, general two-component systems serve two purposes.
First, as a paradigmatic and didactic example that encapsulates the core concepts and principles of our theory in a pedagogic and broadly accessible way.
Second, they provide an elementary basis for further generalizations.
Taken together, the role we envision for 2C-MCRD systems is similar to the role of two-variable systems in dynamical systems theory of ODEs.

The framework we present here has recently been employed and generalized to study the principles underlying coarsening and wavelength selection \cite{Brauns:2020a}, as well as systems with spatially heterogeneous reaction rates \cite{Wigbers:2020a} and the role of advective flow in the cytosol~\cite{Wigbers:2020b}---all in the context of two-component systems.
Moreover, the concepts of local equilibria and regional instabilities have recently proven useful to disentangle the interplay of several distinct instabilities that give rise to Min-protein patterns \emph{in vivo} and \emph{in vitro} \cite{Brauns:2020b}.

Potential future generalizations range from adding more components and more conserved quantities, to going beyond strictly mass-conserving systems (see \hyperref[sec:outlook]{Outlook}).
Mass conservation and, more generally, conserved quantities are inherent to the elementary processes underlying many pattern forming systems. We therefore believe that the local equilibria theory we present here offers a new perspective on a broad class of pattern-forming systems---including intracellular pattern formation, classical chemical systems such as the BZ reaction, and even agent-based active matter systems.

\subsection{Structure of the paper}

Put briefly, the paper is structured as follows.
Section~\ref{sec:two-component-introduction} introduces 2C-MCRD systems, and their applications, most prominently as conceptual models for cell polarization.
The concepts introduced in Sec.~\ref{sec:setting-the-stage} and Sec.~\ref{sec:lateral-instability} form the foundation of the proposed framework and the subsequent analysis.
The following two sections present results that are particularly relevant in the biological context of intracellular pattern formation: A characterization of the possible pattern types exhibited by 2C-MCRD systems (Sec.~\ref{sec:pattern-characterization}) and a simple heuristic for the threshold perturbation required for stimulus-induced pattern formation (Sec.~\ref{sec:stim-induced}).
Section~\ref{sec:bifurcation-structure} delves into a more technical analysis of the generic bifurcation structure of 2C-MCRD systems. Here, we find striking similarities to the phase diagram of phase-separation phenomena (such as liquid-liquid phase separation and motility-induced phase separation). 
This technical section also includes weakly nonlinear analysis in the vicinity of the lateral instability onset that corresponding to a critical point in the language of phase transitions.
Finally, in Sec.~\ref{sec:discussion}, we provide an extensive discussion of the implications of our results and an outlook to future research directions.


\section{Two-component mass-conserving reaction--diffusion systems} \label{sec:two-component-introduction}

Our goal is to find geometric structures in phase space that allow the characterization of mass-conserving reaction--diffusion (MCRD) systems. The simplest system of this type is a two-component reaction--diffusion system with two scalar densities, $m(x,t)$ and $c(x,t)$,
\begin{subequations} \label{eq:two-comp-dyn}
\begin{align}
	\dt m(x,t) &= D_m \gradx^2 m + f(m, c), 											\label{eq:m-dyn} \\
	\dt c(x,t) &= \kern0.26em D_c \gradx^2 c \kern0.5em - f(m, c), 	\label{eq:c-dyn}
\end{align}	
\end{subequations}
on a one-dimensional domain of length $L$ with reflective (no-flux) boundary conditions $\gradx m |_{0,L} = \gradx c |_{0,L} = 0$;  all results can straightforwardly be generalized to periodic boundary conditions. The global average $\bar{n}$ of total density $n(x,t) = m(x,t) + c(x,t)$ is conserved:
\begin{equation} \label{eq:total-mass}
	\bar{n} = \frac{1}{L}\integral{0}{L}{x} \big(m(x,t) + c(x,t)\big) .\tag{\theequation c}
\end{equation}
We chose the above form for its conceptual simplicity. However, the principles that characterize pattern formation for this `minimal' model can be generalized to more complex systems with more components and conserved species \cite{Halatek:2018a, MinInVivo:unpublished}, and even beyond strictly mass-conserving systems~\cite{Brauns:2020a}; see also Sec.~\ref{sec:outlook}.

Two-component systems of the above form were widely studied as conceptual models for cell polarization \cite{Ishihara:2007a, Otsuji:2007a, Goryachev:2008a, Altschuler:2008a, Mori:2008a, Jilkine:2011a, Jilkine:2011b, Edelstein-Keshet:2013a, Trong:2014a, Seirin-Lee:2015a, Chiou:2018a, Diegmiller:2018a, Hubatsch:2019a}, where Eq.~\eqref{eq:two-comp-dyn} describes the dynamics of a protein species that cycles between membrane (slow diffusing, concentration $m(x,t)$) and cytosol (fast diffusing, concentration $c(x,t)$). 
In this biological context, the nonlinear kinetics term $f(m,c)$ is of the form 
\begin{equation} \label{eq:attachment--detachment-kinetics}
	f_\text{attach--detach} (m,c) = a(m) c - d(m) m,
\end{equation}
where the non-negative terms $a(m) c$ and $d(m) m$ characterize the attachment of proteins from the cytosol to the membrane and the detachment back into the cytosol, respectively. 
This functional form results from the fact that in intracellular systems chemical reactions are mainly restricted to the cell membrane.
We will use kinetics of the above form for illustration purposes; for specific examples see Appendix \ref{app-sec:models}.
Importantly however, our results hold for general kinetics $f$, and are not restricted to models for intracellular pattern formation. 
Moreover, 2C-MCRD systems have also been studied for slime mold aggregation \cite{Keller:1970a}, cancer cell migration (glioma invasion) \cite{Pham:2012a}, precipitation patterns \cite{Scheel:2009a, Hilhorst:2006a}, and simple contact processes \cite{Wijland:1998a, Kessler:1998a}. Finally, non-isothermal solidification models \cite{Caginalp:1986a} can also be rewritten in the form Eq.~\eqref{eq:two-comp-dyn}; see e.g.\ Refs.~\cite{Morita:2010a, Pogan:2012a}.

In the mathematical literature, 2C-MCRD systems with a specific form of the reaction kinetics, $f(m,c) = c - g(m)$, have been studied extensively~\cite{Morita:2010a,Goh:2011a, Jimbo:2013a,Latos:2018a}. The dynamics of these systems can be mapped to a variational form (gradient flow of an effective free-energy density). In this form, the properties of the dynamics and the stationary patterns can be analyzed analogously to the Cahn--Hilliard equation which describes phase separation near \emph{thermal} equilibrium (see e.g.\ Ref.~\cite{Pismen:Book}). In particular, one can prove that these systems always exhibit uninterrupted coarsening, i.e.\ the fully phase separated state is the only stable stationary state of the system~\cite{Jimbo:2013a,Latos:2018a}.
The theory we present here is fundamentally different from these previous mathematical approaches. Instead of an abstract mapping to a variational form, our approach is grounded in concepts with clear physical interpretation that are not restricted to specific reaction terms. Local equilibria, the overarching concept of our theory, can be generalized to systems with more than two components and more complex phenomena such as waves, oscillations and chaos~\cite{Halatek:2018a,Brauns:2020b}.

In closing this section we we would like to point out that systems which are not strictly mass conserving may have a mass-conserving subsystem (or `core') that captures essential aspects of the system's pattern formation dynamics. An example is the Brusselator system \cite{Prigogine:1968a}, a widely used paradigmatic model to study pattern formation. It's reaction kinetics has a `core' of the same form as Eq.~\eqref{eq:two-comp-dyn} with additional linear production and degradation terms that break mass conservation. 
This broken mass conservation can give rise to interesting new phenomena, not present in the mass-conserving core. Still, the core dynamics can be useful in understanding these new phenomena by exploiting a time scale separation between (fast) mass-conserving processes and (slow) production/degradation processes~\cite{Kuwamura:2015a,Kuwamura:2017a,Brauns:2020a}.
In the Outlook, Sec.~\ref{sec:outlook}, we briefly discuss this example and the broader prospects of such an approach to non-conservative systems. 

\section{Setting the stage---geometric structures in phase space} \label{sec:setting-the-stage}

In this section we introduce the basic geometric concepts in phase space, which we will later use for a full characterization of pattern formation, including pattern types, bifurcations, and the corresponding characteristic length and time scales.
To this end, we will first study the spatially homogeneous (well-mixed) case where we can use the classical geometric tools for studying ordinary differential equations \cite{Strogatz:Book, Izhikevich:Book, Jackson:Book_Vol_1}.
Subsequently, we will build on the phase-space structures obtained from the well-mixed case to also understand pattern formation in spatially extended systems in terms of flow in phase space.

\subsection{Phase-space analysis of a well-mixed system} \label{sec:well-mixed}

\begin{figure}[tbp]
	\centerline{\includegraphics{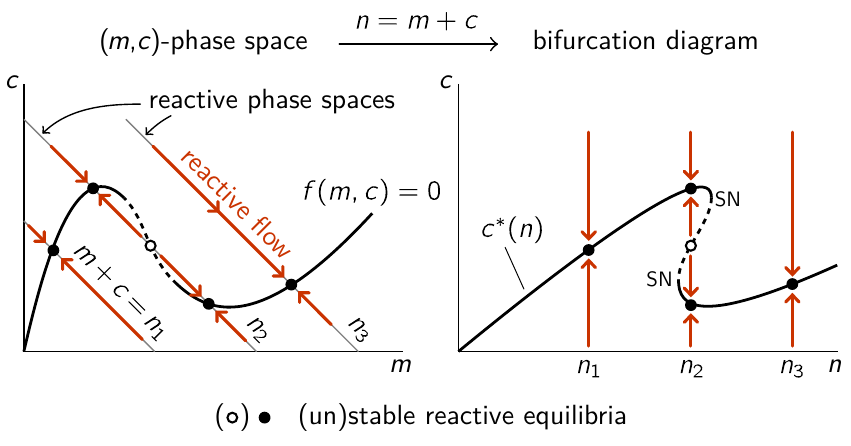}}
	\caption{
    Phase space and bifurcation structure of a well-mixed 2C-MCRD system. 
	The conservation law $m + c = n$ is geometrically represented by 1-simplices in phase space, referred to as \emph{reactive phase spaces}. Local reactions give rise to a flow in phase space (red arrows) which, due to mass conservation, is confined to the reactive phase spaces. The flow vanishes along the \emph{reactive nullcline} ${f(m,c)=0}$ (black line) which is a line of \emph{reactive equilibria}.  
	Each intersection of a reactive phase space with the reactive nullcline is a reactive equilibrium
	$\uvec^*(n)$
	for a given total density $n$ (shown as disks, $\bullet/\circ$, for three different values $n_{1,2,3}$). The $(m,c)$-phase portrait can be transformed into a bifurcation diagram $\ceq(n)$ by the skew transformation $n = m +c$. Because of the conservation law, the well-mixed system has only one degree of freedom, so the only possible bifurcations are saddle-node bifurcations (SN) where the reactive nullcline is tangential to a reactive phase space.
	}
	\label{fig:total-density-bifurcation-structure}
\end{figure}

For a well-mixed system, the dynamics reduces to a set of ordinary differential equations
\begin{align}
	\dt m = f(m, c), \qquad 
	\dt c = -f(m, c).
\end{align}
Since the reaction kinetics conserve total density (protein mass), $n = m+c$, the reactive flow in $(m,c)$-phase space is restricted to the \emph{reactive phase space} where $n$ is a constant of motion (i.e.\ $\dt n = 0$), as illustrated in Fig.~\ref{fig:total-density-bifurcation-structure}a.
The reaction kinetics are balanced at the reactive equilibria (fixed points) $\uvec^* = (\meq,\ceq)$,
\begin{equation} \label{eq:chem-eq}
 	\uvec^*(n) : \left\{
	\begin{array}{rl}
		f(\uvec^*) =& 0,\\[0.2ex]
		\meq + \ceq =& n,
	\end{array}
	\right.
\end{equation}
which are given by the intersection points between the reactive \emph{reactive nullcline} (NC) (or `line of reactive equilibria'), $f(\meq,\ceq) = 0$, and the reactive phase space for a given mass $n$.
The \emph{reactive flow} in the respective phase space is organized by the location and (linear) stability of these fixed points (which are both functions of $n$); see Fig.~\ref{fig:total-density-bifurcation-structure}a and discussion below. 
By varying the total density $n$, i.e.\ by shifting the reactive phase space, one can construct a bifurcation diagram of the (reactive) equilibrium $\ceq(n)$ as a function of the total density $n$ (Fig.~\ref{fig:total-density-bifurcation-structure}b).
The total density is a control parameter of the reactive equilibria. 
When the total density changes, the local equilibria shift. These \emph{shifting (or moving) local equilibria}, introduced in Ref.~\cite{Halatek:2018a}, are the key to understanding the mass-conserving reaction--diffusion dynamics as we will see repeatedly throughout this paper.

In each reactive phase space, we can eliminate (the cytosolic density) $c(t)$ and write the dynamics in terms of (the membrane density) $m(t)$ alone: $\partial_t m(t) = f\big(m(t), n - m(t)\big)$; equally well $m(t)$ could be eliminated. 
In the vicinity of an equilibrium $\meq$ the linearized reactive flow reads $\partial_t m(t) \approx (f_m - f_c)(m(t) - \meq)$, with $f_{i} := \partial_{i} f|_{\uvec^*}, i \in \{m,c\}$. We can read off the eigenvalue $\sigma_\text{loc}(n) := f_m - f_c$ for the local equilibrium and obtain to linear order in the vicinity of the reactive nullcline:
\begin{equation} \label{eq:kinetics-linearization}
\begin{split}
	f(m,c) 	&\approx \sigma_\text{loc}(n) \cdot \big[m-\meq(n)\big] \\
				&= - \sigma_\text{loc}(n) \cdot \big[c - \ceq(n)\big].
\end{split}
\end{equation}
The sign of $\sigma_\text{loc}$---and thereby stability of the reactive equilibria---can be inferred from the slope of the nullcline
\begin{equation} \label{eq:chi-def}
	 \snc(n) 
	 :=  \partial_m c^*(m) \Big|_n 
	  = -\frac{f_m}{f_c}\Big|_n.
\end{equation}
For $f_c > 0$, which is always the case for attachment--detachment kinetics where $f_c = a(m)$, local equilibria are stable, $\sigma_\text{loc}(n) < 0$, if (and only if) the slope of the reactive nullcline is less steep than the slope of the reactive phase space: 
\begin{equation} \label{eq:local-stability-slope-criterion}
	\snc(n) = -1.
\end{equation}
Figure~\ref{fig:total-density-bifurcation-structure}a shows an example for reaction kinetics where the dynamics is monostable except for a window of protein masses exhibiting bistability with one unstable ($\circ$) and two stable fixed points ($\bullet$). 
(Note that the local eigenvalue $\sigma_\text{loc}$ can be rewritten as $\sigma_\text{loc} = f_c \cdot (-\snc - 1)$, which shows why the slope criterion, Eq.~\eqref{eq:local-stability-slope-criterion}, for local stability is reversed for $f_c < 0$.)

\subsection{Stationary patterns are embedded in a flux-balance subspace of phase space} \label{sec:flux-balance-subspace}

To generalize the above approach to spatially extended systems, one has to understand the role of diffusive coupling. 
We start by studying stationary patterns. The insights gained from this analysis will later prove useful for studying the dynamics (instability of the homogeneous state and stimulus-induced pattern formation).

\begin{figure*}[t]
	\centerline{\includegraphics{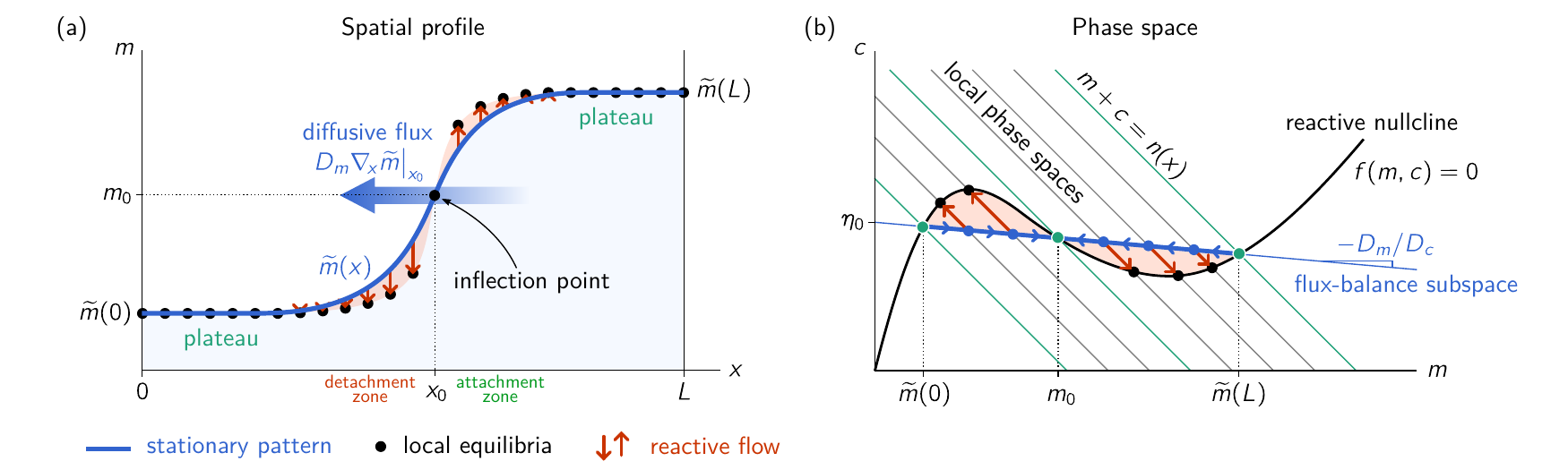}}
	\caption{
	Illustration of a stationary pattern and its embedding in phase space which motivates the flux-balance construction based on the reactive nullcline. (Movie~1 shows the dynamics that lead to such a stationary pattern).
	(a) Local (reactive) equilibria $\meq(\nstat(x))$ ($\bullet$) act as a ``scaffold'' for the pattern profile which is tied to the equilibria by local reactive flows (red arrows). At the inflection point $x_0$ the flux-balance subspace intersects the reactive nullcline $\tilde{f}(m(x_0),\eta_0) = 0$. In a steady state, the diffusive flux on the membrane (blue arrow) is balanced by an equal and opposite flux in the cytosol. Together these fluxes exactly cancel the cumulative reactive turnover on either side of the inflection point (indicated by red shaded areas).
	(b) The phase-space trajectory corresponding to the stationary pattern is embedded in a flux-balance subspace (thin blue line). The local reactive flows (red arrows) are restricted to the local phase spaces where total density is conserved locally. The intersections between the local phase spaces and the reactive nullcline $f(m,c)=0$ yield the local equilibria ($\bullet$). Slow membrane diffusion (blue arrows) balances the reactive flows towards the equilibria in the vicinity of $x_0$ (cf.\ Eq.~\eqref{eq:flux-turnover-balance-dm}). The regions left and right of $x_0$ can be intuitively characterized as attachment and detachment zones based on the direction of the reactive flow in them. A balance of total turnovers (red shaded areas between pattern and local equilibria in (a) and (b) determines $\eta_0$ (cf.\ Eq.~\eqref{eq:Maxwell-construction}). 
	}
	\label{fig:stat-pattern-and-phase-space}
\end{figure*}

A \emph{stationary pattern} $\ustat(x) = [\mstat(x)$, $\cstat(x)]$ is a solution to the steady-state equations,
\begin{subequations} \label{eq:stat-pattern}
\begin{align}
	D_m \gradx^2 \mstat + f(\mstat, \cstat) &= 0, 	\label{eq:m-stat} \\
	D_c \gradx^2 \cstat - f(\mstat, \cstat) &= 0, 	\label{eq:c-stat}
\end{align}	
\end{subequations}
under the constraint of a given average total density $\bar{n}$ (Eq.~\eqref{eq:total-mass}). 
Figure~\ref{fig:stat-pattern-and-phase-space}a shows the sketch of a typical stationary pattern $\mstat(x)$ (solid line) and the corresponding local equilibria $\meq(x)$ (black disks) obtained from a numerical solution of Eq.~\eqref{eq:stat-pattern} with a reaction term as, for instance, in Refs.~\cite{Mori:2008a, Mori:2011a, Chiou:2018a} (see Appendix~\ref{app-sec:models}).
We study patterns with monotonic density profiles, which serve as elementary building blocks for more complex stationary patterns (see Section~\ref{sec:pattern-characterization}). 
Figure~\ref{fig:stat-pattern-and-phase-space}a shows an example for a monotonic pattern profile exhibiting two plateau regions connected by an interface region with an inflection point at $x_0$. (We will later see that this type of pattern, termed `\emph{mesa}', is one of three elementary pattern types found in two-component reaction--diffusion systems; see Sec.~\ref{sec:pattern-types}).

Here we ask what can be learned about the stationary pattern by applying geometric concepts in phase space alone, i.e.\ without relying on an explicit numerical solution.
Observe that Eqs.~\eqref{eq:stat-pattern} imply that the diffusive fluxes of $m$ and $c$ have to balance locally at each position $x$ in the spatial domain $[0,L]$ in steady state:
\begin{equation} \label{eq:diffusive-flux-balance}
	D_m \gradx \mstat(x) = -D_c \gradx \cstat(x). 
\end{equation}
This local \emph{flux-balance condition} is obtained by adding the two steady-state equations, Eq.~\eqref{eq:m-stat} and Eq.~\eqref{eq:c-stat}, integrating over $x$ and employing no-flux boundary conditions. 
Integrating this relation once more from the boundary to any point $x$ in the domain yields that any stationary pattern obeys the linear relation 
\begin{equation} \label{eq:flux-balance-subspace}
	\frac{D_m}{D_c} \, \mstat(x) + \cstat(x) = \eta_0, 
\end{equation}
where $\eta_0$ is a constant of integration. 
(An alternative derivation of this relation that directly generalizes to higher spatial dimensions (cf.\ Eq.~\eqref{eq:stat-pattern-eta} is provided below.)
Equation~\eqref{eq:flux-balance-subspace} defines a family of linear subspaces in the $(m,c)$-phase plane parametrized by $\eta_0$.
We shall call these subspaces the \emph{flux-balance subspaces} (FBS), since they represents the local balance between the diffusive fluxes on the membrane and in the cytosol.
Any stationary pattern is confined to one such subspace; see Fig.~\ref{fig:stat-pattern-and-phase-space}b. 
We will learn later (Sec.~\ref{sec:scaffolding-introduction}), how the value of $\eta_0(\bar{n},L)$ is determined by the balance of reactive processes.

Equation~\eqref{eq:flux-balance-subspace} has been previously used to mathematically simplify the construction and analysis of stationary patterns in two-component systems, by introducing the new phase-space coordinate (orthogonal to the flux-balance subspace)
\begin{equation} \label{eq:eta-def}
	\eta := \frac{D_m}{D_c} \, m + c,
\end{equation}
and describing the spatiotemporal dynamics in terms of the scalar fields $n(x,t) $ and $\eta(x,t)$ (cf.\ Refs.~\cite{Otsuji:2007a, Chiou:2018a}). 
The physical origin (diffusive flux balance) and the geometric interpretation (flux-balance subspace) discussed above explains why Eq.~\eqref{eq:flux-balance-subspace} has proven to be useful before (and why it will be central in our further analysis). 
In particular, note that by adding Eqs.~\eqref{eq:m-dyn} and~\eqref{eq:c-dyn} one finds that gradients in $\eta(x,t)$ drive mass redistribution:
\begin{equation} \label{eq:n-dyn-eta}
	\partial_t n(x,t) = D_c \gradx^2 \eta(x,t).
\end{equation}
We will therefore call $\eta(x,t)$ the \emph{mass-redistribution potential}.
Substituting $c$ using $\eta$, the reaction term reads
\begin{equation} \label{eq:fbs-transformation}
	\tilde{f}(m,\eta) := f\left(m,\eta - m \, D_m/D_c\right),
\end{equation}
and the stationarity conditions, Eqs.~\eqref{eq:m-stat} and~\eqref{eq:c-stat}, are replaced by 
\begin{subequations} \label{eq:stat-pattern_m-eta}
\begin{align} 
	D_m \gradx^2 \mstat + \tilde{f}(\mstat,\widetilde{\eta}\kern0.1em) &= 0, \label{eq:m-stat-fbs} \\
	D_c \gradx^2 \widetilde{\eta} &= 0. \label{eq:stat-pattern-eta}
\end{align}
\end{subequations}
From the second equation, we recover that in steady state, the mass-redistribution potential must be constant in space, $\widetilde{\eta} = \eta_0$, on a domain with no-flux or periodic boundary conditions.
This result also holds in higher spatial dimensions, as one can see by analogy to the electric potential in a charge free space.
The mass-redistribution potential plays a role analogous to the chemical potential in Model~B dynamics \cite{Hohenberg:1977a}. 
However, it does not follow from a free energy density. 
Instead, it is determined by the local concentrations via Eq.~\eqref{eq:eta-def}, and its spatial gradients represent the local imbalance of diffusive fluxes.
Finally, note that the equation for the mass-redistribution dynamics Eq.~\eqref{eq:n-dyn-eta} is not closed. 
Later, in Sec.~\ref{sec:lateral-instability}, we will introduce an approximate ``closure relation'' for Eq.~\eqref{eq:n-dyn-eta}.

The above analysis can be generalized to systems with $N$ components whose total mass is conserved (describing, for instance, a single protein species with $N$ conformational states). The mass-redistribution potential is the sum of the concentrations weighted by their respective diffusion constants. Respectively, the flux-balance subspace in the $N$-dimensional concentration phase space is a $N-1$ dimensional hyperplane orthogonal to the vector of diffusion constants $(D_1, D_2, \ldots, D_N)$.

\subsection{Stationary patterns are ``scaffolded'' by \\ local equilibria} 
\label{sec:scaffolding-introduction}

Whenever the diffusion constants $D_m$ and $D_c$ are unequal, the flux-balance subspace cannot coincide with any reactive phase space (which has slope $-1$). 
Hence, a non-uniform total density profile $\nstat(x) := \mstat(x) + \cstat(x)$ is innate to any stationary pattern (non-uniform $\mstat(x)$) whenever $D_m \neq D_c$.

As we will see next, this non-uniform total density profile is key understand the relationship between the stationary pattern in real space and the reactive nullcline in the phase plane.

Consider the system as being spatially dissected into notional \emph{local} compartments~\cite{Halatek:2018a}. 
Within each such compartment, local reaction kinetics induce a reactive flow $f(m,c)$ that lies in the \emph{local phase space} $\{(m,c) : m+c = \nstat(x)\}$ which is determined by the respective local mass $\nstat(x)$. 
We define the \emph{local equilibria} 
\begin{equation} \label{eq:loc-eq}
\begin{split}
	f\big(\uvec^*(n(x,t))\big) &= 0, \\
	\meq(n(x,t)) + \ceq(n(x,t)) &= n(x,t),
\end{split}
\end{equation}
analogously to Eq.~\eqref{eq:chem-eq}, where we emphasize that the total density $n(x,t)$ is a function of position $x$ and time $t$ here.
The local equilibria are geometrically determined by intersection points of the \emph{local} phase space with the reactive nullcline.
Together with their linear stability the local equilibria serve as proxies for the local reactive flow in each notional compartment (as in the well-mixed system discussed in Sec.~\ref{sec:well-mixed}; cf.\ Eq.~\eqref{eq:kinetics-linearization}). 
Thus, by thinking about a system as dissected into small compartments coupled by diffusion, we can carry over the phase-space structure of the local reaction kinetics to the spatially extended system (Fig.~\ref{fig:stat-pattern-and-phase-space}b).

What is the relationship between the local equilibria $\uvec^*(\nstat(x))$ and the stationary pattern $\ustat(x)$?
To gain some intuition, suppose the compartments were isolated from each other, i.e.\ diffusive coupling between them were shut off.
If we choose the compartments small enough to be well-mixed, then the concentrations $m$ and $c$ in each of them will simply relax to the local equilibrium (black disks in Fig.~\ref{fig:stat-pattern-and-phase-space}) determined by the total density $\nstat(x)$ that varies from compartment to compartment. 
In that sense, the local equilibria act as a \emph{scaffold} to which the pattern is ``tied'' by local reactive flows. 
Because total density must be conserved individually in each of the (now uncoupled) compartments the approach (red arrows) to the local equilibria is confined to the reactive phase space (gray lines) given by the total density in the compartment.

Let us now consider diffusive coupling between these compartments.
In essence, diffusion acts to remove spatial gradients (as indicated by the small blue arrows along the FBS in Fig.~\ref{fig:stat-pattern-and-phase-space}b) and is counteracted by reactive flows towards the local equilibria (indicated by the red arrows from the FBS to the local equilibria in Fig.~\ref{fig:stat-pattern-and-phase-space}a,b). How does this competition play out in detail?
In steady state, the \emph{net} diffusive flux in and out of the compartment is balanced by the deviation from local (reactive) equilibrium (it is instructive to compare Eq.~\eqref{eq:chem-eq} for reactive equilibria and Eqs.~\eqref{eq:stat-pattern} for stationary pattern).
If the gradient does not change across the compartment, such that the flux into and out of the compartment are identical, the net diffusive flux vanishes: $\gradx [D_m \gradx \mstat(x)] = 0$.
(Thanks to the flux-balance condition Eq.~\eqref{eq:diffusive-flux-balance}, the same holds automatically for $\cstat(x)$.)
In turn, the stationary pattern pattern must coincide with the local equilibria:
$\ustat(x) = \uvec^*(\nstat(x))$
when the gradient does not change across a compartment.
This holds exactly at \emph{inflection points} of the pattern. 
For \emph{plateaus}, the gradient is small, $\gradx \ustat(x) \approx 0$, in a spatially extended region, and so is the local net flux, that is, $\gradx^2 \ustat(x) \approx 0$. Hence, for plateaus we have $f(\ustat(x)) \approx 0$, such that the pattern can be locally approximated by the respective local equilibria, that is, $\ustat(x) \approx \uvec^*(\nstat(x))$ in plateau regions.

\subsection{The flux-balance construction on the reactive nullcline}

Combining these insights with the fact that the stationary pattern must be embedded in a flux-balance subspace, we can identify plateaus and the inflection point as intersection points of a flux-balance subspace and the nullcline (FBS--NC intersections). 
As illustrated in Fig.~\ref{fig:stat-pattern-and-phase-space}, these \emph{`landmark points'} in phase space enable us to graphically construct the spatial patterns in real space as two plateaus connected by an interface (\emph{flux-balance construction}).

Near the interface the densities $\ustat(x)$ of a stationary pattern will deviate from the corresponding local equilibria.
The ensuing reactive flows (red arrows) left and right of the inflection point are of opposite sign and correspond to attachment and detachment zones for protein patterns (see Fig.~\ref{fig:stat-pattern-and-phase-space}a) \cite{Halatek:2018b}.
Linearizing the phase-space flow around the landmark points will later enable us to further quantify the spatial profile of stationary patterns, i.e.\ to determine the relevant length scales.

\subsection{Turnover balance determines $\eta_0$}

Integrating one of the stationarity conditions, Eq.~\eqref{eq:m-stat-fbs}, over the whole spatial domain $[0,L]$ yields that in steady state the total reactive turnover must vanish
\begin{equation}
	\integral{0}{L}{x} \tilde{f}(\mstat(x),\eta_0) = 0.	
\end{equation}
This \emph{total turnover balance} determines the position $\eta_0$ of the flux-balance subspace in steady state. 
A mathematically more convenient form of turnover balance is obtained by multiplying Eq.~\eqref{eq:m-stat} with $\partial_x \mstat(x)$ before integrating:
\begin{equation}\label{eq:turnover-balance}
	\integral{\mstat(0)}{\mstat(L)}{m} \tilde{f}(m,\eta_0) = 0.
\end{equation}
In this form, it becomes evident that the total turnover balance does not depend on the full density profile $\mstat(x)$, but only on the densities at the boundaries, $\mstat(0)$ and $\mstat(L)$. Total turnover balance, Eq.~\eqref{eq:turnover-balance}, together with the stationarity condition for $\mstat(x)$, Eq.~\eqref{eq:m-stat-fbs}, fully determine the stationary patterns.

Geometrically, total turnover balance can be interpreted as a kind of (approximate) Maxwell construction in the $(m,c)$-phase plane (balance of areas shaded in red in Fig.~\ref{fig:stat-pattern-and-phase-space}).
This requires the following approximations.
First, we linearize the reactive flow around the reactive nullcline (cf.\ Eq.~\eqref{eq:kinetics-linearization}):
\begin{equation} \label{eq:local-linearization}
	\tilde{f}\big(m,\eta_0\big) \approx \sigma_\text{loc}(\nstat(m)) \cdot \big[m-\meq(\nstat(m))\big],
\end{equation}
where $\nstat(m) := \eta_0 + (1 - D_m/D_c)m$ because the pattern is embedded in the flux-balance subspace, cf.\ Eq.~\eqref{eq:flux-balance-subspace}.
The expression in the square brackets of Eq.~\eqref{eq:local-linearization} is simply the distance of the reactive nullcline from the flux-balance subspace measured along the respective local phase space. 
Further, suppose for the moment that the local eigenvalue $\sigma_\text{loc}(n)$ is approximately constant in the range of total densities attained by the pattern. 
Turnover balance, Eq.~\eqref{eq:turnover-balance}, is then represented by a balance of the areas between nullcline and flux-balance subspace on either side of the inflection point (see areas shaded in red (light gray) in Fig.~\ref{fig:stat-pattern-and-phase-space}b):
\begin{equation} \label{eq:Maxwell-construction}
	\integral{\mstat(0)}{\mstat(L)}{m} \big[m - \meq\big(\nstat(m)\big)\big] = 0. 
\end{equation}

In the characterization of pattern profiles in Sec.~\ref{sec:pattern-characterization}, we will use that for a spatial domain size $L$ much larger than the interface width, one can approximate the plateau concentrations by FBS-NC intersections: 
$\mstat(0) \approx m_-(\eta_0)$ and $\mstat(L) \approx m_+(\eta_0)$. 
In this case, Eq.~\eqref{eq:turnover-balance} is closed and can be solved for $\eta_0$, either numerically or geometrically using the approximate `Maxwell construction,' Eq.~\eqref{eq:Maxwell-construction}.

Multiplying the stationarity condition, Eq.~\eqref{eq:m-stat-fbs}, by $\partial_x \mstat(x)$ (as we did to obtain Eq.~\eqref{eq:turnover-balance} for total turnover balance) and integrating over the spatial subinterval $[0,x_0]$), one obtains a relation that depends only on $\eta_0$ and the boundary concentrations $\mstat(0)$ and $\mstat(L)$:
\begin{equation} 
\label{eq:flux-turnover-balance-dm}
\begin{split}
	\frac{1}{2}D_m \big(\gradx \mstat \big|_{x_0}\big)^2 
	&= \mathop{\hphantom{-}} \integral{\mstat(0)}{m_0(\eta_0)}{m} \tilde{f}(m,\eta_0) \\
	&= - \integral{m_0(\eta_0)}{\mstat(L)}{m} \tilde{f}(m,\eta_0).
\end{split}
\end{equation}
These equations state that the net turnover on either side of the pattern inflection point $x_0$, has to be balanced by the net diffusive flux across that point as illustrated by the blue arrow in Fig.~\ref{fig:stat-pattern-and-phase-space}a.
Because the reactive flow changes sign at the inflection point, the reactive turnover (integrated flow) is extremal there and determines the maximal slope $\mstat'(x_0)$ of the pattern profile. 
Depending on how the reactive turnover saturates on either side of the inflection point, the system exhibits, as we will learn in Sec.~\ref{sec:pattern-types}, three distinct characteristic elementary pattern types, classified by the shape of the concentration profile $\mstat(x)$: 
mesas, peaks/troughs and nearly harmonic (or `weakly nonlinear') patterns. 

\subsection{Summary of geometric structures in phase space}

Let us pause for a moment and briefly summarize our findings so far. 
We have established three major geometric structures in $(m,c)$-phase space: 
First, the \emph{reactive nullcline}, $f(m,c)=0$, along which the local reaction kinetics are balanced; 
second, the \emph{local phase spaces}, $m + c = \nstat(x)$, determined by the local total densities $\nstat(x)$---local equilibria $\uvec^*(\nstat(x))$ are intersections of the reactive nullcline and the local phase spaces; 
third, the family of \emph{flux-balance subspaces}, within which diffusive flows in membrane and cytosol balance each other. 
The position of the flux-balance subspace, $\eta_0$, of a stationary pattern is determined by total turnover balance, Eq.~\eqref{eq:turnover-balance}, which represents a balance of reactive processes. 

This geometric picture underlies the key results we present in the remainder of the paper. 
Up to now, we only discussed the embedding of the pattern in the $(m,c)$-phase plane. 
To study the possible pattern profile shapes $\mstat(x)$ in real space, we need to understand the dynamic process of pattern formation, in particular the factors determining the interface region. 
As we will see below (Sec.~\ref{sec:pattern-characterization}), the interface of a pattern is inherently connected to lateral instability. 
We will therefore first analyze lateral instability and the dynamic process of pattern formation in the following section. With these tools at hand, we will then be able to characterize the distinct pattern types exhibited by 2C-MCRD systems.

\section{Lateral instability} \label{sec:lateral-instability}
 
How can the geometric structures introduced in the previous section help us to understand the physical process of pattern formation? Previous research \cite{Halatek:2018a} suggests that the total densities are the essential degrees of freedom and their redistribution is the key dynamic process. Building on this insight, we systematically connect the geometric structures established above (Sec.~\ref{sec:setting-the-stage}) to the lateral instability, i.e.\ instability against spatially inhomogeneous perturbations, of a homogeneous steady state.

\subsection{Mass-redistribution instability} 

Consider the dynamics of the local total density $n(x,t) = c(x,t) + m(x,t)$. 
Because the kinetics conserve local total density, the time evolution of $n(x,t)$ is driven only by diffusion due to spatial gradients in the concentrations $c(x,t)$ and $m(x,t)$:
\begin{equation} 
\label{eq:n-dynamics}
	\partial_t n(x,t) 
	=  D_c \gradx^2 c(x,t) + D_m \gradx^2 m(x,t). 
\end{equation}
As a result of mass redistribution, the local equilibrium concentrations $\uvec^*(n(x,t))$ change. 
In turn, these locally shifted equilibrium concentrations induce changes in the local reactive flows and thereby result in altered spatial gradients in $\uvec(x,t)$. 
This intricate coupling between redistribution of total mass, reactive flows, and diffusive flows drives pattern formation. 

Qualitatively, this coupling between reactive and diffusive flow can be understood by observing that the dynamics depends mainly on the direction in which the local equilibria shift due to increasing or decreasing local total density.
Let us therefore posit that the relevant diffusive gradients can be (qualitatively) estimated by replacing the local concentrations by the (locally stable) local equilibrium
\begin{equation} 
\label{eq:adiabatic-scaffolding}
	\uvec(x,t) \to \uvec^*\big(n(x,t)\big),
\end{equation}
such that the local mass $n(x,t)$ is the only remaining degree of freedom:
\begin{equation}
	\partial_t n(x,t) \approx  D_c \gradx^2 \ceq(n) + D_m \gradx^2 \meq(n). \label{eq:n-dynamics-slaved}
\end{equation}
We term this the \emph{local quasi-steady state approximation}.
Note that this approximation becomes exact in the long wavelength limit where diffusive redistribution is much slower than chemical relaxation; see Sec.~\ref{sec:diff-react-limited} and Appendices~\ref{app-sec:lateral-instability} and~\ref{app-sec:adiabatic-scaffolding} for a detailed discussion.
Applying the chain rule once, we can rewrite the mass redistribution dynamics as 
\begin{equation} \label{eq:n-dynamics-expanded}
	\partial_t n(x,t) \approx \gradx \big[\big(D_c \, \partial_n \ceq + D_m \, \partial_n \meq\big) \gradx n\big]
\end{equation}
which is simply a diffusion equation for the total density $n(x,t)$.
For locally stable equilibria, the effective diffusion constant will become negative (which entails anti-diffusion) if
\begin{equation} 
\label{eq:slope-criterion}
	\frac{\partial_n \ceq}{\partial_n \meq} 
	= \snc(n) 
	= -\frac{f_m}{f_c} 
	< - \frac{D_m}{D_c}, 
\end{equation}
where $\snc(n) = \partial_m \ceq(m)|_n$ is the slope of the reactive nullcline $\ceq(m)$ (cf.\ Eq.~\eqref{eq:chi-def} in Sec.~\ref{sec:well-mixed}; note that local stability ensures $\partial_n \meq > 0$ when $f_c > 0$, for $f_c < 0$ the inequality Eq.~\eqref{eq:slope-criterion} is reversed). 
Hence, starting from a homogeneous steady state $\uvec^*(\bar{n})$, a \emph{lateral instability} due to effective anti-diffusion takes place if (and only if)
\begin{equation}
 \snc(\bar{n}) < - \frac{D_m}{D_c}.	
\end{equation}
This condition for lateral instability has a simple geometric interpretation in the $(m,c)$-phase plane: A spatially homogeneous steady state with total density $\bar{n}$ is laterally unstable if the slope of the nullcline is steeper than the slope of the flux-balance subspace. 
We term the mechanism a \emph{mass-redistribution instability} to emphasize the underlying physical process and to contrast this mechanism to the ``activator--inhibitor mechanism'' (see Sec.~\ref{sec:AI} in the Discussion).
Importantly, the mass-redistribution instability in reaction--diffusion systems is a Turing instability \cite{Turing:1952a}, in the sense that it is a diffusion-driven instability of a system that is stable in a well-mixed situation (i.e.\ stable against spatially homogeneous perturbations)~\cite{Note:Turing}.
The bifurcation where a homogeneous steady state becomes laterally unstable, i.e.\ Turing unstable, will be referred to as a \emph{Turing bifurcation}.

\begin{figure*}[t]
	\centering
	\includegraphics{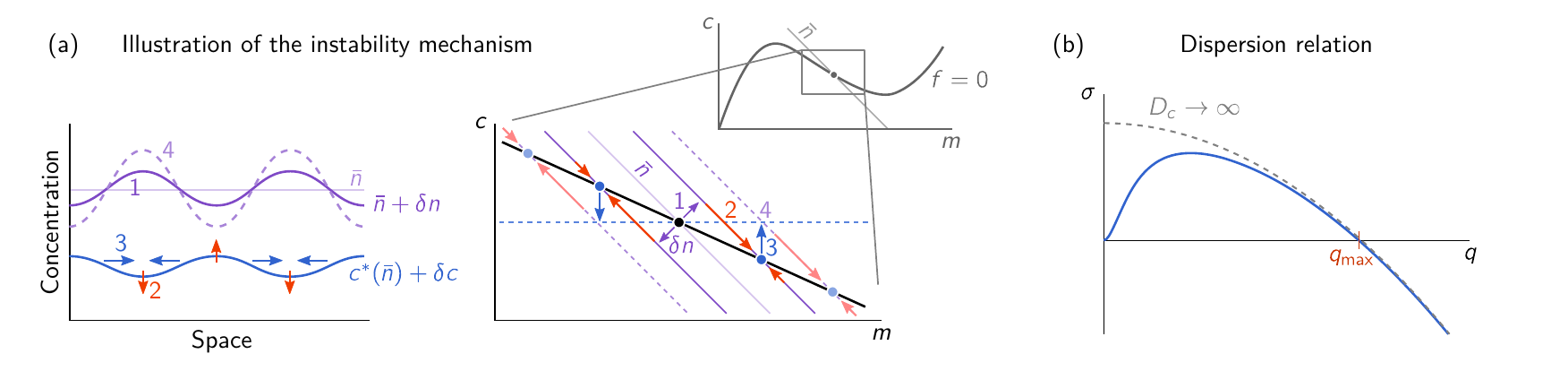}
	\caption{
	Mass-redistribution instability.
	(a) Illustration of the underlying mechanism.
	Consider a small amplitude modulation of the total density (purple line in the real space plot) on a large length scale~(1). 
	As diffusion is slow on large scales, the system will locally relax to its reactive equilibrium (2). The resulting cytosolic density profile $\delta c(x) \approx \partial_n c^*(n)|_{\bar{n}} \delta n(x)$ is shown by the blue line in the real space plot. 
	If the cytosolic equilibrium density decreases due to an increase of total density (i.e.\ if the nullcline slope, $\snc$, is negative), the ensuing diffusive fluxes in the cytosol (3) will amplify the modulation of the total density profile (4), thus driving an instability.
	The membrane gradient is opposite to the cytosolic one, such that membrane diffusion counteracts the instability and suppresses it if $\snc > -D_m/D_c$; cf.\ Eq.~\eqref{eq:slope-criterion}.
	(b) Dispersion relation (solid blue line). In the limit $D_c \to \infty$, the dispersion relation approaches the function $f_m - D_m q^2$ shown as gray, dashed line. This limit is discussed in Appendix~\ref{app-sec:diffusion-limits}.
	}
	\label{fig:MRI} 
\end{figure*}

The mass-redistribution dynamics, Eq.~\eqref{eq:n-dynamics-expanded}, can be rewritten most compactly using the mass-redistribution potential, $\eta$, (cf.\ Eq.~\eqref{eq:n-dyn-eta}):
\begin{equation} \label{eq:slaved-dyn-eta}
	\partial_t n(x,t) \approx D_c \gradx \big[\partial_n \eta^*(n) \, \gradx n\big].
\end{equation}
This implies that a mass-redistribution instability occurs if an increase in total density entails a decrease of the mass-redistribution potential (i.e.\ $\partial_n \eta^*|_{\bar{n}} < 0$).

Importantly the instability condition, Eq.~\eqref{eq:slope-criterion}, can be related directly to an underlying physical mechanism. 
For illustration purposes let us disregard membrane diffusion ($D_m = 0$). 
Following a small modulation  $\delta n$ of the mass on a large length scale, the local reactive equilibrium within each compartment shifts (1 in Fig.~\ref{fig:MRI}a). 
The instability criterion, Eq.~\eqref{eq:slope-criterion}, requires the slope of the reactive nullcline to be negative. 
In this case, the equilibrium shifts to lower cytosolic concentration $c^*(n)$ as total density $n$ is increased. 
In other words, in regions with a higher total density, there will be a reactive flow onto the membrane (red arrows) as the shifted local equilibrium is approached---thus creating cytosolic sinks (2 in Fig.~\ref{fig:MRI}a). 
Conversely, the regions with lower total density become cytosolic sources.
The ensuing cytosolic gradient leads to diffusive mass-redistribution (3), resulting in a further shift of the local equilibria (4), thus sustaining and amplifying the diffusive flux---the cycle feeds itself. 
Taken together, this shows that the \emph{mass-redistribution instability is a self-amplifying mass redistribution cascade}. 
In contrast, when the cytosolic equilibrium density rises due to an increase in total density (i.e.\ for positive nullcline slope $\partial_n \ceq|_{\bar{n}} > 0$) the compartment with more total density will become cytosolic source inducing mass redistribution that brings the system back to a homogenous state.

\subsection{Diffusion- and reaction-limited regimes} \label{sec:diff-react-limited}

On sufficiently large length scales, diffusive relaxation (transfer of mass) is slow compared to chemical relaxation $D_c q^2  \ll |\sigma_\text{loc}|$, such that the local quasi-steady state approximation Eq.~\eqref{eq:adiabatic-scaffolding} becomes exact---the concentrations are \emph{slaved} to the local equilibria.
This is the \emph{diffusion-limited} regime: the growth rate of the lateral instability is limited by cytosolic redistribution via diffusion ($\sigma_\text{lat} \approx - \partial_n \ceq \cdot D_c q^2$).
In contrast, if cytosolic diffusion is much faster than chemical relaxation ($D_c q^2 \gg |\sigma_\text{loc}|$), the lateral instability is limited by the rate at which the shifting equilibria are approached ($\sigma_\text{lat} \approx \partial_n \ceq \, \cdot \, \sigma_\text{loc} = f_m$). This is the \emph{diffusion-limited} regime.
Importantly, the concept of shifting local equilibria still informs about the presence of the lateral instability in this regime. But it does no longer yield the growth rate quantitatively.
A more detailed discussion of the local quasi-steady state approximation is provided in Appendix~\ref{app-sec:adiabatic-scaffolding}.

The principle of shifting local equilibria provides insight into the spatial dynamics of systems with more than two components: 
In a five-component MCRD model for \emph{in vitro} Min patterns, the concept of scaffolding allowed to predict the transition to chaos (qualitative change of the local attractors from stable fixed point to limit cycle) \cite{Halatek:2018a}.
Notably, in this system, the onset of lateral instability is not a long wavelength instability but takes place for a band of unstable modes bounded away from $q = 0$, corresponding to `type~I' instability in the Cross--Hohenberg classification scheme. 
Thus, the principle of shifting local equilibria is not restricted to systems with a long wavelength (`type~I') instability.

\subsection{The marginal mode $\qmax$ reveals the role of membrane diffusion}

Let us compare the instability criterion Eq.~\eqref{eq:slope-criterion} to `classical' linear stability analysis of Eq.~\eqref{eq:two-comp-dyn} around the homogenous steady state (see Appendix~\ref{app-sec:lateral-instability}). 
There one obtains the dispersion relation $\sigma(q)$ for the growth rate $\sigma$ of a mode with wavenumber $q$ (see Fig.~\ref{fig:MRI}b). 
It exhibits a band of unstable modes, $\sigma(q) > 0$ for $0 < q < \qmax$, with 
\begin{align} \label{eq:qmax}
	\qmax^2 := \frac{f_m}{D_m}-\frac{f_c}{D_c} = \frac{\tilde{f}_m}{D_m} 
\end{align}
if and only if $f_m/D_m > f_c/D_c$, i.e.\ exactly when the slope criterion, Eq.~\eqref{eq:slope-criterion}, is fulfilled. 
Equation~\eqref{eq:qmax} can be rewritten as $\qmax^2 = f_c/D_m (-\snc - D_m/D_c)$, which shows why the slope criterion Eq.~\eqref{eq:slope-criterion} is reversed for $f_c < 0$.

The instability condition Eq.~\eqref{eq:slope-criterion} and the expression for the edge of the edge of the band of unstable modes Eq.~\eqref{eq:qmax} inform about the role of membrane diffusion as counteracting the cytosolic mass-redistribution that drives the instability. This is because the membrane gradient will always be opposed to the cytosolic gradient whenever the nullcline slope is negative $(-1 < \partial_n \ceq < 0)$ (because $\delta m = \delta n - \delta c$ and $\delta \ceq = \delta n \, \partial_n \ceq$). 

The expression for $\qmax$ can be found quite easily by utilizing phase-space geometry. 
We start from the observation that $\qmax$ is a non-oscillatory marginal mode; it cannot be oscillatory for a locally stable fixed point, $\sigma_\text{loc} < 0$, as shown in Appendix~\ref{app-sec:lateral-instability}. 
Hence, the eigenvalue $\sigma(\qmax) = 0$, so the mode $\sim \cos(\qmax x)$ must fulfill the steady-state condition, Eq.~\eqref{eq:stat-pattern}, and  
the corresponding eigenvector must point along a flux-balance subspace $\propto (1,-D_m/D_c)^\text{T}$ in phase space. 
The steady state condition in flux-balance subspace is given by Eq.~\eqref{eq:m-stat-fbs}, which, in linearization around a homogeneous steady state reads
\begin{equation} 
	0 = D_m \gradx^2 \delta m(x) + \left[f_m - \frac{D_m}{D_c} f_c\right]_{\bar{n}} \delta m(x).
\end{equation}
This equation is solved by the mode $\delta m(x) \propto \cos(\qmax x)$ with $\qmax^2 = f_m/D_m - f_c/D_c$ (cf.\ Eq.~\eqref{eq:qmax}).
To conclude, the two effects of membrane diffusion are interlinked in the expression for $\qmax$: 
(\textit{i}) The condition $\qmax = 0$ determines the critical NC-slope ($\snc^\text{crit} = -D_m/D_c$) for the (long wavelength) onset of lateral instability.
(\textit{ii}) In the laterally unstable regime, $\qmax$ determines the smallest unstable length scale $\ell = \qmax^{-1}$. In the limit of large $D_c$, this length scale is given by $\ell^2 \approx D_m/f_m$, i.e.\ a balance of membrane diffusion and reactive flows.
In the next section, it will be shown that the marginal mode $\qmax$ \emph{at the pattern inflection point} determines (to leading order) the interface width of a stationary pattern.

\section{Characterization of stationary patterns}
\label{sec:pattern-characterization}

With an intuitive picture of the principles underlying pattern formation in 2C-MCRD systems in hand, we now return to the spatially continuous system. 
We first study the characteristic types of stationary patterns exhibited by 2C-MCRD systems, focusing on \emph{elementary} stationary patterns with monotonic concentration profiles on a domain with no-flux boundaries. 
More complex, non-monotonic stationary patterns (also in domains with periodic boundary conditions) can always be dissected into such elementary patterns at their extrema (recall that due to the diffusive flux-balance condition, Eq.~\eqref{eq:diffusive-flux-balance}, extrema in $\mstat(x)$ and $\cstat(x)$ must coincide). 
Previous studies have observed that 2C-MCRD systems typically exhibit coarsening~\cite{Kang:2007a, Ishihara:2007a, Otsuji:2007a, Chiou:2018a}.
In a follow-up work building on the concepts presented here, we show that coarsening is indeed generic in all 2C-MCRD systems, independently of the specific form of the reaction kinetics~\cite{Brauns:2020a}.

\subsection{Interface width} \label{sec:interface-width}

The width of the interfacial region, $\Lint$, is the only intrinsic length scale of the elementary patterns. 
Recall that the pattern inflection point, which defines the position of the interface region, is in local reactive equilibrium---geometrically determined by an FBS--NC intersection $(m_0,c_0)$ (cf.\ Fig.~\ref{fig:stat-pattern-and-phase-space} in Sec.~\ref{sec:scaffolding-introduction}). 
The interface is maintained by a balance of diffusion and the reactive flow in the vicinity of the inflection point.
Therefore,the interface $\mstat(x - x_0) \approx m_0 + \delta \mstat(x)$ is to leading order determined by linearizing the steady-state equation, Eq.~\eqref{eq:m-stat-fbs}, around the inflection point,
\begin{equation}  \label{eq:m-stat-interface-linearized}
	0 = D_m \gradx^2 \delta \mstat(x) + \left[f_m - \frac{D_m}{D_c} f_c\right]_{n_0} \delta \mstat(x),
\end{equation}
where $n_0 = m_0 + c_0$, and we used the flux-balance subspace constraint, Eq.~\eqref{eq:flux-balance-subspace}, to substitute the cytosol concentration $\delta \cstat(x)$. 
Equation~\eqref{eq:m-stat-interface-linearized} exactly resembles the equation that determines the marginal mode $\sin(\qmax x)$ in the dispersion relation (right-hand edge of the band of unstable modes). Hence, the interface length scale is determined by the marginal mode of the dispersion relation \emph{at the inflection point}:
\begin{equation} 
\label{eq:Lint-approx}
	\Lint \simeq \pi/\qmax(n_0) 
	= \pi \sqrt{D_m/\tilde{f}_m|_{n_0}}.
\end{equation}
The interface shape is approximated by the corresponding eigenfunction $\delta \mstat(x) \propto \sin\big(\qmax(n_0) x\big)$.

Let us pause for a moment to look at the interface region from the perspective of mass redistribution: 
the total density $n_0$ at the inflection point is such that the corresponding reactive equilibrium is \emph{laterally} unstable, because the nullcline slope is steeper than the FBS-slope there; 
see Fig.~\ref{fig:regions} below, where we elaborate on this point in terms of spatial \emph{regions}. 
From the spectrum of modes, only the marginally stable one, $\qmax(n_0)$, fulfills the (linearized) stationarity condition. 
Thus, intuitively it must be the $\qmax$-mode that determines the interface length scale. 
Importantly, because the pattern is formed by mass redistribution, the total density $n_0$ at the inflection point does not coincide in general with the average total density $\bar{n}$. The interface width is determined $\qmax(n_0)$, not by $\qmax(\bar{n})$. 
This also implies that the interface width depends on the  FBS-position $\eta_0$ because the inflection point $(m_0, c_0)$, and hence $n_0 = m_0 + c_0$, is determined geometrically as FBS-NC intersection point.
We explicitly denote the interface width by $\Lint(\eta_0)$ when we use this relationship in the following.

Finally, to approximate the stationary concentration profile of the interface, we use that its maximal slope $\mstat'(x_0)$ is attained at the pattern inflection point $x_0$ and can be calculated by flux-turnover balance \eqref{eq:flux-turnover-balance-dm}. 
Together with the harmonic mode $\delta \mstat(x) \propto \sin(\pi x/\Lint)$ obtained by linearizing phase space flow, we find
\begin{equation} \label{eq:interface-profile-approx}
	\mstat(x) \approx m_0 + \mstat'(x_0) \frac{\Lint}{\pi} \sin\left(\pi \frac{x-x_0}{\Lint}\right),
\end{equation} 
in the vicinity of the inflection point.
To go beyond this leading order approximations, one can perform a perturbative expansion of $\tilde{f}(m_0 + \delta m,\eta_0)$ and $\mstat(x_0+\delta x)$ in Eq.~\eqref{eq:m-stat-fbs} around the pattern inflection point $(m_0,\eta_0)$. 
The solution of this expansion can then be matched to the plateaus to obtain an approximation of the interface profile shape. 
Linearization around the plateaus yields exponential decay towards the plateaus (``exponential tails'') ${\sim} \exp(- x/\ell_\pm)$, where the decay lengths are given by $\ell_\pm^2 = D_m/\tilde{f}_m(n_\pm)$.

\subsection{Regions generalize the concept of local compartments} \label{sec:regions}

\begin{figure*}
	\centerline{\includegraphics{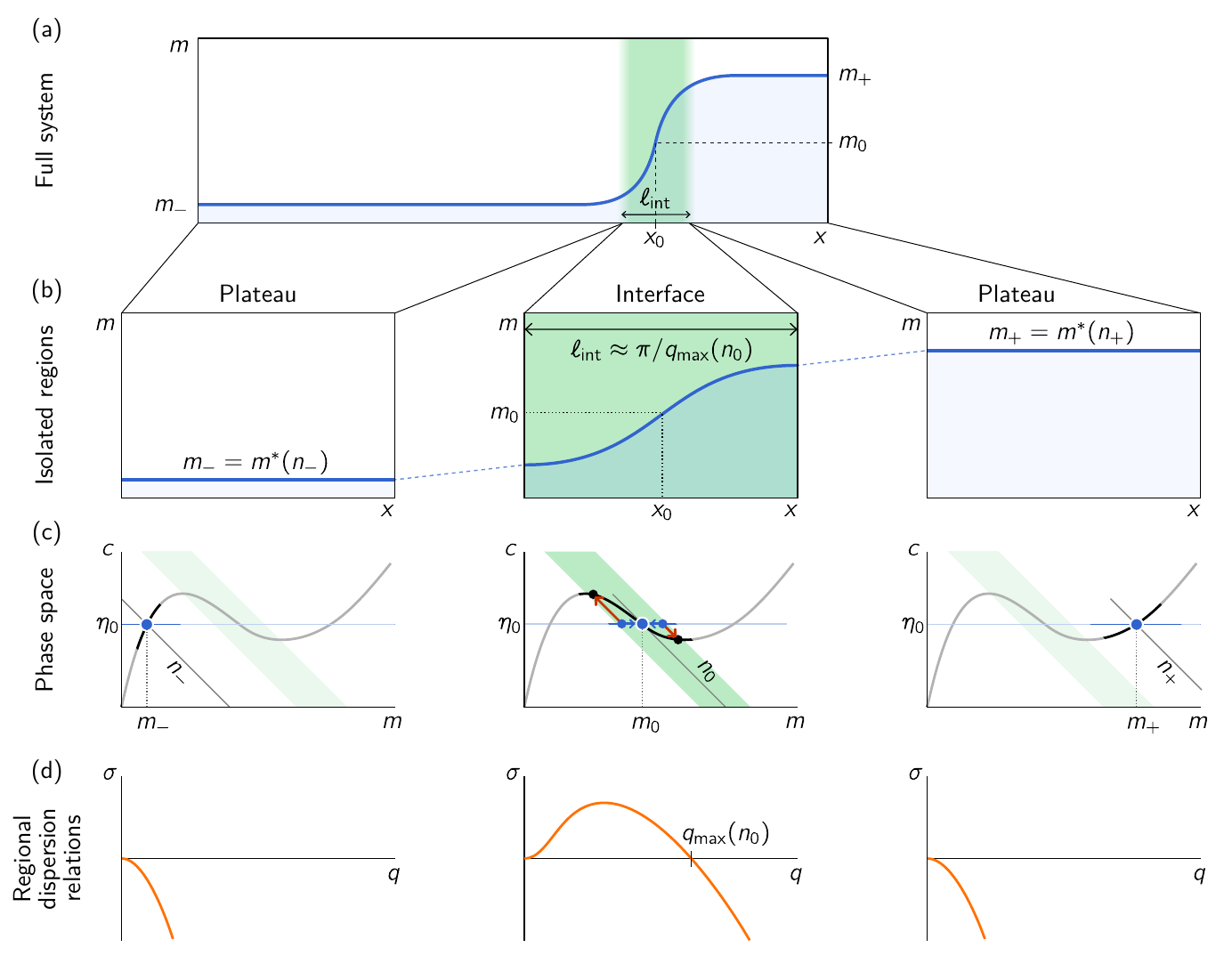}}
	\caption{
	Decomposition of a stationary pattern (a) into spatial regions that correspond to $(m,c)$-phase-space regions in the vicinity of landmark points.
	(b) Three characteristic spatial regions of the pattern (plateaus and the interface) can be notionally isolated.
	(c) The different average total densities $n_-$, $n_0$, and $n_+$ in three spatial regions determine the phase space regions corresponding to these spatial regions. The phase space region associated to laterally unstable nullcline segment is shaded in green.
	(d) Linearization of the reaction--diffusion dynamics around the reactive equilibria at $n_-$, $n_0$, and $n_+$ yields \emph{regional dispersion relations} that determine the properties of the regions. The plateaus are laterally stable regions, while the interface region is necessarily laterally unstable. The interface width can be estimated by the marginally stable mode $\qmax(n_0)$ at the right-hand edge of the dispersion relation of the interface region: $\Lint \simeq \pi/\qmax(n_0)$.
	}
	\label{fig:regions} 
\end{figure*}

Not only the interfaces ($n_0$) but also the plateaus ($n_\pm$) of patterns correspond to FBS-NC intersection points in phase space (Fig.~\ref{fig:regions}).
However, in contrast to the interface, the plateaus lie on laterally stable sections of the nullcline where $\snc(n_\pm) > -D_m/D_c$; recall the slope criterion for lateral instability Eq.~\eqref{eq:slope-criterion}.
Put more precisely, the pattern profile becomes flat in the vicinity of $n_\pm$ \emph{because} of regional lateral stability \cite{Note:Splitting}.
	\phantom{\cite{Ivanov:2010a,Crampin:2002a}}

Thus, the FBS-NC intersection points act as ``landmark points'' that enable us to (notionally) dissect the pattern profile into spatial regions (plateaus and interface) in such a way that these spatial regions can be associated with regional phase spaces.
The (linearized) properties of the reaction--diffusion dynamics---encoded in the \emph{regional dispersion relations} (Fig.~\ref{fig:regions}d)---in the vicinity of these landmark points can be used to determine the regional properties in real space. 
Within each of these regions, we can ask what would happen there if we were to isolate it from the rest of the system, akin to the question we asked in the context of local equilibria.
Just as the local equilibria scaffold the interface, these regional ``attractors'' serve as scaffolds for the global pattern.
The regional properties depend on the average regional mass which is redistributed between regions by diffusion, and the properties of the full pattern can be pieced together by (characteristically distinct) isolated regions (plateaus and interfaces).  

Taken together, the nonlinear kinetics is encoded by the nullcline shape. The internal properties of the spatial regions are determined by \emph{regionally linear} properties of phase space flow, encoded in the regional dispersion relations. 
We will therefore refer to this as the \emph{method of regional phase-spaces and regional attractors}.
This method bridges the gap between the linear and the highly nonlinear regime.

Finally, let us note that based on the region decomposition (cf.\ Fig.~\ref{fig:regions}), the interface position---determining the global spatial structure---can be pictured as a collective degree of freedom. 
A conceptually similar, but technically more involved approach to study interfaces (also called `kinks' or `internal layers') and their dynamics is singular perturbation theory (specifically matched asymptotic expansion) where one uses an asymptotic separation of spatial scales, see e.g.\ Ref.~\cite{Ward:2006a} and references therein. Such methods also facilitate a phase-space geometric analysis~\cite{Jones:1994a}.

\subsection{Pattern classification} \label{sec:pattern-types}

\begin{figure}
	\centerline{\includegraphics{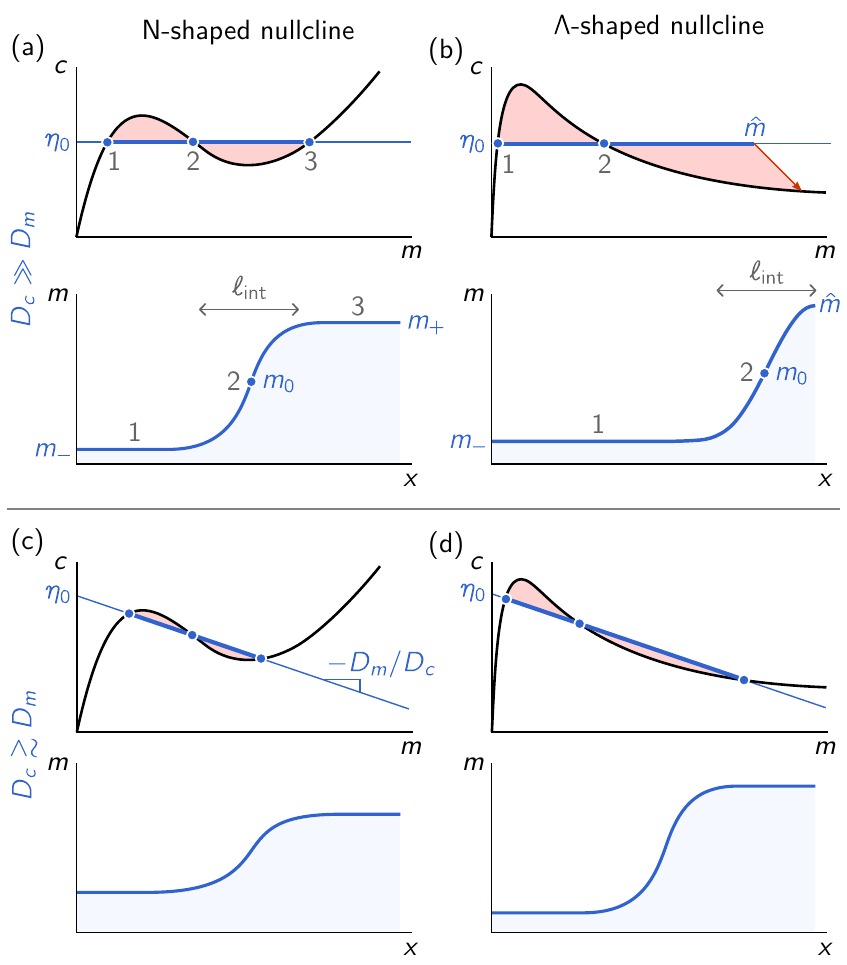}}
	\caption{
	Nullcline shape, and FBS-slope (diffusion constant ratio), and average total density determine wether a mesa or a peak pattern forms. Top row: fast cytosol diffusion; bottom row: slow cytosol diffusion. Each panel shows the pattern profile (top) and the respective phase portrait (bottom). 
	(a,c and d) Mesa patterns; the average total density, $\bar{n}$, determines the position of the interface; see Eq.~\eqref{eq:interface-pos-approx}.
	(c) A peak pattern forms if the pattern amplitude does not saturate in a third FBS-NC intersection point. The peak amplitude is determined by the average total density, $\bar{n}$, via the interface width $\Lint$; see Eq.~\eqref{eq:peak-approx}. As $\bar{n}$ is increased, the peak amplitude will grow, eventually reaching the third FBS-NC intersection point such that a mesa pattern forms.
	}
	\label{fig:profile-types}
\end{figure}
 
Employing the concept of regions we will now turn to the classification of patterns.
We distinguish two generic pattern types: mesas and peaks.
Mesa patterns are composed of plateaus (low density and high density) connected by an interface (Fig.~\ref{fig:profile-types}a, c and d), while the term peak refers to an interface concatenated to a plateau only on the low density site (Fig.~\ref{fig:profile-types}b) \cite{Note:Trough}.
For small systems, close to onset, there is an additional pattern type comprising only an interface that spans the whole system; see Sec.~\ref{sec:sub-supercrit}.

What are the conditions for the formation of a peak pattern versus the formation of a mesa pattern?
A mesa pattern requires two plateau regions, each characterized by an FBS-NC intersection point, one at low density and one at high density.
The low density plateau is generically present because the densities must be positive and thus are bounded from below. In contrast, the position of the FBS-NC intersection point at high density depends sensitively on the shape of the nullcline and the slope of the diffusive flux-balance subspace $-D_m/D_c$.
Let us first consider the case of fast cytosol diffusion $D_c \gg D_m$.
For an N-shaped nullcline, i.e.\ one that has an ``upwards-pointing'' tail (see Fig.~\ref{fig:profile-types}a), the flux-balance construction presented in Sec.~\ref{sec:scaffolding-introduction} yields a mesa pattern.
The situation is different for a ``$\Lambda$-shaped'' nullcline that has an asymptotically flat tail for large $m$ (e.g.\ approaching $c^*(m) \to 0$ for $m \to \infty$); see Fig.~\ref{fig:profile-types}b.
In that case the third FBS-NC intersection point generically is far away from the first two; in Fig.~\ref{fig:profile-types}b it lies out of frame.
The requirement of total turnover balance (approximated by a balance of the areas shaded in light red in the Fig.~\ref{fig:profile-types}) limits the maximum membrane concentration $\hat{m}$, such that there is no high density plateau and the pattern assumes a peak profile instead (Fig.~\ref{fig:profile-types}b bottom).
In the detailed analysis of peak patterns below, we will show that the FBS-position $\eta_0$, and thus the peak amplitude, is determined by the total mass in the system 

Let us now consider what happens if the FBS is made steeper by lowering the ratio of diffusion constants $D_c/D_m$.
As the FBS becomes steeper, the third FBS-NC intersection point moves towards lower membrane concentration. 
Eventually, this limits the total turnover on the right hand side of the inflection point, such that the peak amplitude saturates in a plateau, i.e.\ a mesa pattern forms (Fig.~\ref{fig:profile-types}d).
In the case of an N-shaped nullcline, lowering $D_c$ reduces the concentration difference between the two plateaus, because the FBS-NC intersection points move closer together (Fig.~\ref{fig:profile-types}c).

Taken together, the phase-plane analysis reveals how the interplay of nonlinear reactions (encoded in the nullcline shape) and diffusion (encoded in the FBS-slope) determine the pattern type and pattern amplitude.

\subsubsection{Mesa patterns}

To characterize mesa patterns in the limit $L \gg \Lint$, we first determine the FBS-position, $\eta_0$, using total turnover balance, Eq.~\eqref{eq:turnover-balance} (cf.\ Sec.~\ref{sec:scaffolding-introduction}). 
Since the plateaus are scaffolded by (laterally stable) local equilibria we can approximate the boundary concentrations
\begin{equation} 
\label{eq:plateau-approximation}
	\mstat(0) \approx m_-(\eta_0), \quad \mstat(L) \approx m_+(\eta_0),
\end{equation}
where the \emph{plateau scaffolds} $m_\pm(\eta_0)$ are  geometrically determined in phase space as intersection points $m_\pm(\eta_0)$ of FBS and NC:
\begin{equation}
\label{eq:plateau-intersections}
	 \tilde{f}(m_\pm,\eta_0)
	 =0. 
\end{equation}
With the approximation, Eq.\,\eqref{eq:plateau-approximation}, the total reactive turnover balance condition, Eq.~\eqref{eq:turnover-balance}, becomes:
\begin{equation} 
\label{eq:plateau-turnover-balance}
	\etaInfty : \, \integral{m_-^\infty}{m_+^\infty}{m} \tilde{f}(m,\etaInfty) = 0,
\end{equation}
where $m_\pm^\infty = m_\pm(\etaInfty)$ and $\etaInfty$ denotes the FBS-position in the large system size limit. 
Equation~\eqref{eq:plateau-turnover-balance} is closed and can be solved for $\etaInfty$. 
Once one has determined $\etaInfty$, the interface width $\Lint(\etaInfty)$ can be estimated with Eq.~\eqref{eq:Lint-approx}.

This \emph{total turnover balance} condition implicitly determines the FBS-offset $\eta_0 = \etaInfty$. 
Note that equation Eq.~\eqref{eq:plateau-turnover-balance}, and hence $\etaInfty$ depends only on the function $f$ and the ratio of the diffusion constants, but not on the average mass $\bar{n}$ or the domain size $L$ in the limit $L \gg \Lint$.
Instead, the average total density $\bar{n}$ determines the position $x_0$ of the interface. 
Again assuming an interface much narrower than the domain size, the contribution of the interface region can be neglected, $L \bar{n} \approx n_-^\infty x_0 + n_+^\infty (L-x_0)$, which yields
\begin{equation} 
\label{eq:interface-pos-approx}
	x_0 \approx L \frac{n_+^\infty - \bar{n}\;}{\;n_+^\infty - n_-^\infty},
\end{equation}
where $n_+^\infty$ and $n_-^\infty$ are the average total densities in the plateau regions:
\begin{equation} 
\label{eq:plateau-masses}
	n_\pm^\infty := \etaInfty + (1-D_m/D_c) \, m_\pm(\etaInfty).
\end{equation}
This shows that the amplitude of mesa patterns is geometrically determined by the reactive nullcline alone and does not sensitively depend on average mass $\bar{n}$ or system size $L \gg \Lint$. Moreover, far away from the critical point $D_c^\mathrm{min}$, cf.\ Sec.~\ref{sec:n-Dc-monostable} the mesa-pattern amplitude becomes independent of the ratio of the diffusion constants. Adding mass to a mesa pattern shifts the interface position $x_0$ as the additional mass is redistributed between the two plateau regions.

Notably, a geometric argument shows that mesa patterns are the generic pattern for $L \to \infty$ when the ratio of the diffusion constants is nonzero $D_m/D_c > 0$, and $m \geq 0$, $c \geq 0$ (as must be the case for concentrations): 
The FBS intersects the $m$-axis  ($c = 0$) at $(D_c/D_m)\eta_0$, and hence must intersect the nullcline at some finite value $m < (D_c/D_m)\eta_0$. 
For $L \to \infty$ keeping the average mass $\bar{n}$ constant, the pattern profile will eventually reach this third FBS--NC intersection point, and thus become a mesa pattern.
Next, we will discuss the conditions under which peak/trough patterns occur.

The approximation Eq.~\eqref{eq:plateau-approximation} for the plateau densities, and in turn also Eq.~\eqref{eq:interface-pos-approx} for the interface position, will break down when the distance of the interface to one of the system boundaries becomes smaller than the interface width $\Lint(\etaInfty)$. 
Then the stationary pattern no longer exhibits a plateau on that side and instead becomes a plateau--interface pattern, forming either a peak when $\bar{n}$ is close to $n_-^\infty$, or a trough (`anti-peak') when $\bar{n}$ is close to $n_+^\infty$. 
An estimate for these transition from mesa to peak/trough patterns can be obtained based on the approximated interface position, Eq.~\eqref{eq:interface-pos-approx}:
\begin{equation} \label{eq:peak-mesa-transition}
	L \, \big|\bar{n} - n_\pm^\infty\big| \lesssim \Lint(\etaInfty) \, \big(n_+^\infty - n_-^\infty\big).
\end{equation}

\subsubsection{Peak patterns}

Let us now study these peak/trough patterns. 
Their defining characteristic is that a plateau, corresponding to laterally stable FBS-NC intersection point, forms only only on one side of the interface. Correspondingly, the reactive turnover saturates on the side where the plateau forms, while it depends on the variable pattern amplitude on the other side.
For specificity, we focus on peak patterns here.
As explained above, such a peak pattern forms when the nullcline is $\Lambda$-shaped, flux-balance subspace is very shallow ($D_c \gg D_m$); see Fig.~\ref{fig:profile-types}b.

Suppose for a moment that we can freely choose the FBS-position, $\eta_0$, and that the total mass is not fixed. 
Given some $\eta_0$, the FBS-NC intersection point (1) determines the low-density plateau at the foot of the interface and thus the total turnover on this side, corresponding to the enclosed area in the interval between (1) and (2).
This turnover must be balanced by an equal and opposite turnover on the right. Using again the (approximate) correspondence to the enclosed phase-plane area, it becomes obvious that this balance of areas determines the peak amplitude $\hat{m}$ (3).
Using that the interface profile can be approximated as $\propto \sin((x - x_0)/\Lint)$, where $\Lint$ is determined by the steady state equation linearized around the pattern inflection point (see Sec.~\ref{sec:interface-width}), we can roughly estimate the total mass in a peak as
\cite{Note:RollingBallPeak}
\begin{equation} 
\label{eq:peak-approx}
	N_\text{peak}(\eta_0) 
	\approx \frac{1}{2} \Lint(\eta_0) \, [\hat{m} - m_-(\eta_0)].	
\end{equation}
The sinusoidal shape of the interface furthermore mandates that the inflection point $m_0$ is approximately half-way between the plateau $m_-$ and the maximum $\hat{m}$, such that we can eliminate $\hat{m} \approx m_- +2\, (m_0 - m_-)$ in Eq.~\eqref{eq:peak-approx}. 
The remaining unknowns $m_0$ and $m_-$ are determined geometrically (FBS--NC intersections) as functions of $\eta_0$.
Thus we obtain a relation for the average total density $\bar{n}(\eta_0) \approx n_-(\eta_0) + N_\text{peak}(\eta_0)/L$ as a function of $\eta_0$. 
The inverse of this relation yields the FBS-position $\eta_0(\bar{n})$ as a function the control parameter $\bar{n}$. 
This estimate will hold until the peak density $\hat{m}$ reaches the third FBS-NC intersection point $m_+(\eta_0)$, where a second plateau will start to form, such that the peak pattern transitions to a mesa pattern.
In Appendix~\ref{app-sec:peak-patterns} we present the details of the peak approximation, and a comparison to numerical solutions.

Our estimate for the peak mass Eq.~\eqref{eq:peak-approx} and the resulting relation $\eta_0(\bar{n})$ show that, in contrast to mesa patterns, the amplitude of peak patterns sensitively depends on the total mass $N = L \, \bar{n}$ and the membrane diffusion constant (via $\Lint^2 \sim D_m/f_m$, cf.\ Eq.~\eqref{eq:Lint-approx}). 
In addition, the position of the third FBS-NC intersection point $m_+(\eta_0)$ that limits the maximum peak density, sensitively depends on the FBS slope $-D_m/D_c$.
In the limit $D_m/D_c \to 0$, the third FBS-NC intersection point $m_+$ moves to infinity. Hence, in this limit, a system with an asymptotically flat nullcline tail never exhibits mesa patterns.

\subsubsection{More general nullcline shapes}

Here we considered two types of nullcline shapes---N- and $\Lambda$-shaped---that both have a single maximum in the $(m,c)$-phase plane, but differ in their `tail behavior'.
Beyond these two prototypical nullcline shapes, more general nullcline shapes are possible. For instance, reaction kinetics of the attachment--detachment form, Eq.~\eqref{eq:attachment--detachment-kinetics}, with higher order nonlinearities (e.g.\ 5th-order polynomials) may exhibit nullclines with multiple maxima in the $(m,c)$-phase plane. For general reaction kinetics $f(m,c)$, more exotic shapes of the nullclines, e.g.\ with multiple disconnected branches, are possible. Our findings equally apply to such nullclines since our phase-space analysis is based on simple geometric properties such as slopes and intersection points with the FBS. Additional care is required if $f_c$ changes sign along the nullcline, since the slope criteria for local and lateral stability, Eqs.~\eqref{eq:local-stability-slope-criterion} and~\eqref{eq:slope-criterion}, are reversed for $f_c < 0$.
Conveniently, for reaction kinetics of the attachment--detachment form, Eq.~\eqref{eq:attachment--detachment-kinetics}, one has $f_c = a(m)$ which is generally positive for systems of biochemical origin.

\subsection{Generic bifurcation structure under variation of the average mass $\bar{n}$} \label{sec:n-bifurcations}

\begin{figure}
	\centerline{\includegraphics{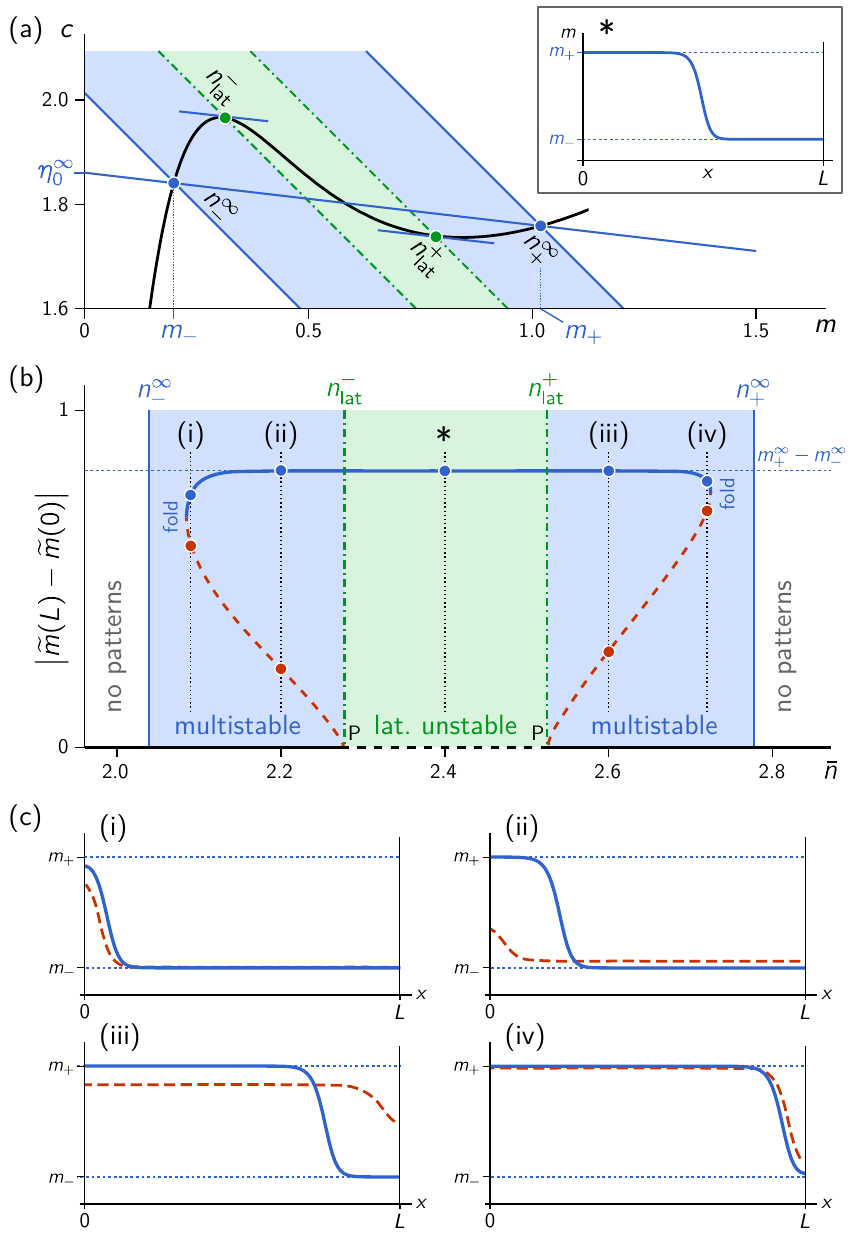}}
	\caption{
	Bifurcations of mesa patterns in the large system size limit ($L \to \infty$) can be constructed geometrically using the reactive nullcline.
	(a) Geometric construction of pattern bifurcation points for an example two-component system: Eq.~\eqref{app-eq:wave-pinning} with $k = 0.07$, $D_m=1$, and $D_c= 10$. The laterally unstable regime (shaded in green) is delimited by Turing bifurcations where the FBS is tangential to the NC (green dots, $n_\pm^\text{lat}$). FBS-NC intersection points (blue dots, $n_\pm^\infty$), delimit the range of pattern existence (shaded in blue), where the FBS-position, $\etaInfty$, is determined by global turnover balance, Eq.~\eqref{eq:turnover-balance}. Inset in the top right corner: membrane density $\mstat(x)$ of a stable mesa pattern for $\bar{n} = 2.4$ (see star in (b)).
	(b) Bifurcation structure of the pattern amplitude $|\mstat(L)-\mstat(0)|$ for the control parameter $\bar{n}$ obtained by numerical continuation (cf.\ Appendix~\ref{app-sec:continuation}) for a system size of $L = 100$. The branch of stable patterns (solid blue line) and the branches of unstable patterns (dashed red line) meet in fold bifurcations of patterns. Due to the finite system size, these fold bifurcations are offset from $n_\pm^\infty$ (vertical solid blue lines) by an amount ${\sim}\Lint(\etaInfty)/L$. The unstable patterns emerge in subcritical pitchfork bifurcations (P) from the homogeneous steady state (black line) at the Turing bifurcations (vertical dash-dotted green lines). 
	(c) Profiles $\mstat(x)$ of stationary patterns (solid blue: stable, dashed red: unstable) for the average total densities $\bar{n} = 2.09, 2.2, 2.6$, and $2.72$ (see thin dashed lines in (b)). The plateau scaffolds $m_\pm(\etaInfty)$ are shown as thin black lines. 
	}
	\label{fig:n-bifurcation-mesa-patterns}
\end{figure}

Now that we have classified the different types of stationary patterns exhibited by 2C-MCRD systems, we turn to study bifurcations where the patterns change structurally or in stability. 
The bifurcation parameter we study first is the average total density $\bar{n}$. This parameter does not affect the phase-space geometry (NC and FBS), which makes it particularly easy to study. 
Later, in Sec.~\ref{sec:bifurcation-structure}, we generalize our findings to bifurcation parameters that change the phase-space geometry: diffusion constants change the FBS-slope, whereas kinetic rates affect the nullcline shape. 
For biological systems, the average total density $\bar{n}$ is a natural parameter as it can be tuned by up- or down-regulating the production of a protein.

Let us begin with the bifurcations where the homogenous steady state becomes laterally unstable. 
We already learned in Sec.~\ref{sec:lateral-instability} that there is a band of unstable modes, $[0,\qmax]$, if the NC-slope $\snc(\bar{n})$ is negative and steeper than the FBS-slope, $-D_m/D_c$ (cf.\ Eq.~\eqref{eq:slope-criterion} and Fig.~\ref{fig:MRI}a). 
Hence, a band of unstable modes exists if $\bar{n}$ is in the range $(\nLat^-,\nLat^+)$, bounded by the points $\nLat^\pm$ where the flux-balance subspace is tangential to the reactive nullcline (dash-dotted green lines in Fig.~\ref{fig:n-bifurcation-mesa-patterns}a.
(Note that a system of finite size $L$, is unstable if the longest wavelength mode lies in the band of unstable modes $\pi/L < \qmax(\bar{n})$, where $\qmax^2 = \tilde{f}_m/D_m$, as defined in Eq.~\eqref{eq:qmax} and $\tilde{f}_m = f_m - f_c \, D_m/D_c = (-D_m/D_c - \snc) f_c$.)

What about the range where stationary patterns exist? 
The plateau scaffolds $m_\pm(\eta_0)$ are geometrically determined by the reactive nullcline via the FBS--NC intersection points. The position $\eta_0$ of the flux-balance subspace generally depends on $\bar{n}$ and $L$ via total turnover balance, Eq.~\eqref{eq:turnover-balance}. 
However, in the large system size limit ($L \to \infty$), the FBS position $\etaInfty$ is independent of $\bar{n}$ and $L$ (cf.\ Eq.~\eqref{eq:plateau-turnover-balance}). 
For patterns to exist, the average total density $\bar{n}$ must lie in-between the plateau densities $n_\pm^\infty$; see Fig.~\ref{fig:n-bifurcation-mesa-patterns}a, cf.~Eq.~\eqref{eq:interface-pos-approx}. 
Hence, in the limit $L \to \infty$, stationary patterns exist in the range $n_-^\infty < \bar{n} < n_+^\infty$.

Importantly the range of pattern existence generically extends beyond the range of lateral instability ($\nLat^- > n_-^\infty$ and $\nLat^+ < n_+^\infty$) by geometric necessity for N-shaped nullclines; see Fig.~\ref{fig:n-bifurcation-mesa-patterns}a. 
This implies, that generically there are regions of multistability in parameter space, where stable stationary patterns exist, and the homogeneous steady state is stable (regions shaded in blue in Fig.~\ref{fig:n-bifurcation-mesa-patterns}).

To gain some intuition on the steady states in the multistable regimes, we performed numerical continuation (see Appendix~\ref{app-sec:continuation} for details) of the stationary patterns for an example 2C-MCRD system using the attachment--detachment kinetics Eq.~\eqref{app-eq:wave-pinning} from Ref.~\cite{Mori:2008a} which exhibit an N-shaped nullcline. 
Figure~\ref{fig:n-bifurcation-mesa-patterns}b shows the numerically obtained bifurcation structure where we plot the pattern amplitude $|\mstat(L) - \mstat(0)|$ against the bifurcation parameter $\bar{n}$. 
The star marks a typical stable mesa pattern (see inset in Fig.~\ref{fig:n-bifurcation-mesa-patterns}a) in the central region of the branch of stable patterns (solid blue line). 
As the plateaus are scaffolded by the FBS-NC intersections $m_\pm(\etaInfty)$, the pattern amplitude stays approximately constant ($m_+{-}m_-$, dotted blue line) across the whole range of $\bar{n}$ where patterns exist. 
Changing total average density simply shifts the interface position (cf.\ panels $(ii)$ and $(iii)$ in (c)). 
When the interface position is in the vicinity of a boundary, mesa pattern transitions to peak patterns (see. panels $(i)$ and $(iv)$ in (c)) as we learned in the previous section (Eq.~\eqref{eq:peak-mesa-transition} in Sec.~\ref{sec:pattern-types}). 
The numerical continuation shows, that the peak/trough patterns are then annihilated in saddle-node bifurcations (SN), where the branch of stable patterns meets a branch of unstable patterns (dashed, red line). 
Due to the finite system size, the exact positions of the SN-bifurcation points are sightly offset from $n_\pm^\infty$ (by an amount ${\sim} \, \Lint/L$). The branches of unstable patterns emerge from homogenous steady state in subcritical pitchfork bifurcations (P) at the Turing bifurcations ($\nLat^\pm$). 
(In a finite sized system, the onset of lateral instability is offset by an amount ${\sim}\,L^2$ from the geometrically defined points $\nLat^\pm$, because the system is unstable only if the longest wavelength mode lies within the band of unstable modes; see Sec.~\ref{sec:sub-supercrit}). 
In the multistable regions (shaded in blue), patterns can be triggered by a finite amplitude perturbation. 
The unstable patterns are ``transition states'' (or ``critical nuclei'') that lie on the separatrix separating the basins of attraction of the stable patterns and the stable homogeneous steady state. 
The actual separatrix is a complicated object in the high-dimensional PDE phase space. 
In the next section (\ref{sec:stim-induced}), we will show that a heuristic can be inferred from the nullcline shape to estimate the threshold for stimulus-induced pattern formation for a prototypical class of spatial perturbation profiles.

Because the unstable patterns are peak/trough patterns, they can be approximated by the `peak approximation' introduced in Sec.~\ref{sec:pattern-types} (see also Appendix~\ref{app-sec:peak-patterns}). 
Thus the qualitative structure of the branches of unstable patterns is determined by $(m,c)$-phase-space geometry independently of the details of the reaction term $f(m,c)$, as long as the reactive nullcline $f(m,c) = 0$ is N-shaped.

In summary, we conclude that the qualitative form of the bifurcation structure shown in Fig.~\ref{fig:n-bifurcation-mesa-patterns}b is determined by geometric relations in $(m,c)$-phase space. 
In particular, we find that 2C-MCRD systems generically have regions of multistability and that the onset of lateral instability is generically subcritical for large system size $L \gg \Lint$. 
Further, this implies that such systems exhibit stimulus-induced pattern formation, and that there is hysteresis of stationary patterns when the total average density is varied.

\section{Perturbation threshold for stimulus-induced pattern formation} \label{sec:stim-induced}

\begin{figure*}
	\centerline{\includegraphics{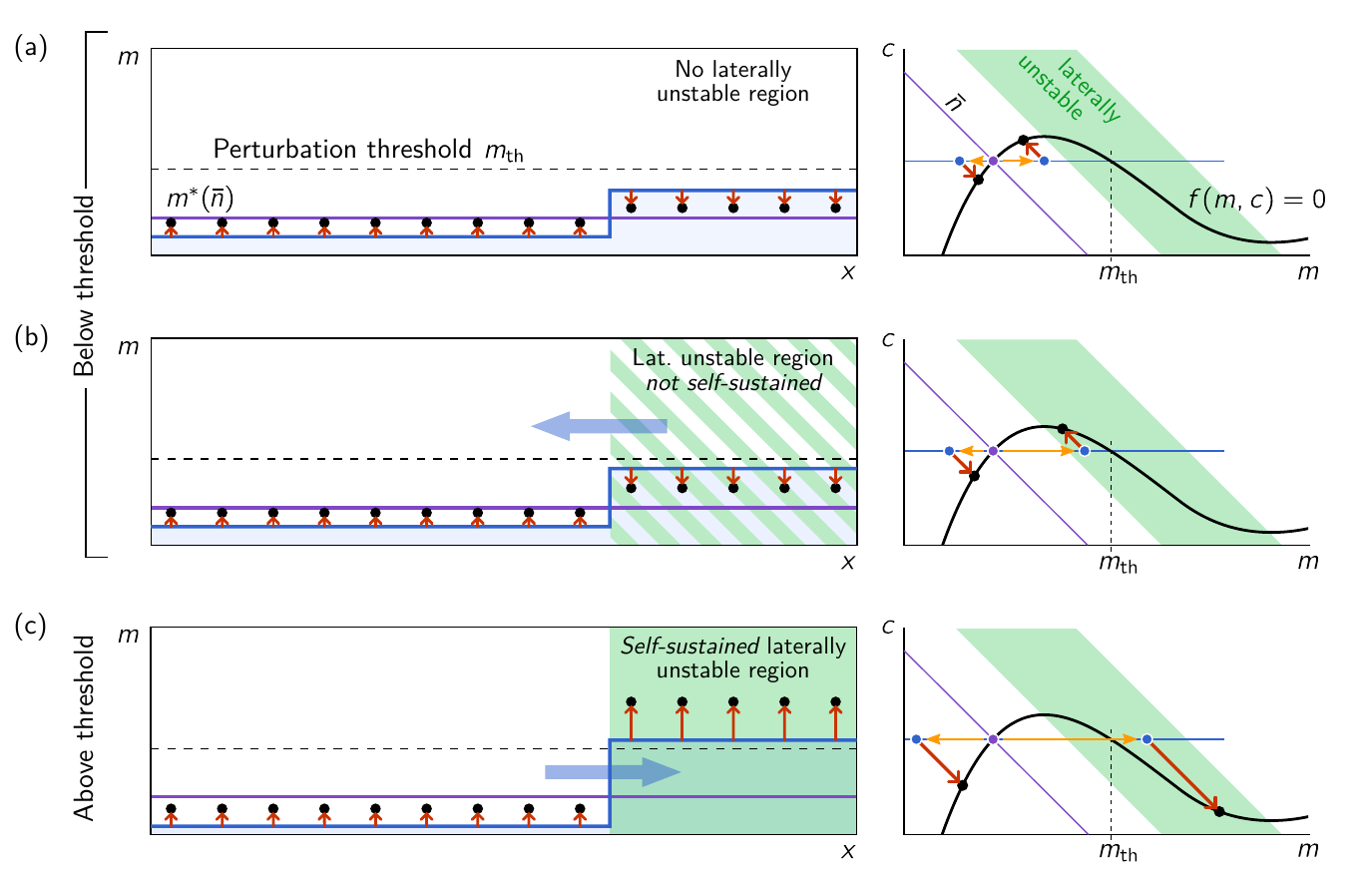}}
	\caption{
	Subcritical stationary patterns can be induced by perturbations above a threshold that can be heuristically estimated from the reactive nullcline.
	(a) After a small perturbation (blue profile, yellow arrows in phase space), that does not induce a laterally unstable region, the system returns to its uniform steady state (purple line). 
	(b) A perturbation that creates a laterally unstable region, but does not cross the nullcline. Because the cytosolic concentration is lower in the laterally stable region, mass-redistribution (illustrated by the blue arrow) will disband the laterally unstable region.  
	(c) A perturbation that crosses the nullcline will not only induce a laterally unstable region but also shifts the cytosolic equilibrium concentration in this region such that the lateral instability is \mbox{(self-)}sustained by mass redistribution from the stable into the unstable region (blue arrow). 
	}
	\label{fig:stim-induced} 
\end{figure*}

Before we delve into the more technical analysis of bifurcation structures, we would like to discuss one more important aspect of pattern formation: stimulus-induced pattern formation, i.e.\ the ability to induce the transition from one stable attractor (homogeneous steady state) to another (stationary pattern) by a large enough perturbation (stimulus). 
(In the context of phase separation, this is called \emph{nucleation and growth}).
Stimulus-induced pattern formation is a particularly important aspect of 2C-MCRD systems, because, as we have shown above, these systems generically have regions of multistability. 
Furthermore, biologically it is often desirable to be able to form a pattern following an external or internal stimulus that exceeds a certain threshold (``nucleation threshold'') . 
As of yet, this threshold could only be determined numerically~\cite{Jilkine:2011a}. 
In the following, we will show how simple heuristic reasoning---based on \emph{regional} lateral instability---yields a geometric criterion for the perturbation threshold in the $(m,c)$-phase plane.

As we have shown in the previous section, the hallmark of a stationary pattern is a laterally unstable region surrounding the pattern inflection point $x_0$ (even if the homogeneous state of the system is laterally stable). 
In the proposed framework, the phase-space dynamics are simply represented by the expansion of the system in the $(m,c)$-phase plane due to mass redistribution.
Hence, to lead to a stationary pattern, a trajectory in the (high-dimensional) phase space of a partial differential equation (PDE) must enter and remain in a (linearly) laterally unstable region in the $(m,c)$-phase plane (shaded in green in Fig.~\ref{fig:stim-induced}). The laterally unstable region in $(m,c)$-space corresponds to a respective region in real space.
If the homogenous state is laterally stable then a finite perturbation (stimulus) is required to create a laterally unstable region. 
Let us study a prototypical perturbation able to induce a laterally unstable region: 
a step function that represents moving a `block' of protein mass (total density) from one end of the system to the other; 
for an illustration of the spatial perturbation and the resulting flows in phase space see Fig.~\ref{fig:stim-induced}.
Generalization to other perturbations is straightforward and based on analogous arguments. 
Such perturbations can be created by various means of `active' mass redistribution, e.g.\ active transport in the cell cortex, along microtubules, and hydrodynamic cytosolic flows; see for instance Ref.~\cite{Goehring:2011a}.

Following a (large amplitude) perturbation, there are two distinct processes that are triggered in phase space as shown in Fig.~\ref{fig:stim-induced}.  
On the one hand, in the laterally unstable region (green shaded area), a mass-redistribution instability will start to form a pattern, thus further amplifying the perturbation. 
On the other hand, because the perturbation shifts the regional reactive equilibria (black disks), there will be reactive flows (red arrows) in the regions that induce a cytosolic gradient which leads to mass redistribution between the regions by cytosolic diffusion (large blue arrows). 
If the cytosolic density in the laterally stable region is lower than in the laterally unstable one, the regional instability may not be sustained and the system returns to homogenous steady state (Fig.~\ref{fig:stim-induced}b). 
Conversely, if the cytosolic density is lower in the laterally unstable region than in the laterally stable region, then the cytosolic flow between the regions (blue arrow) will sustain the regional instability (Fig.~\ref{fig:stim-induced}c). 
Because the mass-redistribution instability creates a self-organized and self-sustaining cytosolic sink, the laterally unstable region can be self-sustained. 
The heuristic criterion for a \mbox{(self-)sus}\-tained laterally unstable region is that the perturbation must cross the nullcline (see Fig.~\ref{fig:stim-induced}c). 
Then, the overall cytosolic concentration in the laterally unstable region is decreased by reactive flows (red arrows) such that cytosolic diffusion (blue arrow) between the regions will sustain the laterally unstable region.

In Appendix~\ref{app-sec:stim-induced} we show that this simple criterion already provides a very good approximation for the threshold in comparison to full numerical simulation.
We conclude that the reactive nullcline provides the key information for understanding pattern formation dynamics, in a similar way as for the characterization of stationary patterns (Sec.~\ref{sec:pattern-characterization} and the analysis of the linear mass-redistribution instability (Sec.~\ref{sec:lateral-instability}). 
Specifically it enables one to estimate the basins of attraction of the uniform steady state and the polarized pattern. 
We further learned that regional lateral instability underlies stimulus-induced pattern formation from laterally stable homogeneous steady states.

The threshold estimate provided here might help to understand this ``nucleation'' of patterns from laterally stable homogeneous steady states. The unstable peak/trough patterns (dashed red lines in Fig.~\ref{fig:n-bifurcation-mesa-patterns} are part of the separatrix between to the basin of attraction of stable stationary patterns, and can be pictured as canonical critical nucleus \cite{Bates:1993a}. The peak approximation described in Sec.~\ref{sec:pattern-types} and compared to numerical continuation in Appendix~\ref{app-sec:peak-patterns} provides a simple estimate for this critical nucleus.

\section{Complete bifurcation structure} \label{sec:bifurcation-structure}

Bifurcation diagrams of 2C-MCRD systems were previously studied for specific choices of the reaction kinetics $f(m,c)$ using numerical methods~\cite{Mori:2011a}. 
Furthermore, based on numerical studies of various models, it was hypothesized that there might be a general bifurcation scenario underlying cell polarity systems~\cite{Trong:2014a}. 
Here we use the insight gained on phase-space geometry to systematically build the complete general bifurcation structure of 2C-MCRD systems. 
Our findings generalize previous results and unify them in the context of phase-space geometry. 
For large system size, the bifurcation structures are fully determined geometrically. 
We illustrate the effect of finite system size using numerically computed bifurcation diagrams shown in Appendix~\ref{app-sec:continuation}. 

Above, we studied the bifurcation diagram of stationary patterns for the bifurcation parameter $\bar{n}$ in a system with monostable kinetics; see Sec.~\ref{sec:n-bifurcations} and Fig.~\ref{fig:n-bifurcation-mesa-patterns} therein. 
Recall, that in large systems ($L \to \infty$), the bifurcation points in $\bar{n}$ can be found based on geometric reasoning in phase space:
(\textit{i}) Lateral instability is identified by a criterion on the nullcline slope: $\snc(\bar{n}) < -D_m/D_c$. 
Hence, the range of lateral instability is bounded by points $\nLat^\pm$ where the FBS is tangential to the NC: $\snc(\nLat^\pm) = -D_m/D_c$. 
(\textit{ii}) FBS--NC intersection points $m_\pm(\etaInfty)$ provide the scaffold for the plateaus of mesa patterns, where the FBS-position, $\etaInfty$, is determined by total turnover balance, Eq.~\eqref{eq:plateau-turnover-balance}. 
Mesa patterns exist as long as the average total density can be distributed between two plateaus $n_\pm^\infty$, i.e.\ in the range $n_-^\infty < \bar{n} < n_+^\infty$; 
recall that $n_\pm^\infty = n_\pm(\etaInfty)$ depend on the position $\etaInfty$ and slope $-D_m/D_c$ of the FBS (cf.\ Eq.~\eqref{eq:plateau-masses}).

Both of these geometric bifurcation criteria depend on the diffusion constants via the slope of the flux-balance subspace $-D_m/D_c$. We keep $D_m$ fixed---thus fixing the smallest characteristic length scale $\ell = \sqrt{D_m/f_m}$ where spatial structures can be maintained against membrane diffusion---and vary $D_c$ to rotate the FBS in $(m,c)$-phase space.
 
\subsection{Generic bifurcation structure of stationary patterns for monostable reaction kinetics} \label{sec:n-Dc-monostable}

\begin{figure}
	\centerline{\includegraphics{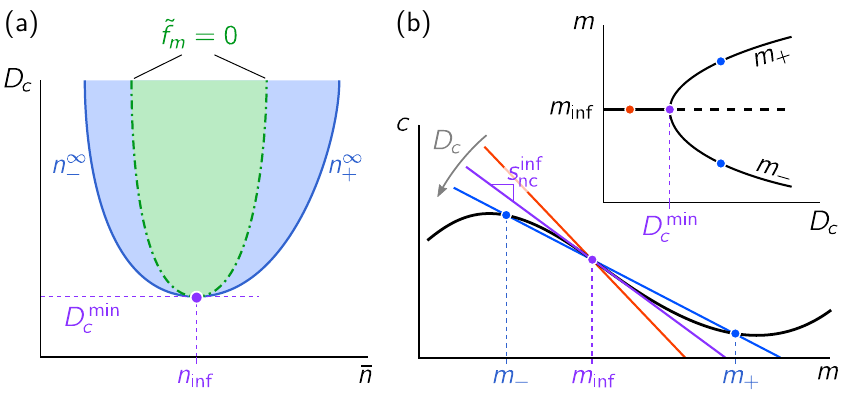}}
	\caption{
	$(\bar{n},D_c)$-bifurcation diagram of stationary patterns for a system with monostable kinetics (same color code as in Fig.~\ref{fig:n-bifurcation-mesa-patterns}).
	(a) The bifurcation diagram for a large system ($L \to \infty$) is obtained by tracking the geometrically constructed bifurcation points $\nLat^\pm$ and $n_\pm^\infty$ as $D_c$, and thus the slope and position of the FBS, are varied (cf.\ Fig.~\ref{fig:n-bifurcation-mesa-patterns}).  The onset of lateral instability (Turing bifurcation shown as green dash-dotted line) is generically subcritical since there exist stationary patterns outside the range of lateral instability $(\nLat^-,\nLat^+)$; in the blue regions, the system is multistable (both the stationary patterns and the homogenous steady state are stable). The scaffolds for the low and high density plateaus ($n_\pm^\infty$) bifurcate supercritically from the homogenous steady state at the critical point $(\nInf,D_c^\text{min})$ (purple point). 
	(b) The critical point in the $(\bar{n},D_c)$-bifurcation diagram corresponds to the inflection point of the nullcline $\nInf$, where the nullcline slope $-f_m/f_c$ reaches its extremal value $\snc^\text{inf}$ and thus determines the minimal cytosolic diffusion $D^\text{min}_c$ (purple line); cf.\ Eq.~\eqref{eq:Dc_min}. At $D^\text{min}_c$, the scaffolds of the plateaus, $m_\pm$, bifurcate in a supercritical pitchfork bifurcation from the nullcline inflection point $m_\text{inf}$ (see inset).
	}
	\label{fig:n-Dc-diagram-monostable}
\end{figure}

We construct the $(\bar{n},D_c)$-bifurcation diagram by inferring $\nLat^\pm$ and $n_\pm^\infty$, as described above, as functions of $D_c$. 
 Qualitatively, this can even be done manually with pen and paper in the spirit of a graphical construction (see e.g.\ Ref.~\cite{Strogatz:Book}) based on the geometric criteria (\textit{i}), (\textit{ii}) above, as shown in Fig.~\ref{fig:n-bifurcation-mesa-patterns}. 
 Figure~\ref{fig:n-Dc-diagram-monostable}a shows the qualitative structure obtained by this graphical construction. 
 A quantitative construction of the bifurcation diagram can be performed with simple numerical implementation of the bifurcation criteria described above, e.g.\ in Mathematica (see SM\ File: ``flux-balance-construction.nb'', and Fig.~\ref{app-fig:geometry-bifurcations} for figures of quantitative bifurcation structures). 
 As we will see in the following, the structure of the bifurcation diagram is qualitatively the same for all monostable, N-shaped nullclines, independently of the details (nonlinearities and kinetic rates) of the reaction term $f(m,c)$. 
 The bifurcation diagram is qualitatively different when the nullcline has a segment of bistability (where $\snc < -1$, cf.\ Fig.~\ref{fig:total-density-bifurcation-structure}). 
 We will analyze this case and in particular the role of bistability further below in Sec.~\ref{sec:n-Dc-bistable}.

As $D_c$ is decreased, the flux balance subspace becomes steeper, and thus the bifurcation points $\nLat^\pm$ and $n_\pm^\infty$ start to converge (see Fig.~\ref{fig:n-Dc-diagram-monostable}; cf.~Fig.~\ref{fig:n-bifurcation-mesa-patterns}). 
They meet in the inflection point of the reactive nullcline, $\nInf$, where the nullcline slope, $\snc^\text{inf}$, is extremal ($\partial_n \snc |_{\nInf} = 0$). 
The extremal nullcline slope at the nullcline inflection point determines the minimal cytosolic diffusion constant,
\begin{equation} 
\label{eq:Dc_min}
	D_c^\text{min} := \frac{D_m}{-\snc(\nInf)},
\end{equation}
above which there are three FBS--NC intersection points. 
When the `critical' point $(\nInf, D_c^\text{min})$ is traversed in $D_{\kern-0.075em c}\kern0.075em$-direction, the FBS--NC intersections bifurcate in a (supercritical) pitchfork bifurcation; 
see Fig.~\ref{fig:n-Dc-diagram-monostable}b.
Since the FBS--NC intersection points $m_\pm$ are the scaffolds for the plateaus (in short: \emph{plateau scaffolds}; 
cf.\ Fig.~\ref{fig:stat-pattern-and-phase-space}), this bifurcation at the critical point $(\nInf, D_c^\text{min})$ is a bifurcation of the scaffold itself. Importantly, the actual pattern is bounded by the plateau scaffolds. 
Thus, if there are no plateau scaffolds (i.e.\ only one FBS-NC intersection point), there cannot be stationary patterns.
For $L \to \infty$, patterns emerge slaved to the plateau scaffold, such that the pattern bifurcation is supercritical at the nullcline inflection point ($\bar{n} = \nInf$). 
Away from the nullcline inflection point ($n \neq \nInf$), the lateral instability bifurcation is always subcritical for $L \to \infty$ because the range $(n_-^\infty,n_+^\infty)$ where patterns exist always exceeds the range $(\nLat^-,\nLat^+)$ of lateral instability, as we learned above in Sec.~\ref{sec:n-bifurcations} (cf.\ Fig.~\ref{fig:n-bifurcation-mesa-patterns}).

As we will see below in Sec.~\ref{sec:sub-supercrit}, for finite $L$, the bifurcation is supercritical in the vicinity of the nullcline inflection point. The transition from super- to subcriticality depends on a subtle interplay of diffusive and reactive flow together with geometric factors like nullcline curvature.

Interestingly, the regimes and their interrelation in the $(\bar{n},D_c)$-bifurcation diagram, as shown in Fig.~\ref{fig:n-Dc-diagram-monostable}a, are phenomenologically similar to the phase diagram of (near equilibrium) phase separation kinetics for binary mixtures, described by Cahn--Hilliard equation \cite{Cates:2018a}. 
In a previous study using the amplitude equation formalism, Ref.~\cite{Bergmann:2018a}, a mapping from 2C-MCRD models to Model~B has been found for the vicinity of the critical point, where the pattern emerges from the Turing bifurcation in a supercritical or weakly subcritical pitchfork bifurcation (see Sec.~\ref{sec:sub-supercrit}).

Strikingly, our geometric reasoning shows that the physics implied by the bifurcation diagram is the same as in phase separation kinetics (binodal and spinodal regimes) \emph{for all N-shaped nullclines}, and far away from the critical point.
We discuss this finding in Sec.~\ref{sec:discussion-non-equilibrium}.

\subsection{Locally bistable kinetics} \label{sec:n-Dc-bistable}

Changing the kinetic rates deforms the nullcline shape. When the nullcline slope becomes smaller than $-1$, a regime of locally bistable reaction kinetics emerges (cf.\ Fig.~\ref{fig:total-density-bifurcation-structure}). 

\subsubsection{Fronts in bistable media}
\label{sec:fronts-bistable-media}

\begin{figure}
	\centerline{\includegraphics{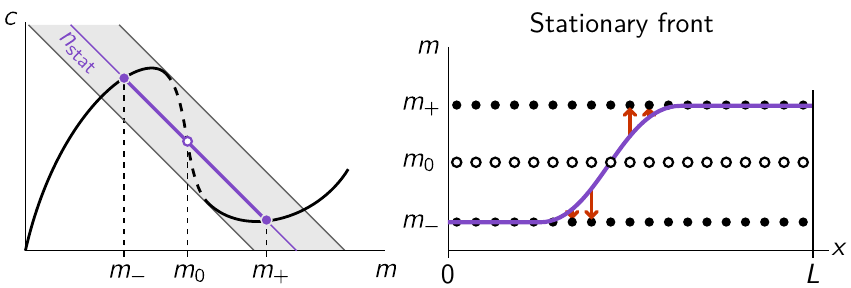}}
	\caption{Phase-space structure for bistable local kinetics in the case of equal diffusion constants, $D_m = D_c$, where there is no mass redistribution such that the system can be reduced to a one-component system, Eq.~\eqref{eq:bistable-medium}. The stable equilibria $m_\pm$ form a \emph{static}, spatially homogenous scaffold; the equilibrium $m_0$ is locally unstable. Only if the local kinetics are fine-tuned such that the total turnover vanishes (e.g.\ by tuning $\bar{n} = n_\text{stat}$), the front is stationary (marginally stable).}
	\label{fig:static-bistable-scaffold}
\end{figure}

To elucidate the role of bistability, let us first consider the case of equal diffusion constants $D_c = D_m =: D$. 
(Although this does not make sense in the intracellular context anymore, where typically $D_c > D_m$, we stick to the notation with concentrations $m$ and $c$.)
Then mass redistribution decouples from the kinetics $\partial_t n = D \gradx^2 n$, i.e.\ the total density becomes uniform by diffusion (see the mass-redistribution dynamics, Eq.~\eqref{eq:n-dyn-eta}, and note that $\eta(x,t) = n(x,t)$ for equal diffusion constants). 
As a consequence, the system can be reduced to one component, for instance the membrane density
\begin{equation} 
\label{eq:bistable-medium}
	\partial_t m = D \gradx^2 m + f(m,\bar{n}-m),
\end{equation}
where the local kinetics is bistable at every point in space (Fig.~\ref{fig:static-bistable-scaffold}). 
This corresponds to a (classical) one-component model for bistable media which generically exhibits propagating fronts~\cite{Mikhailov:1994}. 
A standard calculation, commonly performed by `Newton mapping' (briefly described in Sec.~\ref{sec:scaffolding-introduction}) or by phase-space analysis (in $(m,\partial_x m)$-phase space), shows that the propagation velocity $v$ of a front is proportional to the imbalance in reactive turnover~\cite{Mikhailov:1994}: 
$v \propto \integral{m_-}{m_+}{m} f(m, \bar{n}-m)$. 
Hence, a stationary front can only be realized by fine-tuning of parameters, e.g.\ the average total density $\bar{n}$, such that total turnover is balanced:
\begin{equation} 
\label{eq:bistable-mass-tuning}
	n_\text{stat} : \integral{m_-}{m_+}{m} f\kern-0.2em \left(m, n_\text{stat}-m\right) = 0.
\end{equation}
The balanced (fine-tuned) case corresponds to the Allan--Cahn equation ~\cite{Allen:1972a} (also called `model A' dynamics~\cite{Hohenberg:1977a}).
(In a finite size system, there will be logarithmic coarsening, see Ref.~\cite{Ward:2006a} and references therein.)

\begin{figure*}
	\centerline{\includegraphics{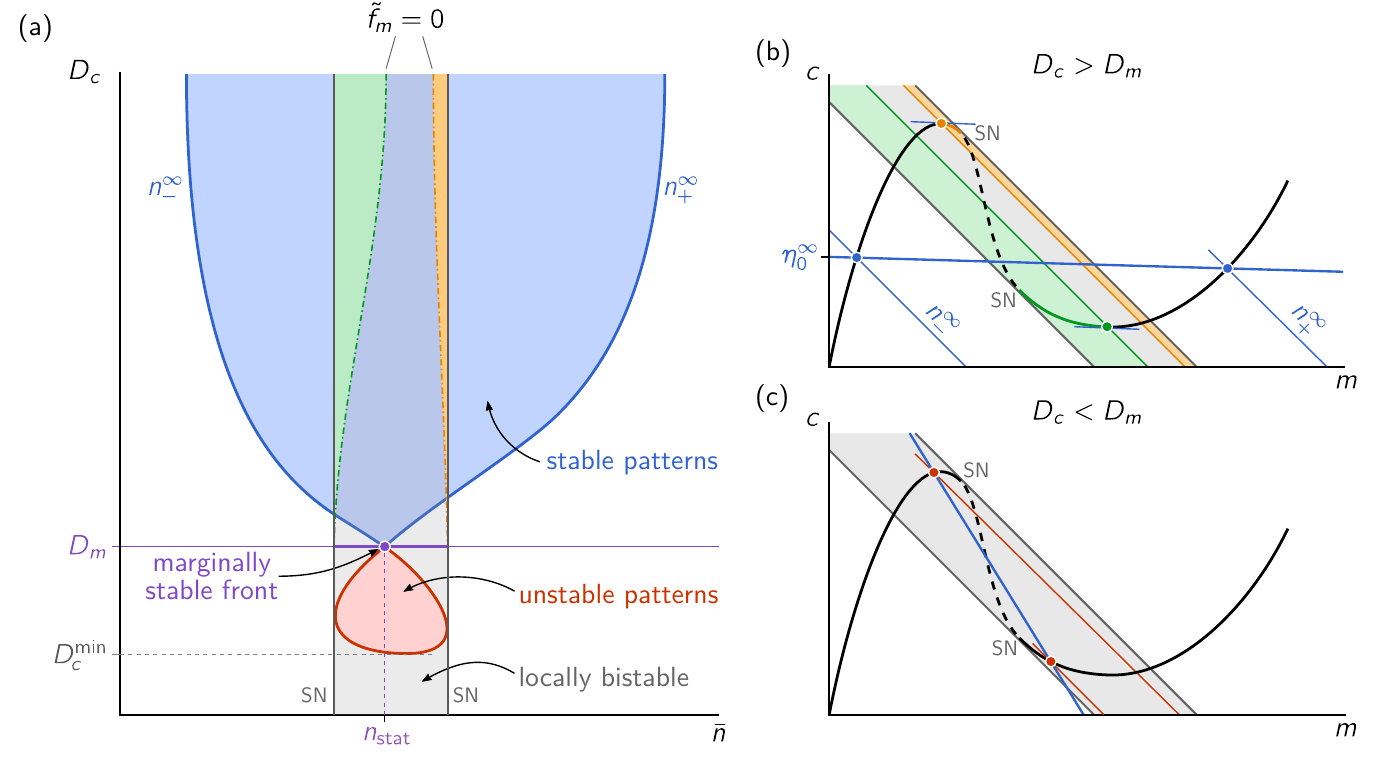}}
	\caption{Geometrically determined, schematic $(D_c, \bar{n})$-bifurcation diagram in the large system size limit for a system with bistable kinetics (the locally bistable region is shaded in gray in both the bifurcation diagram (a) and the phase-space plots (b)). (a) Bifurcation diagram: the regions where stationary patterns exist (shaded in red and blue) and where a homogenous steady state is laterally unstable (shaded in green and orange) are constructed based on the same geometric criteria as in the case of monostable kinetics (cf.\ Fig.~\ref{fig:n-bifurcation-mesa-patterns}). Along the purple line $D_c = D_m$ there is no mass redistribution, and the system exhibits classical traveling fronts within the bistable regime (cf.\ Fig.~\ref{fig:static-bistable-scaffold}). A marginally stable front exists at the singular point at $(D_c = D_m, n_\text{stat})$ where total reactive turnover is balanced by fine-tuning $\bar{n}$ (cf.\ Eq.~\eqref{eq:bistable-mass-tuning}). Outside the regions shaded in red and blue, there are no \emph{stationary} patterns. There might however be non-stationary patterns like the traveling fronts in the bistable medium for $D_c = D_m$. Non-stationary patterns for $D_c \gtrless D_m$ are outside the scope of this study.
	(b) Phase-space plot showing the reactive nullcline (black line, dashed in the locally unstable region). The sections of the nullcline where homogenous steady state is laterally unstable (shaded in green and orange) are delimited by points where the FBS is tangential to the NC. Intersection points (blue dots) of the flux-balance subspace (thick blue line) with the reactive nullcline determine the range $n_\pm^\infty$ where stationary patterns exist.
	(c) Phase-space plot for the case $D_c < D_m$, where the slope of the FBS (thick red line) is more negative than $-1$. The plateau scaffolds of stationary patterns can be constructed via FBS-NC intersection points (red dots), as long as $D_c > D_c^\text{min}$ (cf.\ Eq.~\eqref{eq:Dc_min}). These patterns are unstable though (cf.\ Eq.~\eqref{eq:eta-int-linearized}).
	}
	\label{fig:n-Dc-diagram-bistable}
\end{figure*}

With respect to the concept of local equilibria as scaffolds for patterns (cf.\ Sec.~\ref{sec:scaffolding-introduction}), the bistable local equilibria (fixed points $m_\pm$) can be regarded as a \emph{static scaffold} for front solutions; see Fig.~\ref{fig:static-bistable-scaffold}. 
Because there is no mass-redistribution, the scaffold must remain static and can not adapt to balance the total reactive turnover. 
Instead, fine-tuning of parameters (e.g.\ $\bar{n}$), is required to obtain a balance of total turnover and thus a stationary front. 
In the $(\bar{n},D_c)$-bifurcation diagram, Fig.~\ref{fig:n-Dc-diagram-bistable}a, the stationary bistable front with a static scaffold appears only at a singular point $(\bar{n},D_c) = (n_\text{stat},D_m)$. 

What happens when the diffusion constants are unequal $D_c \neq D_m$? 
Then, mass will be redistributed, leading to shifting of the local equilibria that scaffold the pattern. 
As we know from our analysis for monostable reaction terms, this dynamic scaffold is able to self-balance the total reactive turnover---fine-tuning of $\bar{n}$ is no longer required to obtain stationary patterns. 
Interestingly, for a bistable reaction term, stationary patterns can be constructed both for $D_m > D_c$ and for $D_m < D_c$, as we discuss next; see Fig.~\ref{fig:n-Dc-diagram-bistable}b,c. 
To determine the stability of these patterns, we will examine below (after the description of the bifurcation diagram) how the scaffold self-balances via mass redistribution.
 
\subsubsection{Bifurcation diagram for locally bistable reaction kinetics}

The bifurcation diagram, Fig.~\ref{fig:n-Dc-diagram-bistable}a, for the large system size limit ($L \to \infty$) is obtained using the same geometric criteria as for the case of locally monostable reaction kinetics (see Fig.~\ref{fig:n-Dc-diagram-bistable}b, cf.\ Fig.~\ref{fig:n-bifurcation-mesa-patterns}a). 
The presence of a bistable nullcline segment does not affect the feasibility of the geometric construction itself. 
However, the bifurcation diagram one obtains is qualitatively different from the monostable case, as we will see next. 
We discuss the regimes of stationary patterns (shaded in blue and red) first, before we analyze the regions of lateral instability (shaded in green and orange). 

The region where stable stationary patterns exist (shaded in blue) is delimited by lines $n_\pm^\infty(D_c)$. 
These lines converge in singular point $(n_\text{stat},D_c = D_m)$, where a marginally stable front exists in a bistable medium without mass redistribution (Sec.~\ref{sec:fronts-bistable-media}). 
Along the entire line, $D_c = D_m$, in the phase diagram Fig.~\ref{fig:n-Dc-diagram-bistable}a the dynamics can be reduced to a classical one-component system (Eq.~\eqref{eq:bistable-medium}).
Such a system exhibits propagating waves within the region of bistability located between the two saddle-node (SN) bifurcations of local equilibria (gray area in Fig.~\ref{fig:n-Dc-diagram-bistable}a); compare the bistable (gray) area in Fig.~\ref{fig:static-bistable-scaffold} and Fig.~\ref{fig:n-Dc-diagram-bistable}b,c, where the nullcline slope is more negative than $-1$.
Only for the fine-tuned value right at $\bar{n} = n_\text{stat}$, the front velocity is zero (purple dot). 
At $(n_\text{stat},D_m)$, the dynamic scaffold that self-adapts via mass redistribution for $D_c \gtrless D_m$ bifurcates from the static scaffold $m_\pm(n_\text{stat})$ of the marginally stable front.

For bistable kinetics the slope at the inflection point of the nullcline, $\snc^\text{inf}$, is necessarily more negative than $-1$ so stationary patterns may also exist for $D_c^\text{min} < D_m$ (cf.\ Eq.~\eqref{eq:Dc_min}), since they can be constructed from FBS-NC intersection points as shown in Fig.~\ref{fig:n-Dc-diagram-bistable}c. 
(We stick to the notation with concentrations $m$ and $c$, although they are not meaningful as ``membrane'' and ``cytosolic'' concentrations, in the case $D_c < D_m$. Instead they should be understood as abstract concentrations~\cite{Note:Correlations}).
The corresponding region where such stationary patterns exist is shaded in red in the bifurcation diagram shown in Fig.~\ref{fig:n-Dc-diagram-bistable}a. In the bottom half of this ``balloon''-shaped region the equilibria that form the plateau of the constructed pattern are locally unstable. Hence these patterns cannot be stable. 
As we will see in the next subsection, \emph{all} stationary patterns for $D_c < D_m$ are unstable (even if their plateaus are locally stable) since they are destabilized by the imbalance of reaction turnover induced by any (infinitesimal) perturbation.
In contrast, stationary patterns for $D_c > D_m$ are stable because the self-adapting scaffold re-balances the reactive turnover.

The regions with a laterally unstable homogeneous steady state (NC-slope steeper than FBS-slope, cf.\ Eq.~\ref{eq:slope-criterion}) are shaded in green and orange to distinguish in the bistable region which of the two locally stable reactive equilibria is laterally unstable; see Fig.~\ref{fig:n-Dc-diagram-bistable}b
\cite{Note:SaddleNode}. 

In conclusion, we found the generic $(\bar{n},D_c)$-bifurcation diagram of stationary patterns for an N-shaped nullcline with a bistable segment using the same geometric arguments as for the case of a monostable nullcline. Since our analysis crucially relies on the flux-balance subspace, it is limited to stationary patterns. For the special case $D_c = D_m$ the existence of non-stationary patterns (traveling fronts) in the locally bistable regime is well known. By continuity, we expect that there will be traveling fronts also $D_c \neq D_m$. Furthermore, we speculate that there will be non-stationary patterns outside the regime of local bistability, because mass redistribution may dynamically create a \emph{region} of bistability that travels through the system.

\subsubsection{The dynamic scaffold self-balances by shifting the flux-balance subspace} \label{sec:FBS-stability}

In the following, we will assess the stability of the stationary (mesa) patterns found by the geometric construction above. (We only discuss the stability of mesa patterns, which are generic in the limit of large system size, $L\to\infty$.) The phase-portrait analysis in the phase space of chemical reactions facilitates a simple heuristic approach to study pattern stability: Instead of a full stability analysis of the stationary pattern, we consider only the stability of the FBS-position (mass-redistribution potential $\eta(x,t)$), as a proxy for the pattern stability. Intuitively, in the direction along the FBS, the pattern is quickly stabilized by due to scaffolding by local equilibria. In the following we present a simple stability criterion for stationary patterns, derived from this intuition. Details of the (ad hoc) derivation and a comparison to numerical analysis is presented in Appendix~\ref{app-sec:FBS-stability}. A mathematically rigorous stability analysis of stationary patterns (using, for instance, the ``Singular Limit Eigenvalue Problem'' introduced by \cite{Nishiura:1987a}, see \cite{Ward:2006a} for a survey), is outside the scope of this paper.

\begin{figure}
	\centerline{\includegraphics{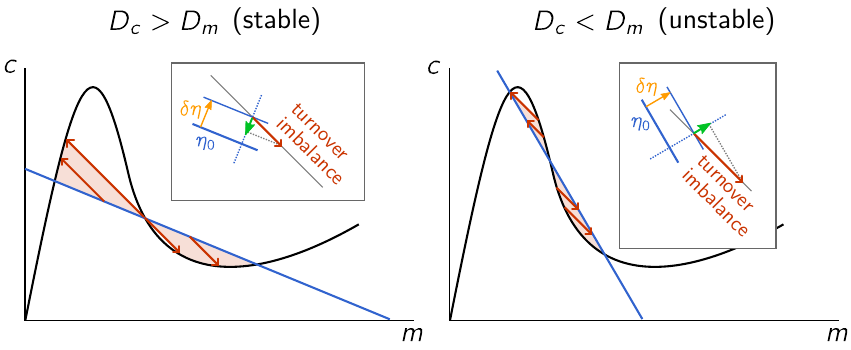}}
	\caption{
		Geometric construction of the stability of the FBS-position $\eta_0$ due to total turnover imbalance which serves as proxy for the stability of the pattern that is embedded in the FBS.
		In steady state, the reactive turnovers (illustrated by red arrows) on either side of the inflection point must be balanced.
		Inset boxes: A perturbation $\delta \eta$ (yellow arrow) that shifts the FBS-position $\eta_0$ will induce an imbalance of reactive turnovers (red arrow). Projecting this net reactive turnover onto the $\eta$-axis (dashed blue line) yields the movement of the FBS induced by the turnover imbalance (green arrow). For $D_c > D_m$, the FBS returns to its steady-state position $\eta_0$. Conversely, for $D_c < D_m$, the FBS is driven further away from its steady-state position, such that the initial perturbation is further amplified in a destabilizing feedback loop. 
	}
	\label{fig:FBS-instability}
\end{figure}

Recall that in steady state $\widetilde{\eta}(x) = \etaInfty$ is spatially uniform and determined by total turnover balance, Eq.~\eqref{eq:turnover-balance}, that can be geometrically interpreted as a Maxwell construction (balance of the red-shaded areas in Figs.~\ref{fig:stat-pattern-and-phase-space}b and~\ref{fig:FBS-instability}). 
How does the system evolve following a perturbation of the FBS-position $\eta$? Consider a spatially uniform shift, $\delta \bar{\eta} > 0$. Then the area between NC and FBS to the left of $m_0$ (inflection point of the pattern) will decrease, while the area on the right will increase; see Figs.~\ref{fig:stat-pattern-and-phase-space}b and~\ref{fig:FBS-instability}. The net reactive flow (``sum of the arrows in the two areas'') leads to a change of the average concentrations $m$ and $c$ (in the interface region) which amounts to a shift of the FBS-position, as shown in the insets in Fig.~\ref{fig:FBS-instability}. Because the net reactive flow points along a reactive phase space (slope $-1$), the direction in which the FBS shifts due to turnover imbalance depends on its slope. 
For $D_c > D_m$, i.e.\ an FBS-slope larger than $-1$, it will move down and, hence, relax back to $\etaInfty$ (recall that we considered an upwards shift as perturbation, the arguments work analogously for a downwards shift).
In other words, the scaffold will adapt until total reactive turnover is balanced again. 
We conclude that the scaffold is \emph{self-balancing} when $D_c > D_m$.
Conversely, when $D_c < D_m$, the FBS will move further in the direction of the perturbation, thus destabilizing the pattern.

This qualitative stability argument can be expressed mathematically, to obtain a quantitative approximation for the growth rate of perturbations $\delta \eta(x_0, t) := \eta(x_0,t) - \etaInfty$ in the vicinity of a stationary mesa pattern $(\mstat(x),\etaInfty)$ (see Appendix~\ref{app-sec:FBS-stability}):
\begin{equation} \label{eq:eta-int-linearized}
	\partial_t \delta \eta(x_0, t) 
	\approx \delta \eta(x_0, t) \, \frac{D_m/D_c - 1}{m_+ \,{-}\, m_-} \integral{m_-}{m_+}{m} \tilde{f}_\eta(m,\etaInfty) 
\end{equation}
where $\tilde{f}_\eta = \partial_\eta \tilde{f}$. 
Comparison to numerically computed linear stability (dominant eigenvalue) of stationary patterns confirms shows that Eq.~\eqref{eq:eta-int-linearized} is a good lowest order approximation for pattern stability; see Appendix~\ref{app-sec:FBS-stability} and Fig.~\ref{app-fig:pattern-stability}.
The integral over $\tilde{f}_\eta$ is the turnover imbalance due to a shift of the FBS, and thereby captures the geometric intuition based on the `Maxwell construction' (area balance) we outlined above. 
 The prefactor $(D_m/D_c - 1)$ determines the direction in which FBS will shift because the integrand is always positive for $f_c > 0$ ($\tilde{f}_\eta (m,\etaInfty) = f_c(m,\etaInfty - m \, D_m/D_c) > 0$). We hence recover the stability criterion from our geometric argument above.

\subsection{The cusp scenario is generic} \label{sec:cusp-scenario}

In some previous literature, it was argued that bistability of the reaction kinetics is an essential prerequisite for polarization to emerge in 2C-MCRD systems~\cite{Jilkine:2011a, Mori:2008a}. This claim was questioned recently \cite{Goryachev:2017a}.

\begin{figure}[bt]
	\centerline{\includegraphics{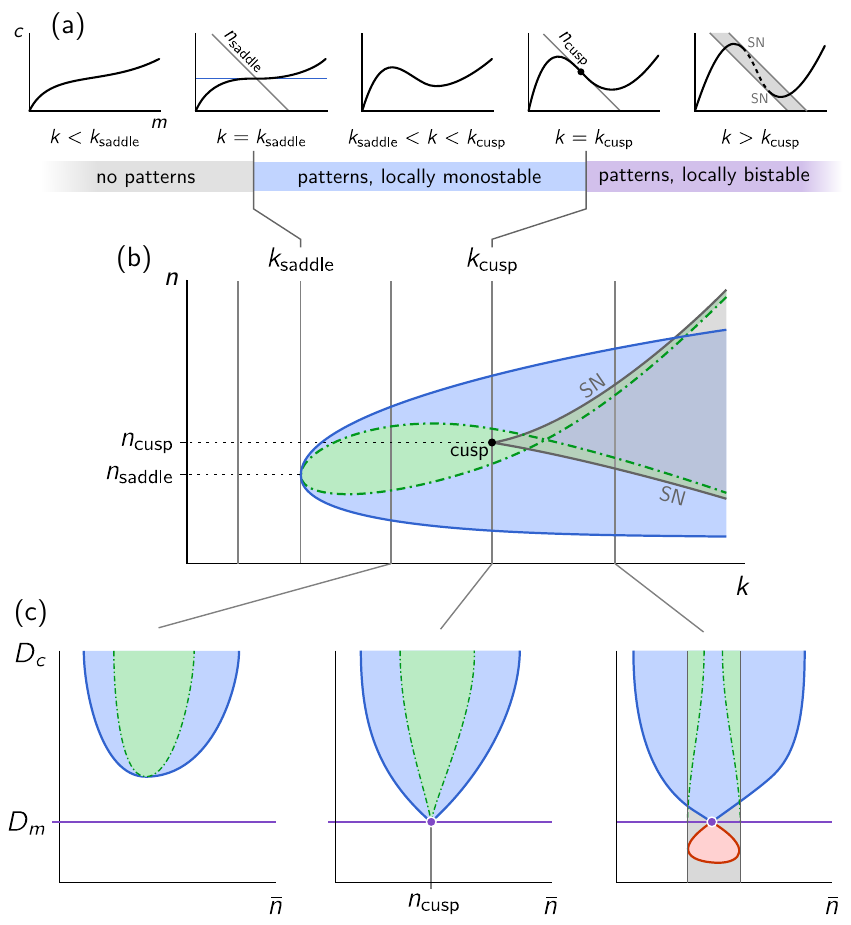}}
 	\caption{
	Schematic of the cusp bifurcation scenario (``unfolding'') in the $(k,\bar{n},D_c)$-bifurcation diagram (same color code as Figs.~\ref{fig:n-bifurcation-mesa-patterns}, \ref{fig:n-Dc-diagram-monostable}, and~\ref{fig:n-Dc-diagram-bistable}, Fig.~\ref{app-fig:geometry-bifurcations}).
	(a) We analyze the effect of a series of deformations of the reactive nullcline parametrized by the (notional) kinetic rate $k$. The respective (schematic) $(k,\bar{n})$-bifurcation diagram of stationary patterns (for $D_m/D_c \to 0$ and $L\to\infty$) is shown in (b). Initially, the nullcline is monotonic, and hence does not facilitate pattern formation.
	At $k_\text{saddle}$ a segment of negative nullcline-slope, $\snc$, emerges in a saddle point ($\snc = 0$ at the inflection point), such that patterns can form for $D_m/D_c \to 0$ (generally, the critical nullcline slope $\snc^\text{crit}$ for pattern formation is simply the ratio $-D_m/D_c$, cf.\ Eq.~\eqref{eq:Dc_min}). The regimes of lateral instability (green) and pattern existence (blue) emanate from this critical point.
	At $k_\text{cusp}$, a region of bistability (shaded in gray, bounded by saddle-node bifurcations) emanates from a cusp bifurcation ($\snc = -1$ at the nullcline's inflection point). At this point, the topology of the $(\bar{n},D_c)$-bifurcation diagram changes (see (c)).
	(In the small gray triangular region in the top-right corner of the $(k,n)$-bifurcation diagram, the system is locally bistable but does not exhibit stable stationary patterns.)
	(c) Schematic $(\bar{n},D_c)$ bifurcation diagrams for $k_\text{saddle} < k < k_\text{cusp}$ (monostable kinetics, Fig.~\ref{fig:n-Dc-diagram-monostable}), at the cusp ($k = k_\text{cusp}$), and for $k > k_\text{cusp}$ (bistable kinetics, Fig.~\ref{fig:n-Dc-diagram-bistable}).
	}
	\label{fig:cusp-scenario}
\end{figure}

Above, we conclusively showed that bistability is \emph{not} necessary for pattern formation. Instead, in systems with conserved quantities, a (non-homogeneous) pattern scaffold can generically self-organize by due to shifting local equilibria when there is mass-redistribution ($D_c \neq D_m$). 
However, there is an interesting and more subtle connection between bistability and the ability to form patterns. 
This connection is revealed by studying the transition from monostable to bistable kinetics due to variation of kinetic parameters.

Variations in the kinetic parameters will change the shape of the reactive nullcline. 
This may not only lead to quantitative but also qualitative changes in the $(\bar{n},D_c)$-bifurcation diagram, namely if there is a transition from a monostable to a bistable reaction kinetics:
Imagine that variation of some rate $k$ in the reaction kinetics generates nullcline deformations as shown in Fig.~\ref{fig:cusp-scenario}a.
Let us start with a nullcline that is strictly monotonically increasing with $m$.
Then according to the geometric criterion, Eq.~\eqref{eq:slope-criterion}, there is no lateral instability and hence no stationary patterns; 
recall that we have shown in Sec.~\ref{sec:interface-width} that an interface, the elementary element of a pattern, must be a laterally unstable region.
Upon further changing the kinetic rate $k$ there may eventually be a threshold value $k_\text{saddle}$ beyond which the nullcline shows a region with a negative slope.
A regime of lateral instability, and with it a regime where patterns can exist, emerges once the nullcline slope first becomes steeper than the slope of the FBS: $\snc^\text{crit} = -D_m/D_c$.
Eventually, at $k = k_\text{cusp}$, a further deformation of the nullcline may create a segment with slope $\snc < -1$ where there is bistability of local equilibria.
The bistable regime (shaded in gray) emanates in a cusp bifurcation where the two saddle-node (SN) bifurcations of the reactive equilibria meet in a single point, and the nullcline inflection point has slope $\snc^\text{inf} = -1$ such that it is tangential to the reactive phase space $m + c = \nInf = n_\text{cusp}$; see Fig.~\ref{fig:cusp-scenario}b. 
The surface of reactive equilibria in the two-parameter bifurcation structure starts to fold over itself at the cusp point. Because the Turing bifurcations lie on different sheets of the folded surface of local equilibria, The Turing-bifurcation line (dash-dotted green line) crosses over itself in the bistable region of the bifurcation diagram.

At the cusp point, the topology of the $(\bar{n},D_c)$-bifurcation diagram changes from the topology characteristic for mono\-stable reaction kinetics (leftmost panel in Fig.~\ref{fig:cusp-scenario}c, cf.\ Fig.~\ref{fig:n-Dc-diagram-monostable}) to the one characteristic for bistable reaction kinetics (rightmost panel; cf.\ Fig.~\ref{fig:n-Dc-diagram-bistable}). 
At $k = k_\text{cusp}$, the bifurcation diagram has the singular topology shown in the center panel, with the cusp critical point at $(n_\text{cusp},D_m)$.

\begin{figure}
	\centerline{\includegraphics{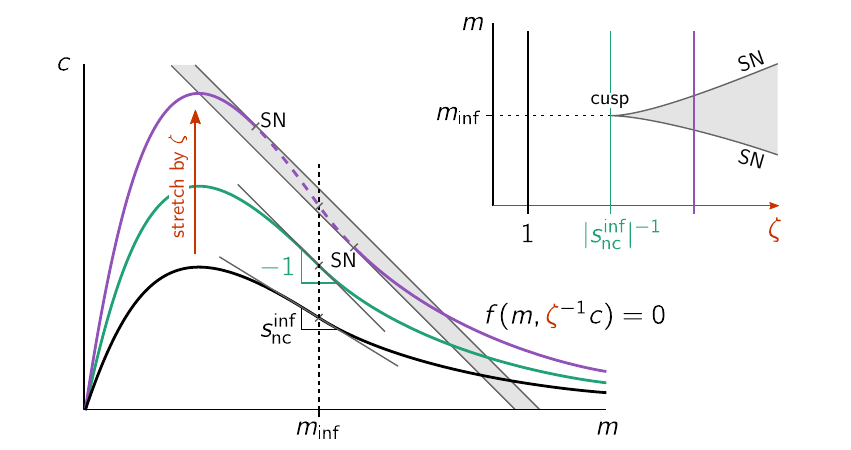}}
	\caption{
	Any nullcline with a segment of negative slope can be deformed by simple dilation to a exhibit cusp bifurcation where two saddle-node (SN) bifurcations of the reactive equilibria meet (see inset). The generic deformation to achieve this is a dilation in the $c$-direction by a scale factor $\zeta$ (i.e.\ replacing $c$ in the reaction term $f(m,c)$ by $\zeta^{-1} c$, such that the dilated nullcline is given by: $f(m,\zeta^{-1} c) = 0$). Say the original nullcline (black curve, $\zeta = 1$) has a slope $\snc^\text{inf} < 0$ at the inflection point $m_\text{inf}$. Then, the nullcline dilated by $\zeta = |\snc^\text{inf}|^{-1}$ (green curve), will have the slope $-1$ at its inflection point, thus yielding a cusp point of reactive equilibria (see inset, cf.\ Fig.~\ref{fig:cusp-scenario}). Upon further dilation, the nullcline (purple curve) will have a segment with slope more negative than $-1$ (dashed line, region shaded in gray), where the reactive equilibria are bistable. Each of the three nullclines in the phase space corresponds a vertical line of the same color in the bifurcation diagram (inset).
	}
	\label{fig:nullcline-deformation}
\end{figure}

Interestingly, by changing kinetic rates, the reactive nullcline of any system that is able to form stationary patterns, can be deformed to undergo such a cusp bifurcation of (local) reactive equilibria. 
To see how this works, first recall that to obtain stationary patterns, there must be a segment where the nullcline slope is $\snc < -D_m/D_c < 0$, i.e.\ necessarily negative. 
This segment can be deformed into a bistable region (slope $\snc < -1$) by a dilation of the nullcline in the $c$-direction in phase space; see Fig.~\ref{fig:nullcline-deformation}. 
Let us denote the scale factor by $\zeta$, then the dilated nullcline is determined by $f(m,\zeta^{-1} c) = 0$. 
A slope $\snc$ of the original nullcline $f(m,c)=0$ will become a slope $\zeta \cdot \snc$ for the dilated nullcline. 
In particular, dilation by a factor $\zeta = |\snc^\text{inf}|^{-1}$ leads to a nullcline with slope $-1$ at its inflection point, hallmark of a cusp bifurcation of the reactive equilibria (cf.\ Fig.~\ref{fig:nullcline-deformation}). 
Any further dilatation leads to a bistable region (nullcline segment with slope more negative than $-1$). 
We conclude that the generic bifurcation scenario underlying all pattern-forming 2C-MCRD systems is a cusp bifurcation of reactive equilibria.
Our geometric reasoning thus explains previous numerical observations~\cite{Trong:2014a}.

\subsection{Sub- and supercriticality of lateral instability in finite sized systems} \label{sec:sub-supercrit}

So far, we have focused on the large system size limit ($L \to \infty$) where bifurcation diagrams can be constructed from phase-space geometry. 
In particular, we found that the onset of a mass-redistribution instability is generically subcritical. To analyze sub- vs.\ super-criticality in the case of finite system size, we use a perturbative approach (weakly nonlinear analysis, see e.g.\ \cite{Manneville:1990a}) for the pattern close to onset. 
In the vicinity of the homogenous steady state $(\meq,\ceq)$, we expand a stationary state $(\mstat(x), \eta_0)$ in harmonic functions (eigenmodes of the Laplace operator under no-flux boundary conditions):
\begin{equation} \label{eq:harmonic-expansion}
	\mstat(x) 
	= \meq + \sum_{k = 0}^\infty \delta m_k \cos(k \pi x/L).
\end{equation}
As we have learned in Sec.~\ref{sec:lateral-instability}, the band of unstable modes always extends to long wavelengths ($q \to 0$) in a 2C-MCRD system (`type~II' instability, cf.\ Fig.~\ref{fig:MRI}b). 
Therefore, in a finite size system, the mode that becomes unstable first at the onset of the lateral instability is the longest wavelength mode: $\cos(\pi x/L)$. 
We want to study the steady-state amplitude of this mode in the vicinity of the onset bifurcation. 
It is not sufficient, though, to keep only this first harmonic in the mode expansion, Eq.~\eqref{eq:harmonic-expansion}, since higher harmonics couple to it through the nonlinear terms.
For an expansion to third order in the first mode amplitude $\delta m_1$, one needs to include only the first and second harmonics in Eq.~\eqref{eq:harmonic-expansion}. 
Higher harmonics are not needed because they couple to $\delta m_1$ through higher nonlinearities $\mathcal{O}(\delta m_1^4)$. 
To leading order the ansatz thus reads
\begin{subequations} \label{eq:weakly-nonlinear-ansatz}
\begin{align}
	\mstat(x) \approx {} & \meq + \delta m_0 + \delta m_1 \cos(\pi x/L) \\
	& + \delta m_2 \cos(2\pi x/L), \notag \\
	\eta_0 \approx {}& \eta^* + \delta \eta_0,	
\end{align}	
\end{subequations}
where $(\meq,\eta^*)$ denotes the homogenous steady state: $\tilde{f}(\meq,\eta^*) = 0$. 
Using this ansatz in Eq.~\eqref{eq:m-stat-fbs} and keeping terms up to third order yields for the steady state pattern amplitude $\delta m_1$ (see Appendix~\ref{app-sec:weakly-nonlinear} for details):
\begin{equation} \label{eq:steady-state-amplitude}
	0 =  F_1 \, \delta m_1 + F_3 \, \delta m_1^3 + \mathcal{O}(F_1 \delta m^3) + \mathcal{O}(\delta m_1^5).
\end{equation}
The first order and third order coefficients read
\begin{subequations}
\begin{align}
	F_1 &= \tilde{f}_m - D_m \pi^2/L^2,  \label{eq:F1} \\
	F_3 &= \frac{\tilde{f}_{mmm}}{8} + 
		\frac{\tilde{f}_{mm}^2}{24} \frac{L^2}{\pi^2 D_m}
		- \frac{\tilde{f}_{mm}}{4} \frac{\tilde{\partial}_{m} \sigma_\text{loc}}{\sigma_\text{loc}}, \label{eq:F3}
\end{align}
\end{subequations}
where $\tilde{\partial}_m = \partial_m - \frac{D_m}{D_c} \partial_c$ is the derivative along the direction of the flux-balance subspace.

\begin{figure}
	\centerline{\includegraphics{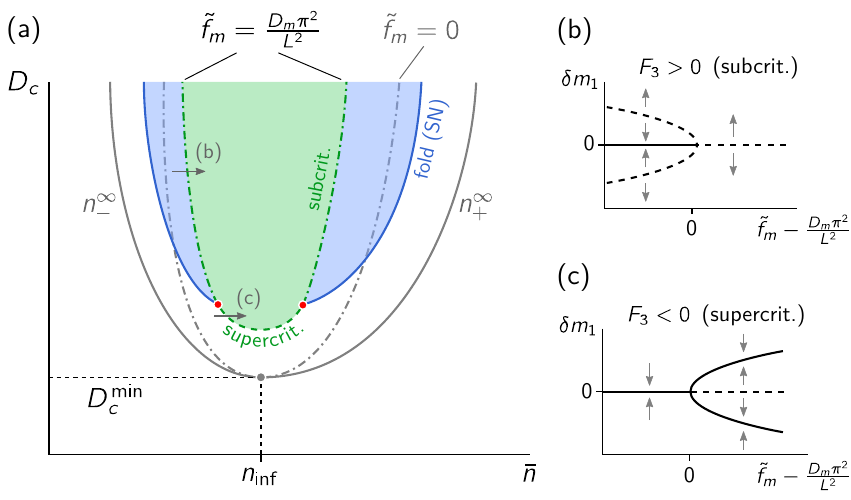}}
	\caption{Pattern bifurcation for finite domain size. (a) Schematic bifurcation structure for a system with monostable reaction kinetics (an analogous plot obtained by numerical continuation for a concrete reaction term, Eq.~\eqref{app-eq:wave-pinning}, is shown in Fig.~\ref{app-fig:n-Dc_finite-size}, Appendix~\ref{app-sec:weakly-nonlinear}). The gray lines indicate the bifurcation structure for $L \to \infty$ for comparison (cf.\ Fig.~\ref{fig:n-Dc-diagram-monostable}a). The Turing bifurcation (green dashed line, $\tilde{f}_m=D_m \pi^2/L^2$) is supercritical in the vicinity of $\nInf$, i.e.\ the nullcline inflection point. In the vicinity of onset (small pattern amplitude $\delta m_1$) the third order coefficient, $F_3$ (cf.\ Eq.~\eqref{eq:F3}), of a weakly nonlinear expansion (cf.\ Eq.~\eqref{eq:weakly-nonlinear-ansatz}), determines if the bifurcation is super- or subcritical. The schmatica bifurcation diagrams (b) and (c), correspond to the small gray arrows in (a). Unstable patterns that emerge in a subcritical pitchfork bifurcation (see (b)) meet the stable patterns in a fold bifurcation (saddle-node of patterns, cf.\ Fig.~\ref{fig:n-bifurcation-mesa-patterns}). These fold bifurcations terminate in points where the lateral instability bifurcation switches from sub- to supercritical (red points), indicated by a vanishing third order coefficient $F_3 = 0$ in the weakly nonlinear expansion (cf.\ Eq.~\eqref{eq:steady-state-amplitude}). Along the line of supercritical lateral instability (dashed green line), stable patterns emerge directly in a supercritical pitchfork bifurcation; see (c).
	\medskip
	}
	\label{fig:sub-vs-supercrit}
\end{figure}

The first harmonic amplitude $\delta m_1$, solution to Eq.~\eqref{eq:steady-state-amplitude} undergoes a pitchfork bifurcation at $F_1 = 0$. 
This bifurcation is simply the Turing bifurcation as it coincides with the onset of lateral instability: the homogenous steady state, $\delta m_1 = 0$, is laterally unstable only if the longest wavelength mode, $q_1 = \pi/L$, is within the band of unstable modes: $\pi^2/L^2 < \qmax^2 = \tilde{f}_m/D_m$, i.e.\ if $F_1 > 0$ (cf.\ Eq.~\eqref{eq:F1}). 
Hence, the sign of the third order coefficient $F_3$, evaluated at the bifurcation point $F_1 = 0$, determines wether the bifurcation is supercritical ($F_3 < 0$) or subcritical ($F_3 > 0$); 
see Fig.~\ref{fig:sub-vs-supercrit}b.

On the basis of this weakly nonlinear analysis, we can study the bifurcation at $F_1 = 0$ in any control parameter $\mu$ (for instance the average total density $\bar{n}$, the system size $L$, kinetic rates, and diffusion constants). 
With the critical value $\mu_\text{c}$, defined by the condition $F_1(\mu_c) = 0$, we introduce the reduced control parameter $\delta \mu = \mu - \mu_\text{c}$, and linearize Eq.~\eqref{eq:steady-state-amplitude} to lowest order in $\delta \mu$:
\begin{equation}
	0 = \partial_\mu F_1|_{\mu_\text{c}} \, \delta \mu \, \delta m_1 +  F_3(\mu_\text{c}) \, \delta m_1^3.
\end{equation}
To leading order, the branch of the solution that bifurcates at $F_1 = 0$ then reads
\begin{equation}
	\delta m_1 = \sqrt{\frac{\partial_\mu F_1}{-F_3} \Big|_{\mu_\text{c}} \delta \mu \; }.
\end{equation}
At singular points where local stability and lateral stability change simultaneously, i.e.\ $\sigma_\text{loc} = 0$ and $F_1 = 0$, the last term in $F_3$ diverges. 
Such codimension-two points require a more technically involved analysis (unfolding) that is outside the scope of this study (see e.g.\ Refs.~\cite{Arnold:Book,Murdock:Book}).

From the third order coefficient $F_3$ (see Eq.~\eqref{eq:F3}), we can analyze the type of bifurcation in terms of geometric features (nullcline curvature $\kappa(n) \sim -\tilde{f}_{mm}|_n$, see Appendix~\ref{app-sec:curvature-approx}) together with the quantity that characterizes the relaxation rate to local equilibrium ($\sigma_\text{loc}$, cf.\ Eq.~\eqref{eq:kinetics-linearization}). 
The first two terms in $F_3$ encode geometric properties of the reactive nullcline (curvature and its rate of change), i.e.\ how the local equilibria shift as mass redistribution shifts the local phase spaces. 
The last term in $F_3$ represents the rate of change of the timescale for relaxation $\sigma_\text{loc}$ to the local equilibria. In the following, we discuss the various regimes that arise due to the interplay of the three terms in $F_3$. (To simplify notation, we will implicitly assume that all coefficients ($\tilde{f}_m$, etc.) in $F_3$ are evaluated at the bifurcation point $\mu_\text{c}$.)

\begin{romanlist}
	\item At the nullcline inflection point $\tilde{f}_{mm} = 0$, only the first summand of $F_3$ remains: $F_3 = \tilde{f}_{mmm}/8$. The third derivative $\tilde{f}_{mmm}$ is proportional to the rate of change of the curvature ($\tilde{f}_{mmm} \sim -\partial_m \kappa$, see Appendix~\ref{app-sec:curvature-approx}). The curvature of a typical N-shaped nullcline must be positive (bent upwards) to the right of the inflection point and negative (bent downwards) to the left of it, otherwise it is impossible to smoothly connect laterally unstable regions \mbox{($\tilde{f}_m < 0$)} to laterally stable regions \mbox{($\tilde{f}_m < 0$)}. This implies that $\tilde{f}_{mmm} < 0$, and we conclude that the bifurcation at the nullcline inflection point is supercritical. This confirms our geometric argument above (cf.\ Fig.~\ref{fig:n-Dc-diagram-monostable}, and thin gray lines in Fig.~\ref{fig:sub-vs-supercrit}).
	\item In the large system size limit ($L \to \infty$) away from the inflection point ($\tilde{f}_{mm} \neq 0$), the second term in Eq.~\eqref{eq:F3} dominates. Because this term is always positive, the Turing bifurcation in large systems is generically subcritical (cf.\ Fig.~\ref{fig:sub-vs-supercrit}) as we have already concluded from geometric arguments in Sec.~\ref{sec:n-bifurcations} above.
	\item In a finite-sized but still large system, the last two terms of $F_3$ are negligible sufficiently close to the nullcline inflection point where $\tilde{f}_{mm}$ vanishes. Hence, finite sized systems are supercritical in the vicinity of the nullcline inflection point, because $\tilde{f}_{mmm}$ must be negative as we argued above in point (\textit{i}); (cf.\ Fig.~\ref{fig:sub-vs-supercrit}). Solving the condition $F_3 = 0$ to to leading order in $L^{-1}$ yields the estimate
\begin{equation}
	\big|\tilde{f}_{mm}\big| < L^{-1} \sqrt{-\frac{3 \pi^2}{2}  D_m \,  \tilde{f}_{mmm}},
\end{equation}
for the range of supercriticality.
	\item In small systems, the transition from supercriticality to subcriticality will depend also on the last term in $F_3$. It can contribute either positively or negatively to $F_3$, depending on the details of the reaction kinetics.
\end{romanlist}
Importantly, the statements (\textit{i})--(\textit{iii}), regarding large systems, follow from purely geometric arguments as they are determined by the first two terms in $F_3$. The reason for this is that at large wavelength, chemical relaxation is fast compared to diffusion, so the pattern is slaved to the scaffold, i.e.\ the reactive nullcline. For the same reason, the long wavelength onset of lateral instability is determined by a geometric criterion (slope of the nullcline) as we have shown in Sec.~\ref{sec:lateral-instability}.

In conclusion, we comprehensively characterized the Turing bifurcation (sub- vs.\ supercritical) and the bifurcations of stationary patterns, using the $(m,c)$-phase-space geometry.
Because of the inherent link between geometry and the physical concepts of mass-redistribution and shifting equilibria, we are able understand the physics underlying the patterns and their bifurcations.
The bifurcations in the large system size limit ($L\to\infty$)---determined by geometry---provide a good starting point to study the bifurcations in a finite sized system, e.g.\ by numerical continuation (see Appendix~\ref{app-sec:continuation}).


\section{Conclusions and Discussion} \label{sec:discussion}

We have introduced a local equilibria theory that allows to analyze and characterize both the initial growth and the eventual stabilization (saturation) of patterns on the basis of geometric objects in phase space:
the shape of the reactive nullcline and its intersections with the flux-balance subspace.
Within this framework, many properties of the nonlinear dynamics of mass-conserving reaction-diffusion systems can now be directly addressed in terms of  phase space geometry, which would otherwise only be accessible by numerical analysis.
In the following we will summarize the key concepts of local equilibria theory and important findings for 2C-MCRD obtained from this theory.
Subsequently, we address new perspectives for the investigation of reaction--diffusion systems with conserved masses.
Protein-pattern forming systems \emph{in vivo} (intracellular) and \emph{in vitro}  (reconstituted systems) will serve as the specific context for this part of the discussion.
Finally, we will give a brief outlook on upcoming work, on questions that are currently under investigation, and on future research directions.
In particular, we outline how local equilibria theory might provide a unifying geometric perspective on pattern formation in mass-conserving non-equilibrium systems and how it can be generalized to systems that are not strictly mass-conserving but contain a mass-conserving `core'.

\subsection{Summary of key concepts and results}

Phase space analysis of (two-component) MCRD systems has shown that spatial variations in protein density give rise to spatially heterogeneous local equilibria.
The relationship between mass and reactive equilibria is represented geometrically as a line of reactive equilibria (\emph{reactive nullcline}) in phase space. Along the reactive nullcline, the reaction kinetics are balanced.
We have shown that this nullcline is a central geometric object in phase space that organizes the spatiotemporal dynamics.
Furthermore, we have identified a one-parameter family of (one-dimensional) manifolds in phase space on which diffusive fluxes balance.
Any stationary pattern is embedded in one of these so-called \emph{flux-balance subspaces}.
The variable that parameterizes the family of flux-balance subspaces acts as a \emph{mass-redistribution potential}: Its spatial gradients represent a local imbalance of diffusive fluxes that drives mass-redistribution. Thus, the mass-redistribution potential encapsulates the interplay between reactions and diffusion processes.

In this way, the spatiotemporal dynamics of the reaction--diffusion system is fully determined by geometric structures in phase space:
(\textit{i}) We introduced a \emph{flux-balance construction}, based on intersection points between flux-balance subspace and reactive nullcline that enabled us to graphically construct stationary patterns and their complete bifurcation structure in the limit of large system size.
Underlying this construction is the general insight that patterns are \emph{scaffolded} by local equilibria which are, in turn, encoded by the reactive nullcline.
This principle can be generalized to dynamics: (\textit{ii}) The interplay between diffusive redistribution of mass and \emph{shifting local equilibria} drives the pattern-forming instability that we termed \emph{mass-redistribution instability}. Our analysis has shown that the slope of the nullcline provides a simple criterion for the occurrence of this instability.
Importantly, we find that the onset of instability is generically subcritical in 2C-MCRD systems on a large domain.
(\textit{iii}) Generalizing the local equilibria concept, we have introduced a decomposition of the spatial domain into regions, such as plateaus and interfaces, which are characterized in terms of \emph{regional dispersion relations}. This has enabled us to find simple heuristics for many properties of the pattern formation dynamics and stationary patterns. For instance, the width of the interface region can be approximated by the marginal mode of the regional dispersion relation at the interface. Building on this regional decomposition, one can characterize different pattern types and the transitions between them as control parameters are changed (Sec.~\ref{sec:pattern-types}). Furthermore, based on the concept of regional (in)stability we found an inherent connection between lateral (Turing) instability and stimulus-induced pattern formation (``nucleation and growth''), which enabled us to estimate the basin of attraction (``nucleation threshold'') for stationary patterns by a simple heuristic using the reactive nullcline.
As an additional advantage of such a characterization, we note that the reactive nullcline could in principle be determined experimentally for any given system in which the average total density can be controlled in a well-mixed ``reactor''.

Importantly, the concepts of scaffolding by local equilibria (\textit{i}), mass-redistribution instability (\textit{ii}) and regional dispersion relations (\textit{iii}) are directly generalizable to systems with more components and more conserved quantities.
We extensively discuss these future directions in the \hyperref[sec:outlook]{Outlook} below.

\subsection{Reaction--diffusion systems} \label{sec:discussion-protein}

For the sake of specificity, we will discuss the implications and application of our findings on mass-conserving reaction--diffusion systems with respect to protein pattern formation, which operates far from thermal equilibrium, and has received growing interest over the past two-decades.
Intracellular, i.e.\ \textit{in vivo}, protein pattern formation and self-organization have been subject to a large body of research, both experimentally (see Refs.~\cite{Campanale:2017a, Chiou:2017a} for recent reviews)
and theoretically (see Ref.~\cite{Halatek:2018b} and references therein.)
Furthermore, the \textit{in vitro} reconstitution of the MinDE system~\cite{Loose:2008a} has made it possible to study protein pattern formation experimentally under a wide range of externally controllable conditions; see Refs.~\cite{Ivanov:2010b, Zieske:2014a, Vecchiarelli:2014a, Caspi:2016a, Kretschmer:2017a, Denk:2018a, Glock:2018a, Litschel:2018a, Ramm:2018a, Kohyama:2019a, Glock:2019a} and Ref.~\cite{Vendel:2019a} for a recent review.

Taken together, these studies of both \textit{in vivo} and \textit{in vitro} systems have led to many important insights. 
However, many intriguing questions that are relevant to all reaction--diffusion systems far from equilibrium, remain open:
What is the role of the (biomolecular) interaction network, and how can complex models be reduced to their essential components? 
What are the physical mechanisms underlying the pattern-forming instabilities and under which conditions do these instabilities arise?
How can the dynamics of patterns far from the homogeneous steady state be studied systematically, i.e.\ how can we bridge the gap between the linear and the highly nonlinear regime?
A particular question in this context is how different patterns and their characteristic length scales (interface width and wavelength) are selected in the highly nonlinear regime.
In what follows, we discuss the implications of our work to these questions.

\subsubsection{Model classification, network motifs and experimental accessibility}

In recent years, several studies have employed high-throughput computational analyses of reaction--diffusion systems and graph theoretical analysis with the goal to infer the pattern-forming capabilities from the topology of the underlying reaction networks \cite{Chau:2012a, Marcon:2016a, Sugai:2017a, Diego:2018a}. 
Our results offer an entirely new and distinct perspective on model classification and the role of the interaction-network topology for mass-conserving systems. We found a simple condition for the pattern-forming (mass-redistribution) instability in 2C-MCRD systems: 
The slope of the line of reactive equilibria (reactive nullcline) must be (sufficiently) negative. 
Broadly speaking, the reactive equilibrium of the faster diffusing (i.e.\ cytosolic) component has to decrease with increasing total density (cf.\ Sec.~\ref{sec:lateral-instability}). 
Importantly, our approach goes beyond the classification based on linear (in)stability. 
It shows that the effect of nonlinearities on the dynamics is encoded in the curved shape of the nullcline.
In particular, there is a direct connection between the nullcline shape and the characteristic spatial density profile of the pattern (cf.\ Sec.~\ref{sec:pattern-types}).
The reactive equilibria, as represented by the reactive nullcline, might therefore provide an alternative approach to model classification.
Hence, a key challenge for future research will be to study how specific reaction kinetics and model parameters affect the shape of the reactive nullcline.

Moreover, a major advantage of reactive equilibria as the essential criteria for model classification is their experimental accessibility. 
In principle, any line of reactive equilibria can be measured directly in experiments by using a single, isolated, and well-mixed reactor and externally controlling the available conserved quantity (e.g.\ particle number). 
Such experiments would allow one to probe and classify the core mechanism quantitatively without any knowledge of the molecular details (which are irrelevant for such a classification).

The concepts that underlie local equilibria theory---mass redistribution and moving local equilibria---are not restricted to two-component systems with a single conservation law.
They have previously been applied to the model of the (\emph{in vitro}) MinDE system, which has two conserved protein species, MinD and MinE, and five components~\cite{Halatek:2018a,Brauns:2020b}.  
We believe that the results presented here are foundational for the development of a more general theory.
Our analysis constitutes the first step in a long-term project to find a geometric representation of the nonlinear dynamics of spatially extended systems.
In the outlook, Section~\ref{sec:outlook}, we briefly describe various forthcoming works and future projects that build upon the present study and generalize its results. 

\subsubsection{Polarity patterns, bistability and the necessary condition for a Turing instability}

Bistability is a generic feature of nonlinear systems, and its putative relation to polarity patterns has been controversially discussed in the literature \cite{Trong:2014a, Goryachev:2017a,Mori:2008a,Cusseddu:2018a,Gamba:2009a}.
In systems without mass conservation, bistable reaction kinetics facilitate traveling fronts that connect the two stable reactive equilibria (homogeneous steady states). (This scenario appears in mass-conserving systems as the special case of equal diffusion constants, which entails the lack of lateral mass redistribution; see Sec.~\ref{sec:n-Dc-bistable}.)
The stable equilibria of such a bistable medium can be pictured as scaffold for the traveling front. Because this scaffold is static in systems without mass redistribution, fine-tuning is required to achieve a stationary front (cf.\ Sec.~\ref{sec:fronts-bistable-media}).
Our results show that, in mass-redistributing systems (i.e.\ systems with unequal diffusion constants), the scaffold becomes dynamic and thereby supports stable, stationary polarity patterns in an extended parameter regime (cf.\ Sec.~\ref{sec:n-Dc-bistable}).
Most importantly, we found that, even in a monostable system, mass redistribution can facilitate the formation of a scaffold for stationary polarity patterns.
Hence, bistability of the reaction kinetics is neither required nor sufficient for the formation of such patterns. 

A central finding of our work is the physical mechanism by which the scaffold of a pattern emerges dynamically from a homogeneous steady state: the mass-redistribution instability (Sec.~\ref{sec:lateral-instability}).
Diffusive mass-redistribution requires that the diffusivities of the two components are different (``differential diffusion''); already a ratio $D_m/D_c$ slightly different than unity is sufficient.
This mass-redistribution drives an \emph{instability} under the condition that the reactive nullcline in the phase space of the reaction kinetics includes a segment of negative slope.
More precisely, the negative slope must be steeper than the flux-balance subspace, whose slope is determined by the negative ratio of the diffusivities $-D_m/D_c$. 
As membrane--bound proteins are significantly less mobile than cytosolic proteins, i.e.\ $D_m \ll D_c$, a small negative slope of the reactive nullcline is already sufficient for a Turing instability in the intracellular context.
Recall also that the criterion for bistability is a NC-slope more negative (``steeper'') than $-1$ (cf.\ Sec.~\ref{sec:well-mixed}), which is obviously a more restrictive condition than that for lateral instability. Hence, a bistable region is generally surrounded by a larger region of lateral instability in parameter space (cf.\ Fig.~\ref{fig:cusp-scenario}).

One might wonder how generic nullclines with a segment of negative slope are.
In fact, they are frequently encountered as N-shaped nullclines in a broad range of classical nonlinear systems~\cite{Strogatz:Book, Izhikevich:Book, Novak:2008a, Murray:Book}.
Typically, these nullclines encode some nonlinear feedback mechanisms that give rise to widespread phenomena such as relaxation oscillations, excitability and bistability.
Feedback loops in intracellular signaling networks (e.g.\ GTPase and phosphorylation cycles) generically lead to these phenomena \cite{Kholodenko:2006a}.
From this perspective, it appears that N-shaped nullclines should be rather common in biological systems.

Finally, the condition for mass-redistribution instability shows that, contrary to popular belief, Turing instability in a mass-conserving system can already arise for a ratio $D_m/D_c$ only slightly below unity when the reactive nullcline has sufficiently steep segment (i.e.\ in the vicinity of saddle-node bifurcations of the reactive equilibria, cf.\ Fig.~\ref{fig:cusp-scenario}).

\subsubsection{Mass-redistribution instability is mechanistically distinct from the activator--inhibitor paradigm}
\label{sec:AI}

A key finding of our work is the physical mechanism of mass-redistribution instability: Shifting local equilibria induce gradients and hence (diffusive) mass-redistribution which, in turn, leads to further shifting of the local equilibria.
Notably, mass-redistribution instability is a Turing instability in the original sense: a lateral instability due to diffusive coupling of a system that is locally stable in the absence of this coupling.

Importantly, the mechanism underpinning the mass-redistribution instability is distinct from the paradigm of short range activation and long range inhibition \cite{Gierer:1972a,Segel:1972a}.
The latter, termed ``activator--inhibitor'' mechanism, innately relies on some form of ``production'' and ``degradation'' processes, either of ``morphogens'' via gene-regulatory networks (e.g.\ in cell-cell signaling \cite{Karig:2018a} and tissue patterning \cite{Manukyan:2017a, Watanabe:2015a, Kondo:1995a}), or of electric signals coupling excitory and inhibitory neurons (e.g.\ in the visual cortex where lateral instability may underlie hallucinations due to long range coupling of inhibitory neurons \cite{Ermentrout:1979a}). 
In contrast, intracellular pattern formation is often driven by (nearly) mass-conserving dynamics \cite{Halatek:2018b} and, hence, is necessarily based on a mass-redistribution instability.
Because, as we emphasized above, mass-redistribution instability is a Turing instability, self-organized intracellular protein-patterns (see Ref.~\cite{Halatek:2018b} for a review)
are examples for Turing patterns (in the sense of patterns that arise from a Turing instability~\cite{Note:Turing}) in biological systems.

In general, it might not always be obvious whether the instability underlying pattern formation in a given system is either a mass-redistribution instability or one that essentially requires production and degradation. In the outlook Sec.~\ref{sec:beyond-mass-conservation}, we propose a general definition to answer this question.

\subsubsection{Subcriticality and stimulus-induced pattern formation}
\label{sec:discussion-subcrit}

An important results of our analysis is that the onset of pattern formation in 2C-MCRD systems is generically subcritical (cf.\ Secs.~\ref{sec:n-bifurcations} and~\ref{sec:sub-supercrit}).
Subcriticality may be beneficial in a biological context, as it confers robustness: Once a pattern is established, it is robust towards parameter variations due to hysteresis \cite{Eldar:2004a, Goehring:2011a, Trong:2014a}.

Moreover, subcriticality implies the existence of parameter regimes where pattern formation can be triggered by sufficiently large perturbations (akin to ``nucleation and growth'' in the binodal regime of phase-separating systems).
Such stimulus-induced pattern formation has been suggested as a new mechanism for pattern formation (under the term ``wave-pinning''), which---it was argued---is fundamentally distinct from a lateral (Turing) instability \cite{Mori:2008a, Jilkine:2011a, Edelstein-Keshet:2013a}.
This claim has been disputed in recent works \cite{Goryachev:2017a, Verschueren:2017a}.
Indeed, our results show that linear instability and stimulus-induced pattern formation are inherently connected: the latter is possible only where there is an adjacent regime of linear lateral instability and the underlying mechanism is a \emph{regional} (linear) lateral instability; see Sec.~\ref{sec:stim-induced}. 
Concretely, our results show that an interface---the elementary building block of a pattern---must necessarily be a laterally unstable region (cf.\ Sec.~\ref{sec:interface-width}). Hence, the creation of a laterally unstable region is a necessary condition for the formation of a stationary pattern. This implies that, any two-component system that has a regime of stimulus-induced pattern formation, must also exhibit a regime where the homogenous steady state is laterally unstable, and this regime can always be reached by simply changing the average total density.
Conversely, this suggests that subcriticality may be a generic feature of mass conserving systems since regional instability will facilitate stimulus-induced pattern formation adjacent to regimes where the homogeneous steady state is laterally unstable.

Finally, building on the concept of regional instability insights, we provided a simple geometric argument for the perturbation threshold for stimulus-induced pattern formation (``nucleation threshold'') based on the reactive nullcline.

\subsubsection{Length scale selection}
\label{sec:discussion-wavelength}

An important consequence of subcriticality is that well-known mathematical results for systems near a \emph{super}critical instability may not apply anymore.
Potentially the most prominent of such results is the existence of a \emph{characteristic wavelength}, determined by the fastest growing mode in the dispersion relation. 
This is often considered as a defining property of ``Turing patterns''. 
However, for subcritical systems, the wavelength of the pattern cannot be inferred from a linear stability analysis of the uniform steady state in general---not even at onset. 
Indeed, 2C-MCRD systems always exhibit uninterrupted coarsening \cite{Brauns:2020a}, i.e.\ the wavelength selected by the fastest growing mode at onset is observed only transiently and the final steady state is fully ``phase separated.''
 
Note, however, that coarsening is \emph{not} a generic feature of all mass-conserving systems.
Multicomponent MCRD models show (multi-)stable patterns with finite wavelengths~\cite{Halatek:2012a, Murray:2017a, Halatek:2018a}. 
We believe that identifying and understanding mechanisms of nonlinear wavelength selection that bring the coarsening processes to a halt and stabilize patterns with finite wavelength are among the most important tasks for future research on multicomponent models. 
While some rather general criteria have been found for one-component systems~\cite{Politi:2004a, Politi:2006a}, a comprehensive understanding of multicomponent systems remains out of reach for now. 

In addition to the wavelength, patterns have a second characteristic length scale---the width of interfaces. 
While the wavelength of patterns far from the homogeneous steady state is not determined by the dispersion relation, we have shown in Section~\ref{sec:interface-width} that the interface width is determined by the marginal mode ($\qmax$) of the interface's \emph{regional dispersion relation}.
The more general insight underlying this finding is that highly nonlinear patterns can be decomposed into spatial regions and studied in terms of regional phase-spaces and regional attractors (cf.\ Sec.~\ref{sec:regions}).
In a follow-up work \cite{Brauns:2020b}, we use this region decomposition to classify the different types of instabilities that govern Min-protein pattern formation \emph{in vivo} and \emph{in vitro}. In particular, we find that the interface width of standing wave patterns is also determined by the marginal mode of the regional dispersion relation.

\subsubsection{Pattern types are determined by the nullcline shape}

Two-component MCRD systems are found to exhibit two generic pattern types, referred to as mesas and peaks. In a recent numerical study of 2C-MCRD models \cite{Chiou:2018a}, a phenomenological ``saturation point'' was found to mark the peak- to mesa-pattern transition.
Our geometric analysis has now revealed the phase-space structure that underlies this ``saturation point'': It corresponds to an intersection point between reactive nullcline and diffusive flux-balance subspace; hence, its position dictated by the nullcline shape and the ratio of the diffusion constants.
Based on this insight, have shown that the nullcline shape serves as a simple criterion that to predicts the type of patterns formed by a given 2C-MCRD system; see Fig.~\ref{fig:profile-types}.

Put briefly, we distinguish $\Lambda$- and N-shaped nullclines, based on their tail behavior for large densities. 
In the case of a $\Lambda$-shaped nullcline (e.g.\ nullclines that asymptotically approach the $m$-axis for large $m$; see Figs.~\ref{fig:profile-types}b and~\ref{app-fig:peak-model-stat-pattern}) and a shallow flux-balance subspace ($D_m/D_c \ll 1$), a mesa pattern forms only when the average mass is sufficiently large.
For lower densities, the system exhibits peak patterns instead. Their amplitude depends sensitively on the total mass and the membrane diffusion constant.
The approximation of the peak pattern amplitude provided in Sec.~\ref{sec:pattern-types} (details in Appendix~\ref{app-sec:peak-patterns}), provides a simple estimate of the total (protein) mass at which peak patterns transition to mesa patterns.
Notably, the amplitude of peak patterns and the transition to mesa patterns depends sensitively on the ratio of the diffusion constants.
In the case of an N-shaped nullcline (see Figs.~\ref{fig:stat-pattern-and-phase-space} and~\ref{fig:profile-types}a), mesa patterns are generic, because the high-density plateau is already formed at low average mass (even if $D_m/D_c \ll 1$).
(Peak patterns are also possible for such nullclines, but require fine tuning of the average mass to the vicinity of bifurcation points (see Fig.~\ref{fig:n-bifurcation-mesa-patterns}).)

These findings have important biological consequences because the characteristic features of a pattern dictate the positional information it can convey, and therefore the biological function it can facilitate \cite{Wolpert:1969a, Alberts:Book, Lutkenhaus:2007a, Bendezu:2012a, Halatek:2018b}.
For example, cell division in budding yeast (\textit{S.~cerevisiae}) requires the formation of a single, narrow polarity site marked by a high density of the protein Cdc42, to uniquely determine the future bud site~\cite{Park:2007a, Martin:2014a}.
This requirement is met by peak patterns. 
Mesa-like patterns, in contrast, sharply separate two spatial domains. This is, for instance, a feature of PAR-protein patterns in \textit{C.~elegans}~\cite{Goehring:2011a, Hoege:2013a, Gross:2018a, Gessele:2020a}.
Interestingly, the stem cells that polarize via PAR-protein pattern formation become smaller after each cell division during morphogenesis of the \textit{C.~elegans} embryo \cite{Hubatsch:2019a}. As the cell size approaches the interface width, the patterns transition from the mesa to the `weakly nonlinear' type (cf.\ Sec.~\ref{sec:sub-supercrit}).
Finally the system size becomes so small that the Turing instability is suppressed (see discussion of the marginal mode $\qmax$ in Sec.~\ref{sec:lateral-instability}).
This size-dependent loss of cell polarization has been shown to be important for stem-cell fate decision~\cite{Hubatsch:2019a}.
 
An important difference between peak and mesa patterns is how they respond to changes in average total protein density, e.g.\ owing to up- or down-regulation of gene expression, or system size, e.g.\ due to growth. 
An increase in average mass makes peak patterns grow in amplitude, while for mesa patterns it leads to a shift in the interface position. 
Upon an increase in system size, peak patterns grow in amplitude, because the total number of proteins in the system increases. In contrast, amplitude and the relative interface position of mesa patterns remain unchanged. 
Hence, mesa patterns inherently scale with the system size---a property that is desirable in developmental systems.
Note, however, that the interface width of mesa patterns does \emph{not} scale with the system size, and is independent of the average total density (Sec.~\ref{sec:pattern-types}).

Another difference between patterns composed of peaks vs patterns composed of mesas is their rate of coarsening by competition for mass. Peak patterns coarsen significantly faster than mesa patterns~\cite{Brauns:2020a,Chiou:2018a}. Fast coarsening is important if the biological function requires selection of a single polarity site, as for instance cell division of budding yeast~\cite{Park:2007a, Martin:2014a}.

\subsection{Non-equilibrium phase separation} \label{sec:discussion-non-equilibrium}

On the phenomenological level, the dynamics of 2C-MCRD systems closely resembles a phase separation process, as exhibited by binary mixtures that undergo liquid-liquid phase separation near \emph{thermal} equilibrium. 
We showed that the bifurcation diagram of 2C-MCRD systems (Fig.~\ref{fig:n-Dc-diagram-monostable}), obtained by the flux-balance construction on the reactive nullcline, resembles the phase diagram of phase separation with spinodal and binodal lines that meet in a critical point.
The analogous process to spinodal decomposition is the mass-redistribution instability (Sec.~\ref{sec:lateral-instability}). In both cases, the condition for instability is that a potential decreases as a function of the total density. In the former case, it is the \emph{chemical} potential, derived from a free energy functional, while in the latter case it is the mass-redistribution potential $\eta^*(n)$ derived from the reactive nullcline.
Moreover, nucleation in the binodal regime of binary-mixture phase separation corresponds to stimulus-induced pattern formation in 2C-MCRD systems (Sec.~\ref{sec:stim-induced}).
In addition to their equivalent phase diagrams, binary-mixture phase separation and 2C-MCRD systems and both exhibit uninterrupted coarsening~\cite{Brauns:2020a}.

This phenomenological equivalence between binary-mixture phase separation and 2C-MCRD dynamics is quite remarkable, since the former describes systems close to thermal equilibrium, while the latter is inherently far from thermal equilibrium (driven by a chemical fuel like ATP/GTP in the case of proteins).
In fact, segregation into domains of high and low density is observed in many other non-equilibrium systems, most prominently self-propelled particles that exhibit formation of polar waves and nematic lanes \cite{Chate:2020a,Huber:2018a} and motility-induced phase separation (MIPS)~\cite{Cates:2015a}.
Further examples include active contractility~\cite{Bois:2011a}, motile bacteria \cite{Liu:2019a}, and shear banding \cite{Olmsted:2008a}.
Some authors have used the term ``active phase separation'' \cite{John:2005a,Bergmann:2018a,AgudoCanalejo:2019a} for such phenomena.

One interesting feature shared by many phase-separating systems is that they exhibit interrupted coarsening (`micro-phase separation') once the mass conservation is weakly broken by additional production and degradation terms. The strength of these terms determines the length scale where coarsening arrests. This holds true both near thermal equilibrium \cite{Glotzer:1995a,Oono:1987a} and for MIPS of self-propelled particles subject to a birth--death process~\cite{Cates:2010a,Curatolo:2019a}.
In a followup work to the present manuscript, we show that the same holds true for two-component reaction--diffusion systems with weakly-broken mass conservation~\cite{Brauns:2020a}. 
Interestingly, interrupted coarsening can also occur as a consequence of non-reciprocal coupling that destroys the variational nature of the Cahn--Hilliard model~\cite{Saha:2020a}.

Given the phenomenological equivalence to phase separation near thermal equilibrium, one might be tempted to search for a mapping to an effective thermodynamic language or develop an entirely phenomenological thermodynamic description. 
In fact, for 2C-MCRD systems with a specific form of reaction kinetics, $f \,{=}\, c \,{-}\, g(m)$, an effective free-energy functional can be constructed~\cite{Morita:2010a, Goh:2011a, Jimbo:2013a}, giving a gradient-flow structure to the 2C-MCRD dynamics with such a reaction term. In this specific case, some of the results presented here, like the phase diagram with binodals and spinodals, can be inferred directly from the mapping to equilibrium phase separation (see e.g.\ Ref.~\cite{Goh:2011a}).
Moreover, the mapping implies uninterrupted coarsening for this specific form of reaction kinetics.

However, such an approach disregards the actual underlying non-equilibrium physics.
Our analysis of 2C-MCRD systems shows how a framework can be developed that is rooted in the underlying physics (here chemical reactions and diffusion) and not subject to the restrictions of a mapping to an effective thermodynamic description.
Concretely, we show in a followup work that uninterrupted coarsening is generic in 2C-MCRD systems independently of the specific reaction kinetics~\cite{Brauns:2020a}.
Moreover, a thermodynamic description cannot account for the rich phenomenology, including and oscillatory patterns and waves~\cite{Bernitt:2017a,Kuwamura:2015a,Kuwamura:2017a, Radszuweit:2013a, Weber:2017a, Halatek:2018a,Brauns:2020b}, that arises once coupling to additional components or source terms breaking mass-conservation are included.
In contrast, local equilibria theory has proven useful also in these more complex scenarios~\cite{Halatek:2018a, Brauns:2020a, Brauns:2020b}. We further discuss this perspective in the final two subsections of the \hyperref[sec:outlook]{Outlook}.

Instead of describing attachment and detachment of proteins at a membrane, Eq.~\eqref{eq:two-comp-dyn} can be interpreted as the mean field equation for particles undergoing Brownian motion and switching between two states with different velocities/tumbling rates.
Then the reaction term $f = f(n)$ describes the switching dynamics that depends on the local density of particles (e.g.\ by some quorum sensing mechanism), and $m$ and $c$ are the concentrations of slow and fast diffusing particles, respectively.
This system constitutes a minimal example for MIPS and shows that MIPS and Turing instability are analogous on the mean field level.
MIPS is typically studied for particles with continually varying velocity that depends on the local particle density. For particles switching between two states with different velocities, the mean velocity varies continuously as a function of the total density.

Hence, we define the average diffusion constant
\begin{equation}
	\bar{D} = \frac{D_m m + D_c c}{m + c},
\end{equation}
which is directly connected to the mass-redistribution potential via the identity $\eta = n \bar{D}/D_c$.
With this, the nullcline-slope condition for lateral instability, $\partial_n \eta^* < 0$, can be recast as
\begin{equation} \label{eq:MIPS-condition}
	\frac{\partial_n \bar{D}^*}{\bar{D}^*} < -\frac{1}{n}.
\end{equation}
(As always, the star denotes evaluation at the reactive equilibrium; the function $\bar{D}^*(n)$ can be obtained from the reactive nullcline, $\eta^*(n)$, via the relation $\bar{D}^*(n) = D_c \eta^*(n)/n$.)
Notably, the condition for MIPS has exactly the same form as Eq.~\eqref{eq:MIPS-condition}, where the density dependent particle velocity $v(n)$ takes the place of the average diffusivity $\bar{D}^*(n)$.

This reveals the \emph{common underlying principle} of Turing instability in MCRD systems and the instability driving MIPS of self-propelled particles: slowing down of particles in regions of high density. For an instability to occur, particles have to slow down enough in response to an increase in density, cf.\ Eq.~\eqref{eq:MIPS-condition}.

Another phenomenon that can be pictured as a phase separation process is shear banding in complex fluids. 
In Appendix~\ref{app-sec:shear-banding}, show how one can establish an analogy between our flux-balance construction on the reactive nullcline for 2C-MCRD systems and the ``common total stress'' construction on the \emph{constitutive} relation employed to analyze shear banding
~\cite{Olmsted:2008a,Divoux:2016a}.
This analogy shows that these physically distinct systems are \emph{topologically equivalent}, i.e.\ share the same phase space geometry.

\subsection{Outlook} \label{sec:outlook}

Based on the theoretical concepts presented in this work for 2C-MCRD systems there are several promising directions for future research. 
First of all, for 2C-MCRD systems several follow-up works elucidate the principles at work during wavelength selection~\cite{Brauns:2020a} and address how self-organization may be controlled by spatiotemporal ``templates''~\cite{Wigbers:2020a} and by advective flows~\cite{Wigbers:2020b}.
Going forward, it will be interesting to explore various avenues toward generalization to MCRD systems with more components and conserved species, including applications to various specific physical and biological systems as outlined shortly below.
The long-term perspective is a generalization toward a geometric theory of MCRD systems. 
Below, we discuss some possible routes towards such a generalization in more detail.

In a broader context, reaction--diffusion systems are part of a large class of non-equilibrium systems that are able to form self-organized patterns.
This includes models for living matter where molecular motors generate active flows \cite{Bois:2011a,Kumar:2014a} and active visco-/poroelastic deformations of the medium \cite{Radszuweit:2013a,Weber:2018b}, as well as particle-based active matter \cite{Marchetti:2013a,Cates:2015a,Chate:2020a} and granular media~\cite{Aranson:2006a}.
Ultimately, it may be fruitful to apply the concepts presented here to such systems.

\subsubsection{Generalization to more complex phenomena}

The 2C-MCRD system studied here has a comparatively simple phenomenology, exhibiting only steady states that are stationary (no oscillations) and no wavelength selection (uninterrupted coarsening).
In a previous study~\cite{Halatek:2018a}, and recently in Ref.~\cite{Brauns:2020b}, the concepts of mass redistribution and local equilibria have already been successfully employed to analyze a multicomponent multispecies MCRD model (for MinDE \textit{in vitro} pattern formation) that exhibits much more complex phenomena, like spatiotemporal chaos (chemical turbulence) at onset and a transition to order (standing and traveling waves).
These phenomena, observed in numerical simulations, can be understood in terms of the changing local stability of local equilibria due to mass redistribution.

In particular, this study revealed intriguing, highly non-trivial, connections between the nonlinear pattern dynamics far from the homogeneous steady state and the dispersion relation that characterizes the linearized dynamics in the vicinity of the homogeneous steady state.
However, these findings are model specific, and rely on numerical simulations.
In contrast, the characterization of the 2C-MCRD systems presented here is independent of the specific model (reaction term) and enables us to predict the pattern formation dynamics from a simple graphical analysis, without the need to perform numerical simulations. 
The obvious next step is to generalize this level of understanding to more complex phenomena by studying three-component MCRD systems.
Such models have recently been employed to model various (bio-)physical phenomena in numerical studies~\cite{Murray:2017a, Alonso:2010a}. 
Studying these models using local equilibria theory might reveal the principles underlying their dynamics and provide a good starting point for a generalization of the theory presented here.

For ODE dynamics, the seemingly small step from two to three variables increases the diversity of phenomena dramatically.
We expect a similar situation for spatially extended systems. 
Note in particular that, with three components and one conserved quantity, the reactive phase space will be two-dimensional. This allows for more complex local dynamics and attractors, such as limit cycle oscillations.
A fully general study of three-component MCRD systems will therefore probably not be possible from the outset. 
Instead we propose to focus on cases where a time-scale separation enables one to build on the results for two-component systems. 
A good starting point is to study cases where the coupling to the additional third component is slow (the model investigated in Ref.~\cite{Murray:2017a} is of that form) using, for instance, a singular perturbation analysis.
Closely related to such three-component MCRD systems are \emph{nearly} mass-conserving two-component systems that contain production/degradation terms on a slow timescale (see next subsection).
Making use of such timescale separations is has proven to be a powerful strategy to study complex phenomena in dynamical systems; see for instance Ref.~\cite{Izhikevich:2000a} for a comprehensive overview of three-variable ODE systems in the context of neural excitability.

The benefit of such an approach is that the effects of strong nonlinearities may be captured by a geometric phase-portrait analysis of the fast two-variable subsystem. More complex behavior arises as the slow dynamics modifies the fast subsystem, driving it through bifurcations.
Analogously, in reaction--diffusion systems, the effects of the strong nonlinearities are encoded in the shape of the nullcline, which enables one to construct the elementary patterns (mesas, peaks).
Much richer phenomenology can arise by the modification of these elementary patterns due to additional \emph{linear} (or weakly nonlinear) terms, whose effects could be studied using the method of regional phase spaces (Sec.~\ref{sec:regions}).
This can be viewed as a dual (``strong coupling'') approach to the amplitude equation formalism (weakly nonlinear analysis), where the elementary pattern originates in the narrow band of unstable modes, and the modification of the pattern due nonlinearities that couple these modes perturbatively. 

Another promising direction of study is the role of noise in stochastic MCRD systems.
Noise-induced phenomena in reaction--diffusion systems have previously been studied for models with (e.g.\ \cite{Altschuler:2008a, Jilkine:2011b}) and without conserved quantities (e.g.\ \cite{Sagues:2007a, Schumacher:2013a, Biancalani:2017a, Karig:2018a, Adamer:2018a}).
Importantly, noise in reaction--diffusion systems is not constrained by the fluctuation-dissipation theorem, but must be inferred from the stochasticity of the chemical processes by explicit coarse graining, using, for instance, path integral approaches~\cite{Weber:2017a}.

\subsubsection{Model reduction and classification}

While multicomponent, multispecies models can exhibit complex phenomenology, as discussed above, this is not \emph{necessarily} the case.
For instance, cell polarization of eukaryotic cells is often brought about by a large number of interacting protein species.
Well-studied examples are the Cdc42 system of \textit{S.\ cerevisiae}~\cite{Wedlich-Soldner:2003a, Goryachev:2008a, Klunder:2013a, Chiou:2018a} and the PAR system of \textit{C.\ elegans}~\cite{Goehring:2011a, Trong:2014a, Gessele:2020a, Gross:2018a}.
Despite the complexity of these systems, their phenomenology---cell polarization---is simple and can already be captured by two-component systems as studied here. 
This raises the question whether complex models can be reduced to an underlying minimal `core' that captures their essential phenomenology.
The finding that redistribution of conserved quantities is the essential driver of pattern formation in MCRD systems suggests that such a reduction might be possible in the phase space of conserved quantities---the \emph{control space}. 
Note that such reduced `core' systems may comprise more than two components and more than a single conserved mass.
As we detail in the next subsection, even non-conservative models can have a mass-conserving `core'.

The central theme of local equilibria theory is that conserved quantities are control parameters for reactive equilibria. 
Hence, the bifurcation scenario of reactive equilibria in the control space may serve as a criterion for classifying models. 
One example of such a class is the cusp bifurcation scenario, found by numerical analysis of various cell-polarization models in Ref.~\cite{Trong:2014a}, and identified here as the general bifurcation scenario underlying pattern formation in 2C-MCRD systems. 
In a forthcoming work, we will use our theory to elucidate the control-space geometry underlying the pole-to-pole oscillations of the \emph{in vivo} MinDE system~\cite{MinInVivo:unpublished}). 
As there are two conserved quantities, the total densities of MinD and MinE respectively, the control space is two-dimensional and there are surfaces (instead of lines) of reactive equilibria. 
Using the local quasi-steady state approximation, these \emph{nullcline surfaces} allow for a geometric analysis of the \emph{in vivo} MinDE dynamics.
We propose this geometrically motivated approach as an alternative to algebraically motivated model reduction methods, such as the quasi-steady-state approximation of slowly diffusing components \cite{Smith:2018a} or (extended) center-manifold reduction \cite{Sewalt:2015a, Doelman:2015a}. 
Such an approach offers the advantage that it does not require abstract mathematical calculations and instead enables one to gain physical intuition from elementary geometric objects and graphical constructions. 
Furthermore, as pointed out earlier, reactive equilibria in principle allow one to assess phase-space geometry experimentally.
This could ultimately make it possible to infer theoretical models from experimental data at a mesoscopic level, especially in situations where access to more molecular information (at the protein level) is not available yet.

\subsubsection{Beyond strict mass conservation} \label{sec:beyond-mass-conservation}

What can we learn from local equilibria theory about systems without strict mass conservation?

\begin{figure}
	\centerline{\includegraphics{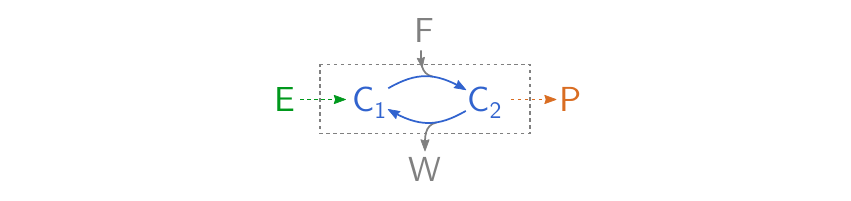}}
	\caption{
	Basic elements of a (chemical) system with a mass-conserving `core'. The cycling of the `core' between two states, $\mathrm{C}_1$ and $\mathrm{C}_2$, is driven out of thermal equilibrium by consumption of a fuel F and production of a waste W. 
		If the fuel is abundant, and replenished from the outside it can be assumed to be present at constant concentration.
		The cycling components are produced from a precursor E and irreversibly degraded to a product P. When production and degradation are slow compared to the rate of cycling between $\mathrm{C}_1$ and $\mathrm{C}_2$, the system inside the dashed box may be treated as mass-conserving on the fast timescale of cycling.
	}
	\label{fig:beyond-mass-conservation}
\end{figure}

Non-equilibrium systems are dissipative, that is, they consume some sort of chemical fuel (e.g.\ ATP in biological systems and malonic acid in the Belousov--Zhabotinsky (BZ) reaction \cite{Zhabotinsky:1964a}) that drives them far from thermal equilibrium.
The chemical fuel often drives cycling of components between different states, as illustrated in Fig.~\ref{fig:beyond-mass-conservation}. Examples are cycling of \mbox{NTPase} proteins between active and inactive states \cite{Neves:2002a}; phos\-pho\-ry\-la\-tion--de\-phos\-pho\-ry\-la\-tion \cite{Hansen:2019a,Alonso:2014a} and membrane attachment--detachment of proteins \cite{Halatek:2018b}; and cycling between molecular bromine and bromide ions and cycling of a metal catalyst (e.g.\ cerium) in the BZ reaction \cite{Belousov:1959a, Zhabotinsky:1964a, Field:1972a}.
While such cycles consume a fuel (and produce a waste), they conserve the total density of the cycling components ($\mathrm{C}_1$ and $\mathrm{C}_2$ in Fig.~\ref{fig:beyond-mass-conservation}).
The chemical fuels are quite generally assumed to be supplied from a large reservoir and hence are not explicitly modeled.
Several recent works study explicitly the role of a finite fuel supply in the framework of stochastic thermodynamics \cite{Rao:2018a,Avanzini:2019a,Ehrmann:2019a}. In particular, in Ref.~\cite{Ehrmann:2019a}, the bifurcations of bistable and oscillatory well-mixed systems are studied as they approach thermal equilibrium where detailed balances holds (vanishing chemical potential difference between fuel and waste).

In addition, there may be production and (irreversible) degradation processes that break the (strict) conservation of the cycling components.
To dissect the roles of conserving and non-conserving processes, any reaction kinetics can be split up into the respective functional terms.
For a general two-component reaction--diffusion system, such a splitting would take the form
\begin{subequations} \label{eq:two-comp-dyn_non-cons}
\begin{align}
	\dt m(x,t) &= D_m \gradx^2 m + f(m, c) \; + \;  \gamma \, s_1(m,c), 
	\label{eq:m-dyn_non-cons} \\
	\dt c(x,t) &= \kern0.26em D_c \gradx^2 c \kern0.5em - f(m, c) \; + \; \gamma \, s_2(m,c).
	\label{eq:c-dyn_non-cons}
\end{align}	
\end{subequations}
where $\gamma$ describes the ratio of the timescale of mass-conserving reactions encoded in $f$ and the timescale of production/degradation processes encoded in the source terms $s_{1,2}$ \cite{Note:GammaInterpolation}.
If $\gamma$ is small, one can study the mass-conserving subsystem (limit $\gamma \to 0$) using the theory presented here. In particular the capacity for pattern formation of the mass-conserving subsystem can be directly read off from the shape of the $f$-nullcline.
The remaining effect of the source terms in this limit is to set the total average mass $\bar{n}$, which is no longer a control parameter when $\gamma \neq 0$, but is controlled by the parameters in $s_{1,2}$ instead.
Geometrically, this corresponds to setting the center of mass of a pattern in the $(m,c)$-phase plane.

A decomposition into mass-conserving and non-conserving terms of the form Eq.~\eqref{eq:two-comp-dyn_non-cons} can be carried out for systems with any number of components. (Note that this should be done \emph{before} model reductions like non-dimensionalization and variable eliminations that affect the mass-conserving terms.)
Linear stability analysis can then be used to determine whether the mass-conserving subsystem is able to form patterns by setting $\gamma = 0$.
This defines two classes of systems, (\textit{i}) those where pattern-forming instability is preserved in the mass-conserving case $\gamma = 0$, and (\textit{ii}) those that inherently depend on production and degradation (i.e.\ don't exhibit lateral instability for $\gamma = 0$).
Put differently, class (\textit{i}) systems do have a \emph{pattern-forming}, mass-conserving core, while class (\textit{ii}) systems don't.
Of course, the distinction between these two classes will be limited if $\gamma$ is too large such that production-degradation dynamics dominate over the mass-conserving core.

The mass-conserving `core' of systems (or regimes) in class (\textit{i}) can be analyzed using the local equilibria theory.
An immediate conclusion is that their lateral instability is driven by a mass-redistribution cascade.
This insight may inform model reduction of many-component models as we have discussed in the previous subsection.
Furthermore, one can then study how the `core patterns' are qualitatively and quantitatively modified by $0 \neq \gamma \ll 1$. 
Importantly, $\gamma \neq 0$ might be a singular perturbation, i.e.\ weakly broken mass conservation might cause a qualitative change in the dynamics, such as interrupted coarsening and oscillations; see example ($i$) below. Importantly, for $\gamma \ll 1$ these qualitatively new phenomena will play out at large length- and timescales, whereas the behavior at short scales is still determined by the mass conserving core; see e.g.\ Ref.~\cite{Brauns:2020a}.
In the language of bifurcation theory, the mass-conserving `core' would take the role of an ``organizing center'' from which the various dynamical regimes of the system ``unfold'' (see e.g.\ section~30 in Ref.~\cite{Arnold:Book} and chapter~3 in Ref.~\cite{Murdock:Book}).
Let us provide examples for the two classes of systems defined above.

\emph{Example for class \textup{(}i\kern0.12em\textup{):} Brusselator model.}\;--- Let us exemplify the reduction to a mass-conserving core for a classical model, the Brusselator \cite{Prigogine:1968a}. This early, qualitative model for the BZ reaction can be written in the form Eq.~\eqref{eq:two-comp-dyn_non-cons} where $f(m,c) = m^2 c - m$, $s_1 = k_\mathrm{in} - k_\mathrm{out} m$, $s_2 = 0$.
The nullcline of $f$ has two segments, $m = 0$ and $\ceq(m) = 1/m$, which is a singular case of an N-shaped nullcline.
Thus, it is immediately clear that pattern formation of the Brusselator is driven by a mass-conserving core and that its elementary stationary patterns are mesa patterns when $D_c \gtrsim D_m$ and peak patterns when $D_c \gg D_m$.
Slow production and degradation lead to interrupted coarsening and splitting of the mesa patterns at length scales that depend on $\gamma$ as has been studied using singular perturbation methods in the limit $\gamma \ll 1$  \cite{Kolokolnikov:2006a,Kolokolnikov:2007a,McKay:2012a,Tzou:2013a}.
Local equilibria theory, in particular the phase-portrait analysis of spatially extended systems that it facilitates, provides a new, intuitive approach that explains the physics underlying interrupted coarsening and mesa splitting~\cite{Brauns:2020a}.
Moreover, in the oscillatory regime of the Brusselator, the limit cycle oscillations can be constructed as relaxation oscillations on the basis of the nullcline of the mass-conserving core ($f$-nullcline) in the limit $\gamma \ll 1$~\cite{Kuwamura:2017a}. The oscillation period depends on $\gamma$ and diverges in the limit $\gamma \rightarrow 0$.

\emph{Example for class \textup{(}ii\kern0.12em\textup{):} Gierer--Meinhardt model.}\;--- An example for a two-component system without a mass-consering core (class (\textit{ii})) is the ``Gierer--Meinhardt'' model \cite{Gierer:1972a}. This model describes an ``activator'' which enhances its own production and the production of an ``inhibitor'' which impedes the ``activator's'' production. Both ``activator'' and ``inhibitor'' are degraded at constant rates.
Written in the form Eq.~\eqref{eq:two-comp-dyn_non-cons}, using $\{a,h\}$ for the ``activator'' and ``inhibitor'' concentrations instead of $\{m,c\}$, one has $f = 0$, $s_1 = k_a^+ + k_\mathrm{fb} a^2/h - k_a^- a$, $s_2 = k_h^+ + k_\mathrm{fb} a^2 - k_h^- h$.
Clearly, this system does not possess a mass-conserving core capable of pattern formation.
Generally, systems in class (\textit{ii}) are those where production (from a reservoir or substrate) and irreversible degradation are the dominant processes; for instance, during tissue patterning (morphogenesis) \cite{Manukyan:2017a, Watanabe:2015a, Kondo:1995a} or cell-cell signaling in bacterial colonies~\cite{Karig:2018a}.

\emph{The role of (nearly) conserved quantities in classical pattern-forming systems.}\;---
Many chemical systems contain a mass-conserving `core' of components that rapidly cycle between different states and are produced and degraded only on a much slower timescale.
An example is the cycling of bromine between a molecular form and a bromide ion in the BZ reaction.
It is an interesting question whether this core alone is able to produce some non-trivial behavior like oscillations or patterns (assuming that the chemical fuel driving the cycling of the core components is abundant, as e.g.\ ATP in protein-based pattern formation).
Systems where this is the case can then be analyzed using local equilibria theory (potentially extended to account for slow production and degradation, see e.g.\ \cite{Brauns:2020a}).
Above, we showed that this is true for the Brusselator, which is a conceptual model for the BZ reaction.
We hypothesize that a similar approach might/will also work for more detailed models such as the so-called FKN mechanism \cite{Field:1972a}, whose mass-conserving core presumably contains more than two components.

Such a program might eventually lead to a perspective that unifies ``classical'' pattern-forming systems such as the BZ reaction and the more recently discovered biological systems, including the Min system \cite{Loose:2008a,Halatek:2018a,Brauns:2020b}, intracellular actin waves \cite{Khamviwath:2013a,Bernitt:2017a}, and Rho excitability~\cite{Bement:2015a,Graessl:2017a,Michaux:2018a,Tan:2020a}.

\vspace{-0.25\baselineskip}
\subsubsection{Beyond reaction--diffusion systems}
\vspace{-0.25\baselineskip}

In the systems discussed so far (excluding Sec.~\ref{sec:discussion-non-equilibrium}), energy is fed into the system via the reaction kinetics alone, while the spatial transport process, diffusion, is passive.
Such systems are part of a broader class of so-called \emph{active} systems where energy is fed in on the microscopic scale. This comprises systems where the transport processes are active, driven e.g.\ by molecular motors such as myosin. 
Examples include systems with active flows generated by cortical contractions \cite{Bois:2011a,Kumar:2014a}, as well as actively deforming visco- and poroelastic media \cite{Weber:2018b,Radszuweit:2013a}.
On a more conceptual level, several non-equilibrium generalization of the Cahn--Hilliard equation have been studied recently~\cite{Zwicker:2015a, Tjhung:2018a, Weber:2019a, Saha:2020a}.
The phenomena exhibited by the above systems include (micro\mbox{-)}phase separation and more complex phenomena like waves and turbulence.
Another broad sub-class of active systems are self-propelled particles that exhibit a huge variety of collective phenomena, including MIPS (\cite{Cates:2015a}, see Sec.~\ref{sec:discussion-non-equilibrium} above) and flocking \cite{Chate:2020a,Huber:2018a,Denk:2020a}. 

What all these systems have in common with MCRD systems is the presence of conserved quantities that serve both as macroscopic variables and as local parameters of the microscopic dynamics. 
We therefore expect that the ideas on mass redistribution and local equilibria put forward in this article can be broadly applied to understand emergent behavior in these systems.
As a concrete example, the self-organized interplay between total density and emergent orientational order (polar or nematic) was investigated recently for self-propelled particles with microscopic alignment interactions that are continuously tunable between polar and nematic symmetry~\cite{Denk:2020a}.
Here, the \emph{local} particle density controls the emergent orientational order, i.e.\ induces local symmetry breaking. In turn, the orientational order leads to particle currents that redistribute the particle density. Strikingly, this interplay explains the coexistence of different macroscopic structures, such as polar flocks and nematic lanes, and the continual interconversion between them as recently observed in experiments and agent-based numerical simulations~\cite{Huber:2018a}.

\begin{acknowledgments}
We would like to thank B.~Lohner who was involved at early, preliminary stages of this project, and M.~Wigbers and B.~Polovnikov for critical reading of the manuscript and for providing valuable feedback.
E.F. acknowledges financial support from the Deutsche Forschungsgemeinschaft (DFG) via project P03 within TRR174 (`Spatiotemporal Dynamics of Bacterial Cells').
F.B. acknowledges financial support by the DFG Research Training Group GRK2062 (`Molecular Principles of Synthetic Biology').\\ 
F.B. and J.H. contributed equally to this work.
\end{acknowledgments}


\appendix

\section{Models used for illustration and numerical studies} \label{app-sec:models}

\begin{figure}
	\includegraphics{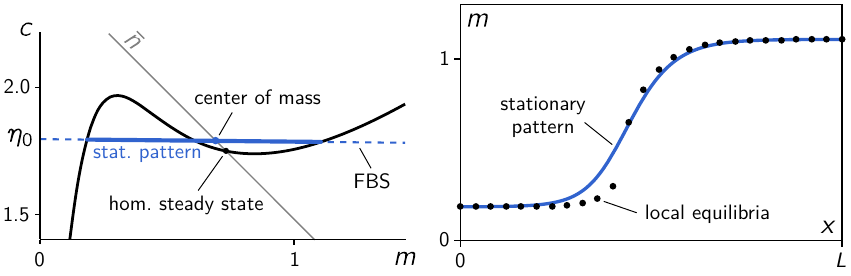}
	\caption{Left: numerically phase-space distribution (thick blue line) of a computed stationary pattern embedded in the flux balance subspace (dashed blue line). The thin gray line shows the reactive phase space corresponding to the average total density $\bar{n}$. Right: spatial profile (solid blue line) together with numerically determined local equilibria ($\bullet$). Parameters: $k = 0.07$, $\bar{n} = 2.48$, $D_m = 0.1$, $D_c = 10$, $L = 10$.}
	\label{app-fig:wp-model-stat-pattern}
\end{figure}

To visualize the our findings on pattern formation in 2C-MCRD systems we use two prototypical reaction terms, $f(m,c)$, that exhibit two distinct nullcline shapes. Both variants effectively model attachment--detachment dynamics as used to describe cell-polarization systems:

(\textit{i}) The reaction kinetics used in Ref.~\cite{Mori:2008a}, to conceptually describe cell-polarization based on autocatalytic recruitment (Michaelis-Menten kinetics with Hill coefficient 2) and linear detachment:
\begin{equation}
	f(m,c) = \left(k_\text{on} + k_\text{fb} \, \frac{m^2}{K_\text{d}^2+m^2}\right) c - k_\text{off} \, m.
\end{equation}
We can non-dimensionalize by expressing time in units of $k_\text{off}^{-1}$ and densities in units of $K_\text{d}$. Furthermore, for specificity we set the (non-dimensional) feedback rate $k_\text{fb}/k_\text{off} =: \hat{k}_\text{fb} = 1$, leaving only $k := k_\text{on}/k_\text{off}$ as free parameter in the non-dimensional reaction term:
\begin{equation} \label{app-eq:wave-pinning}
	f(m,c) = \left(k + \frac{m^2}{1+m^2}\right) c - m. \tag{\theequation*}
\end{equation}
Fig.~\ref{app-fig:wp-model-stat-pattern} shows the nullcline in the $(m,c)$-phase plane of the reaction kinetics Eq.~\eqref{app-eq:wave-pinning} for $k = 0.07$, together with a numerically determined stationary pattern (steady-state solution to Eq.~\eqref{eq:two-comp-dyn}) and the local equilibria (spatial profile on the right).  

\begin{figure}
	\includegraphics{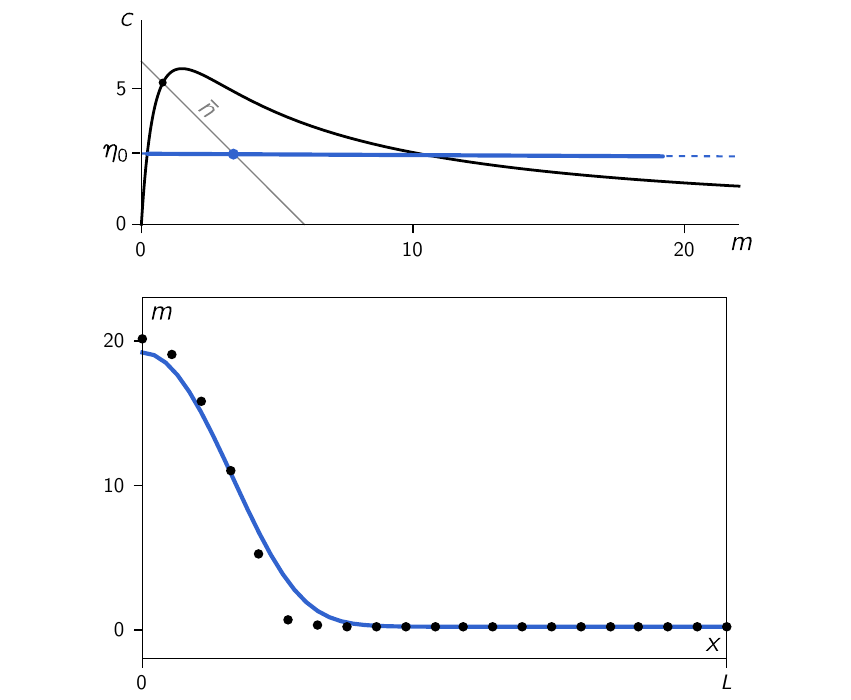}
	\caption{Above: Numerically determined phase space distribution and stationary pattern profile and for an MCRD system with the reaction kinetics Eq.~\eqref{app-eq:peak-model}. Below: stationary pattern together with local equilibria ($\bullet$). Parameters: $k_\text{fb} = 0.45$, $k_\text{off} = 16$, $\bar{n} = 6$, $D_m = 1$, $D_c = 200$, $L = 10$}
	\label{app-fig:peak-model-stat-pattern}
\end{figure}

(\textit{ii}) Dynamics due to attachment together with linear self-recruitment and enzyme driven detachment (described by first order Michaelis--Menten kinetics):
\begin{equation}
	f(m,c) = (k_\text{on}+k_\text{fb} \, m) c - k_\text{off} \, \frac{m}{K_\text{d}+m}.
\end{equation}
We non-dimensionalize by expressing time in units of the attachment rate $k_\text{on}$ and densities in units of the dissociation constant $K_\text{d}$ of the detachment kinetics. The two remaining parameters are the (non-dimensional) feedback rate $\hat{k}_\text{fb} := K_\text{d} k_\text{fb}/k_\text{on}$ and detachment rate $\hat{k}_\text{off} := k_\text{off}/(k_\text{on} K_\text{d})$:
\begin{equation} \label{app-eq:peak-model}
	f(m,c) = (1 + k_\text{fb} \, m) c - k_\text{off} \, \frac{m}{1+m}, \tag{\theequation*}
\end{equation}
where we suppressed the hats. Fig.~\ref{app-fig:peak-model-stat-pattern} shows a typical reactive nullcline for the reaction term Eq.~\eqref{app-eq:peak-model} together with a stationary peak pattern and the local equilibria that scaffold it. 

\paragraph{Length scale.---\kern-0.7em}For convenience, we do not specify a unit of length in the domain size $L$ and the diffusion constants $D_{m,c}$. In an intracellular context a typical size would be $L \sim \SI{10}{\micro m}$, and typical diffusion constants $D_m \sim 0.01{-}\SI{0.1}{\micro m^2.s^{-1}}$ on the membrane and $D_c \sim \SI{10}{\micro m^2.s^{-1}}$ in the cytosol. Rescaling to different spatial dimensions is straightforward. 

\paragraph{Nullcline shape predicts the pattern type.---\kern-0.7em}Important for the distinction between peak-forming vs.\ mesa-forming systems is the behavior of the reactive nullcline for large $n$. If it approaches the \mbox{$m$-axis} monotonically for large $n$, then peak patterns are will form in a large range of $\bar{n}$ for $D_m/D_c \ll 1$ (see Fig.~\ref{app-fig:peak-approximation}). Otherwise, mesa patterns are typical while peak/trough patterns only form in narrow regimes at the edges of the range of pattern existence $[n^\infty_-, n^\infty_+]$. For attachment--detachment kinetics, one can study the nullcline behavior for large $n$ by comparing the largest powers in the denominator and numerator of the functional form $\ceq(m) = m \, d(m)/a(m)$ of the nullcline. For the reaction term Eq.~\eqref{app-eq:wave-pinning}, one obtains $\ceq(m\to\infty) \to m$, i.e.\ typically mesa patterns. For the reaction kinetics Eq.~\eqref{app-eq:peak-model}, one has $\ceq(m\to\infty) \to 0$, favoring peak patterns for $D_c \gg D_m$.

\section{Numerical simulations}  \label{app-sec:simulations}

The reaction--diffusion dynamics were simulated on a domain with no-flux boundaries using the numerical PDE-solver routine \texttt{NDSolve[]} provided by Mathematica (see Supp.\ File ``PDE-solver\_minimal-setup.nb'' for an example setup). In the videos we show the density distribution in phase space and the real space profile $m(x,t)$ together with local equilibria.

\section{Linear stability analysis} \label{app-sec:lateral-instability}

This section provides the technical details of linear stability analysis.

\subsection{Canonical linear stability analysis}

Linear stability analysis of a reaction--diffusion system is performed by expanding a spatial perturbation in the eigenbasis of the diffusion operator (Laplacian) in the geometry of the system. In a line geometry with reflective boundary conditions at $x=0,L$, the eigenfunctions of the Laplacian are the discrete Fourier modes $\cos(k \pi x/L)$ with $k \in \mathbb{N}$. Linearization of the dynamics of a mass-conserving two-component system around a homogeneous steady state $(\meq,\ceq)$ yields the linear dynamics
\begin{equation} 
\label{eq:linear-stability-m-c}
	\dt \begin{pmatrix} \delta m_q(t) \\ \delta c_q(t) \end{pmatrix} = J(q) \begin{pmatrix} \delta m_q(t) \\ \delta c_q(t) \end{pmatrix} 
\end{equation}
with the Jacobian
\begin{equation} \label{eq:jacobian-m-c}
	J(q) = 
	\begin{pmatrix}
		-D_m q_k^2 + f_m & f_c \\
		- f_m & -D_c q_k^2 - f_c
	\end{pmatrix},		
\end{equation}
where we use the abbreviations $q_k = k\pi/L$ for the (discrete) wavenumbers and $f_{m,c} = \partial_{m,c} f\big|_{(\meq,\ceq)}$ for the linearized kinetics at the homogeneous steady state. The eigenvalues of the Jacobian yield the growth rates $\sigma^{(i)}_q$ of the respective eigenmodes such that a perturbation in the spatial eigenfunction $\cos(q x)$ evolves in time as
\begin{equation}
	\begin{pmatrix} \delta m_q(t) \\ \delta c_q(t) \end{pmatrix}  = \sum_{i=1,2} A^{(i)}_q \, \mathbf{e}^{(i)}_q \exp(\sigma^{(i)}_q t) \cos(q x),
\end{equation}
with the eigenvectors $\mathbf{e}^{(i)}_q$. For a given initial condition (perturbation), the coefficients $A^{(i)}_q$ are determined by projecting initial condition the onto the eigenbasis $\mathbf{e}^{(i)}_q \cos(q x)$.

\begin{figure}
	\includegraphics{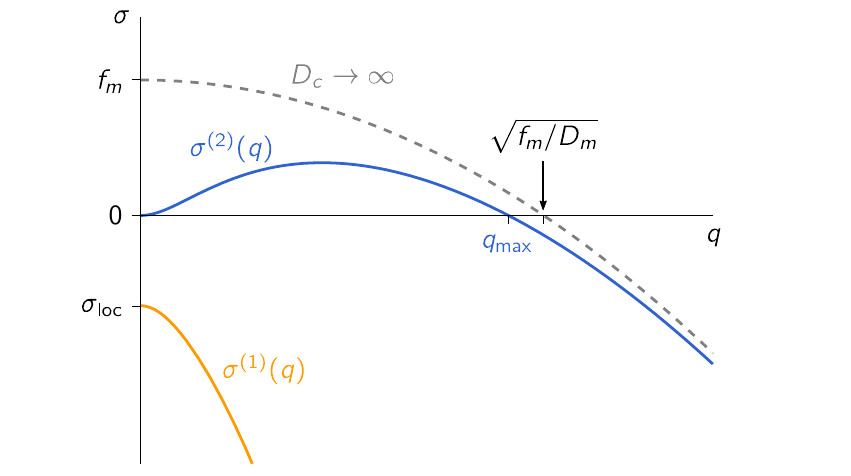}
	\caption{Generic dispersion relation of the 2C-MCRD system in a laterally unstable regime. The two branches $\sigma^{(1,2)}_q$ of the eigenvalue problem for $J(q)$ are shown in blue and yellow. At $q \to 0$, the branches connect to the eigenvalues $\sigma_\text{loc} = f_m - f_c$ and $0$ of the local stability problem. For $D_c \to \infty$, the second branch approaches $\sigma^{(2)}_q \to f_m - D_m q^2$ for $q > 0$. Accordingly, $\qmax$ approaches $\sqrt{f_m/D_m}$. ``Parameters'': $f_m = 0.7$, $f_c = 1$, $D_m = 1, D_c = 10$.
	}
	\label{app-fig:disp-rel}
\end{figure}

To calculate the eigenvalues of the Jacobian, we use that the eigenvalues of a $2\times2$-matrix can be expressed in terms of its trace $\tau$ and determinant $\delta$:
\begin{equation} \label{app-eq:eigenvalues-det-trace}
	\sigma^{(1,2)} = \tau/2 \pm \sqrt{\tau^2/4 - \delta},
\end{equation}
where the indices $\{1,2\}$ correspond to $\{-,+\}$ on the RHS.

The trace and determinant of the Jacobian $J(q)$ can be written as
\begin{subequations} \label{app-eq:Jacobian-trace-det}
\begin{align}
	\tau_q &= f_m - f_c - (D_m + D_c) q^2, \notag \\
	 &= \sigma_\text{loc} - (D_m + D_c) q^2, \label{app-eq:Jacobian-trace}\\
	\delta_q &= q^2 D_m D_c \left(q^2 + \frac{f_c}{D_c} - \frac{f_m}{D_m}\right), \notag \\ 
	&= q^2 D_m D_c \left(q^2 - \qmax^2\right),
\end{align}
\end{subequations}
where we used the expression Eq.~\eqref{eq:qmax} for $\qmax$.

From Eq.~\eqref{app-eq:Jacobian-trace}, it follows that for a locally stable hom.\ steady state ($\sigma_\text{loc} < 0$) the trace $\tau_q$ is negative for all $q$. Hence, the only way to get lateral instability is a negative determinant $\delta_q < 0$. This implies that eigenvalues with positive real part must be purely real, since the term under the square root in Eq.~\eqref{app-eq:eigenvalues-det-trace} is positive. This means that there cannot be oscillatory lateral instability for a locally stable hom.\ steady state. Moreover, the instability condition $\delta_q < 0$ immediately yields the band of unstable modes $[0,\qmax]$.

Figure~\ref{app-fig:disp-rel} shows the two branches of eigenvalues, $\sigma^{(1,2)}_q$, for a laterally unstable case. In the limit $q \to 0$, the first branch connects to the eigenvalue of a well-mixed system $\sigma^{(1)}_0 = \sigma_\text{loc}$ (cf.\ Sec.~\ref{sec:well-mixed}). The corresponding eigenvector lies in the reactive phase space for $q = 0$, and hence fulfills mass conservation. 

The second branch, $\sigma^{(2)}_q$, smoothly approaches zero in the limit $q \, {\to} \, 0$. The eigenvector $\mathbf{e}^{(2)}_0$ corresponding to the marginal eigenvalue $\sigma^{(2)}_0 = 0$ points along the reactive nullcline. It represents a perturbation that changes the total density and thus shifts the reactive equilibrium.
For the stability of a closed, well-mixed system (i.e.\ the stability against homogeneous perturbations) such a perturbation is not relevant since it breaks mass conservation.
For $q \neq 0$, the perturbation is spatially inhomogeneous, and therefore redistributes mass in the system. This mass redistribution shifts the local equilibria.
That the eigenvector points along the reactive nullcline reflects the fact that , the concentrations are slaved to the local equilibria in the long wavelength limit.
As one goes towards shorter wavelengths (i.e.\ larger $q$) the eigenvector begins to deviate from being tangential to the nullcline (see Supplementary Material of Ref.~\cite{Halatek:2018a}). 
In particular, the eigenvalue of the marginal mode $\qmax$ points along the flux-balance subspace, $\mathbf{e}^{(2)}_{\qmax} \propto (1,-D_m/D_c)^\text{T}$.   

We found that the band of unstable modes for the 2C-MCRD system always extends down to long wavelength ($q \to 0$), a situation called ``type~II'' instability in the Cross--Hohenberg classification scheme \cite{Cross:1993a}. In systems with more components and/or multiple conserved species, this is no longer true---the band of unstable modes can be bound away from zero (``type~I'' in Cross--Hohenberg scheme), see e.g.\ Ref.~\cite{Halatek:2018a}.

\subsection{Approximation close to the onset of lateral instability}

\begin{figure*}
	\centerline{\includegraphics{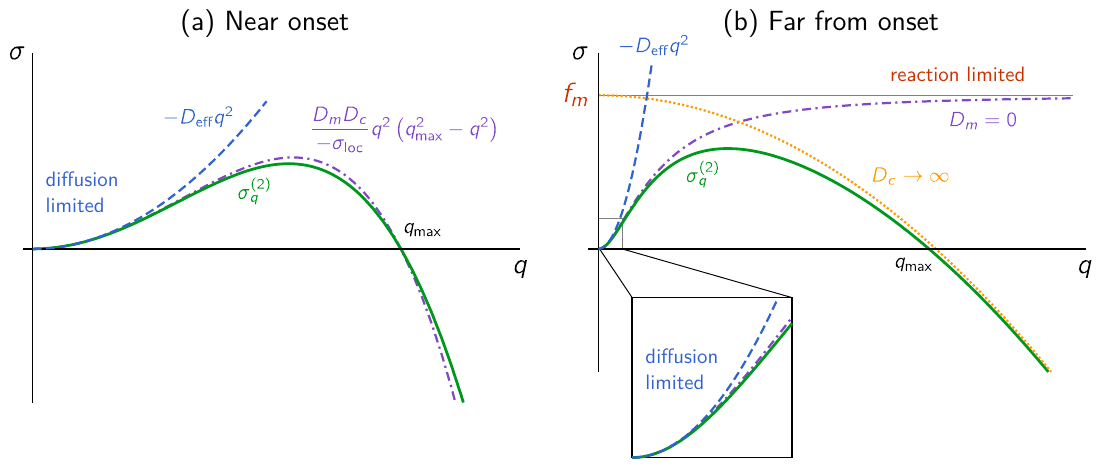}}
	\caption{
		Overview of the regimes of mass-redistribution instability and their interrelations.
		(a) Near onset ($D_m/D_c \approx -\snc$ and therefore $\qmax \ll \sqrt{f_m/D_m}$), the laterally unstable branch ($\sigma^{(2)}_q$, green solid line) of the dispersion relation is approximated by a fourth-order polynomial (purple, dash-dotted line; cf.\ Eq.~\eqref{app-eq:disp-rel-onset}(b)) that corresponds to the dispersion relation of Model~B dynamics. For $q \ll \qmax$, the instability is \emph{diffusion limited} (blue dashed line) and effectively described by anti-diffusion with the effective diffusion constant $D_\text{eff} = D_c \partial_n \eta^*$ (see Sec.~\ref{sec:lateral-instability}).
		(b) Dispersion relation far from onset ($D_m/D_c \ll |\snc|$). The case without membrane diffusion ($D_m =0$, purple dash-dotted line) clearly shows the diffusion-limited and the reaction-limited regimes for $q^2 \ll |\sigma_\text{loc}/D_c|$ and $q^2 \gg |\sigma_\text{loc}/D_c|$, respectively.
		Finite membrane diffusion suppresses the instability at short wavelength in the reaction-limited regime ($q \gg \sqrt{|\sigma_\text{loc}|/D_c}$). There, the dispersion relation is approximated by $f_m-D_m q^2$ (yellow dotted line, note in particular $\qmax \approx \sqrt{f_m/D_m}$, the approximation becomes exact in the limit $D_c \to \infty$). The diffusion-limited regime ($q \ll \sqrt{|\sigma_\text{loc}|/D_c}$, blue dashed line) has the same behavior as in the vicinity of onset (a) with the additional simplification that far from onset, the effective diffusion constant $D_\text{eff} \approx D_c \partial_n \ceq$.
		``Parameters'': $f_m = 0.5, f_c = 1$ (i.e.\ $\snc = -0.5$, $\sigma_\text{loc} = -0.5$), $D_m = 1$, $D_c = 2.1$ for (a) and $D_c = 50$ for (b). 
	}
	\label{app-fig:MRI-regimes}
\end{figure*}

The eigenvalues of a $2\times 2$-matrix, Eq.~\ref{app-eq:eigenvalues-det-trace}, to leading order in $\delta/\tau \ll 1$ are given by
\begin{subequations}
\begin{align}
	\sigma^{(1)} &= \tau - \delta/\tau + \mathcal{O}(\delta^2/\tau^2),\\
	\sigma^{(2)} &= \delta/\tau + \mathcal{O}(\delta^2/\tau^2).
\end{align}
\end{subequations}
We will now use this approximation for the 2C-MCRD dynamics where the trace and determinant of the Jacobian for a mode $q$ are given by Eqs.~\eqref{app-eq:Jacobian-trace-det}. A straightforward calculation shows that $\delta_q$ reaches its extremum at $q^* = \qmax/\sqrt{2}$, with a minimal value of $\delta_{q^*} = -\tilde{f}_m^2 D_c/(4 D_m)$.
Furthermore, the trace $\tau_q \approx \sigma_\text{loc}$ is approximated by the local eigenvalue for $\tilde{f}_m \ll \sigma_\text{loc}$.
Hence, the above approximation is valid in the vicinity of \emph{lateral} instability onset ($\tilde{f}_m \approx 0$), far away from the \emph{local} instability onset (which takes place at $\sigma_\text{loc} = 0$). The dispersion relation then reads
\begin{subequations} \label{app-eq:disp-rel-onset}
\begin{align}
	\sigma_q^{(1)} &\approx \sigma_\text{loc} - (D_m + D_c)q^2,\\
	\sigma_q^{(2)} &\approx \frac{D_m D_c}{- \sigma_\text{loc}} q^2 \left(\qmax^2 - q^2\right), \label{app-eq:disp-rel-onset_s2}
\end{align}
\end{subequations}
for $\tilde{f}_m^2 D_c/D_m \ll |\sigma_\text{loc}|$. The first branch, $\sigma_q^{(1)}$, simply represents relaxation to local equilibrium. The laterally unstable branch, $\sigma_q^{(2)}$, shown in Fig.~\ref{app-fig:MRI-regimes}a, is the identical to the dispersion relation of the Cahn--Hilliard equation \cite{Cahn:1961a} (and the more general class of Model~B dynamics). We can rewrite Eq.~\eqref{app-eq:disp-rel-onset_s2} as 
\begin{equation} \tag{\theequation b*} \label{app-eq:disp-rel-onset_rewritten}
	\sigma_q^{(2)} \approx -D_c q^2 \left( \partial_n \eta^*|_{\bar{n}} - \frac{D_m}{- \sigma_\text{loc}} q^2 \right).
\end{equation}
The two terms of this expression reflect the shifting of local equilibria (dominating at large wavelengths) driving the instability for $\partial_n \eta^*|_{\bar{n}} < 0$ and the competition of local reactive flow towards equilibrium and membrane diffusion that re-stabilizes the system on short length scales. (To recast Eq.~\eqref{app-eq:disp-rel-onset_s2} as \eqref{app-eq:disp-rel-onset_rewritten}, one uses the relation $\partial_n c^*(n) = \partial_m c^*(m)/[1+\partial_m c^*(m)]$.) 

Figure~\ref{app-fig:MRI-regimes} gives an overview of the various regimes of mass-redistribution instability and their interrelation. A typical dispersion relation (solid green line) deep in the laterally unstable regime (i.e.\ far from onset) is shown in Fig.~\ref{app-fig:MRI-regimes}b, together with various limiting cases (see caption for details).
One can distinguish diffusion- and reaction-limited regimes. The former (i.e.\ long wavelength limit) was studied in detail in the main text. 
In the next subsection, we briefly analyze the two limits of fast cytosol diffusion and vanishing membrane diffusion.

\subsection{Limits in the diffusion constants} \label{app-sec:diffusion-limits}

\paragraph{Fast cytosol diffusion.---\kern-0.7em}In the limit $D_c \to \infty$, the dispersion relation approaches the function $f_m - D_m q^2$ for wavenumbers $q \gg |\sigma_\text{loc}/D_c|$; see dashed gray line in Fig.~\ref{app-fig:MRI-regimes}b).
This shape of the dispersion relation reflects the fact that the growth rate of instability is limited by the rate of chemical relaxation to the shifting local equilibria ($\partial_n \ceq \cdot \sigma_\text{loc} = f_m$) and counteracted by membrane diffusion on short scales ($-D_m q^2$).

Somewhat deceptively, this shape of the dispersion relation resembles that of a locally unstable system for $q > 0$, even though the system is locally stable.
In the strict limit, $D_c \to \infty$, the dispersion relation becomes discontinuous at $q = 0$, because the zero eigenvalue at $q = 0$ corresponding to the homogeneous perturbation breaking mass conservation is always present. (Recall that for the stability of a closed system against homogeneous perturbations, this mode is not relevant.)

It is important to keep in mind that the lateral instability is always driven by cytosolic mass redistribution. 
From this perspective the strict asymptotic limit $D_c \to \infty$ (and the common approximation to treat the cytosol as well mixed) is pathological as it masks the core dynamics underlying lateral instability in MCRD systems: 
the formation of gradients in the fast diffusing component(s) and the ensuing diffusive fluxes that redistribute total density. 
Furthermore, in systems with more than two components, the subtle interplay of multiple fast diffusing components might play an important role for pattern formation~\cite{Denk:2018a}. Such aspects would be missed if fast diffusing components are assumed to be well-mixed at all times (which corresponds to setting their diffusion constants to infinity).

\paragraph{Vanishing membrane diffusion.---\kern-0.7em}In the limit $D_m \to 0$, the band of unstable modes extends to arbitrarily small wavelengths (i.e.\ $\qmax \to \infty$); see purple, dash-dotted line in Fig.~\ref{app-fig:MRI-regimes}.
For short wavelengths, the growth rate is reaction limited ($\sigma_q^{(2)} \approx f_m$) because cytosol diffusion is fast.
For long wavelengths, the growth rate is diffusion limited, $\sigma_q^{(2)} \approx D_c q^2 \, \partial_n \ceq$.

\section{Remarks on the local quasi-steady state approximation} \label{app-sec:adiabatic-scaffolding}

The local quasi-steady state approximation (LQSSA), i.e.\ slaving of the chemical concentrations to the local (reactive) equilibria, becomes exact when the timescales of diffusion and local reactions are separated. Specifically, let us scale the reaction terms in the reaction--diffusion dynamics Eq.~\eqref{eq:two-comp-dyn} by $\varepsilon^{-1}$:
\begin{subequations} \label{eq:two-comp-dyn-eps}
\begin{align} 
	\dt m &= D_m \gradx^2 m + \varepsilon^{-1} \, f(m, c), \label{app-eq:m-dyn-eps} \\
	\dt c &= \kern0.26em D_c \gradx^2 c \kern0.5em - \varepsilon^{-1} \, f(m, c).\label{eq:c-dyn-eps}
\end{align}
\end{subequations}
In the limit $\varepsilon \to 0$, relaxation to local equilibria becomes arbitrarily fast compared to diffusive redistribution---the concentrations will be at a local quasi-steady state (a stable local equilibrium). The characteristic spatial scale(s) of the dynamics and of stationary patterns are given by a balance of reaction and diffusion. In particular, in Sec.~\ref{sec:interface-width}, we learned that the interface width is determined by $\pi (D_m/\tilde{f}_m)^{1/2}$. Under the scaling of reaction rates by $\varepsilon^{-1}$, this length scale will go to zero as $\propto\varepsilon^{1/2}$. Hence, in the LQSSA, there is no ``microscopic'' length scale. This behavior is characteristic for a singular perturbation problem where some physics is lost when the small parameter $\varepsilon$ is set to zero \cite{Jackson:Book_Vol_1,Ward:2006a}. A rigorous analysis of Eqs.~\eqref{eq:two-comp-dyn-eps} could be performed in terms of singular perturbation theory. To lowest order in $\varepsilon$, any series of jumps (sharp interfaces) between two plateaus $n_\pm$ that fulfill $\eta^*(n_+) = \eta^*(n_-)$, such that $\eta^*(\nstat(x)) = \eta_0$ is constant in space, is a valid steady state of Eq.~\eqref{eq:n-dynamics-slaved}. However, to be consistent with Eq.~\eqref{eq:two-comp-dyn-eps} in the limit $\varepsilon \to 0$, the FBS-position must be $\eta_0 = \eta_0^\infty$ as determined by total-turnover balance Eq.~\eqref{eq:plateau-turnover-balance}. In addition, the given total density $\bar{n}$ constrains the spatial average $\langle \nstat(x) \rangle_{[0,L]} = L_+ n_+ + L_- n_- = \bar{n}$, where $L_\pm$ are the aggregate lengths of the high and low density regions.

One may compare the LQSSA to the approach used to analyze limit cycle attractors of relaxation oscillators. There the N-shaped nullcline allows an analytic construction of the limit cycle in the asymptotic timescale separation limit. Moreover, even without the timescale separation the qualitative phase-space structure that underlies the oscillations can be deduced from the nullcline shapes. Note that treating such a timescale separation in a mathematically rigorous way requires singular perturbation theory (see Appendix~\ref{app-sec:adiabatic-scaffolding}).

The LQSSA can also be understood as a closure relation. In that picture, $n(x,t)$ corresponds to a ``coarse-grained order parameter'' with the microscopically correct dynamics given by Eq.~\eqref{eq:n-dynamics}. 
This equation is not closed, because $m(x,t)$ and $c(x,t)$ are not known. 
Equation.~\eqref{eq:adiabatic-scaffolding} is a closure for Eq.~\eqref{eq:n-dynamics} at the price of losing the ``microscopic'' length scale. 
One could try construct higher order closures that also take into account deviations from the local equilibria owing to diffusion on short length scales \cite{Mapping:unpublished}.

In a forthcoming publication, we choose a less technical way to qualitatively illustrate the elementary pattern formation dynamics: we consider the dynamics not on a continuous domain but in two diffusively coupled compartments. In this `coarse grained' setting, the LQSSA is well-posed because a ``microscopic'' length scale as been imposed externally by the discretization into two-compartments.

\begin{figure*}
	\includegraphics{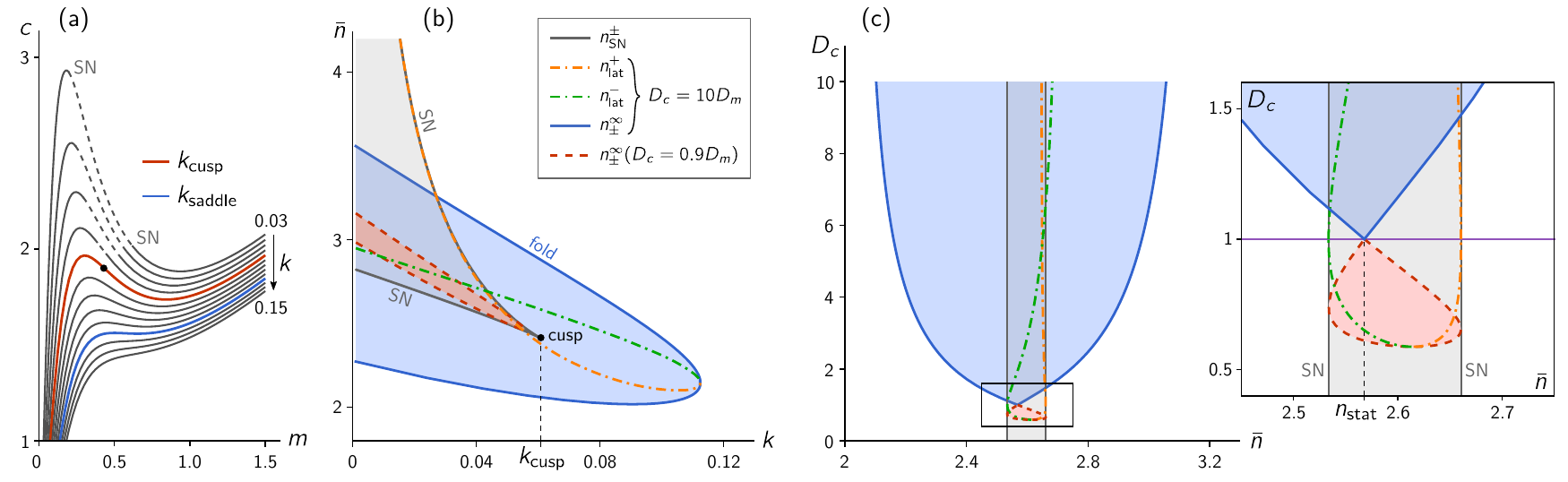}
	\caption{Bifurcation diagrams obtained by the flux-balance construction (reaction term Eq.~\eqref{app-eq:wave-pinning} from Ref.~\cite{Mori:2008a}).
	(a) Deformation of the reactive nullcline under variation of the kinetic rate $k$ from $0.03$ to $0.015$. Dashed sections indicate local instability in the regime of bistability, that emanates from the cusp bifurcation at $k_\text{cusp}$ (red nullcline, inflection point marked by black dot). The nullcline exhibiting a saddle point is shown in blue. 
	(b) $(k,\bar{n})$-bifurcation diagram (compare Fig.~\ref{fig:cusp-scenario}b in the main text) where the lateral instability bifurcation lines ($n_\text{lat}^\pm$, dash-dotted in orange and green) and the regime of pattern existence (shaded in blue, delimited by $n_\pm^\infty$ shown as solid blue line) are shown for $D_c = 10 D_m$. Additionally, the region where unstable stationary patterns exist for  $D_c = 0.9 D_m $ is shaded in red, delimited by a dashed red line. Note the region in the top-right corner, where the local reaction kinetics are bistable, but no stationary patterns exist. 
	(c) $(\bar{n},D_c)$-bifurcation diagram for $k\,{=}\,0.045$, where the reaction kinetics Eq.~\eqref{app-eq:wave-pinning} exhibit a region of bistability; the inset shows a blow-up of the boxed region around $\bar{n} \approx n_\text{stat}$ and $D_c \approx D_m$. 
		The bifurcation diagrams (b) and (c) were constructed based on following the geometric reasoning presented in Sec.~\ref{sec:bifurcation-structure} with the help of a Mathematica script (see Supp.\ File ``flux-balance-construction.nb'').
	}
	\label{app-fig:geometry-bifurcations}
\end{figure*}

\section{Geometric construction of bifurcations} \label{app-sec:geometric-bifurcations}

In Sec.~\ref{sec:n-bifurcations} and Sec.~\ref{sec:bifurcation-structure} we describe how the bifurcations diagrams of stationary pattern can be constructed geometrically using the reactive nullcline and the flux-balance subspace. We implemented this procedure in Mathematica (see Supp.\ File ``flux-balance-construction.nb'' to find quantitative bifurcation structures. As an illustrative example, we present the results for the reaction kinetics Eq.~\eqref{app-eq:wave-pinning}; see Figs.~\ref{app-fig:geometry-bifurcations} and~\ref{app-fig:n-Dc_finite-size}. Figure~\ref{app-fig:geometry-bifurcations}a shows the shape of the reactive nullcline for a range of the kinetic rate parameter $k$ (non-dimensional attachment rate). For $k > k_\text{saddle}$, the nullcline is monotonic, such that pattern formation is impossible. (Recall that we set the non-dimensional feedback rate to 1, so $k$ effectively describes the relative strength of basal attachment vs.\ feedback due to recruitment.) At $k = k_\text{saddle}$ a section of negative slope emerges on the nullcline, giving rise to lateral instability for $D_c/D_m \to \infty$. Further lowering $k$ increases the range of negative nullcline slope and increases the maximal negative nullcline slope (thus decreasing $D_c^\text{min}$, cf. Eq.~\eqref{eq:Dc_min}). Figure~\ref{app-fig:geometry-bifurcations}b shows the regimes of lateral instability and pattern existence for $D_c = 10 D_m$. At $k = k_\text{cusp}$ the maximal negative nullcline slope becomes $-1$, indicating a cusp bifurcation of the reactive equilibria. From this cusp point (black dot on the red nullcline in (a)), a regime of bistability emerges, section of unstable equilibria shown as dashed line in (a)). The locally bistable regime (shaded in gray in (b) is delimited by two saddle-node bifurcations (SN) which emerge from the cusp point, shown as black dot in (b). In the locally bistable regime, there exist \emph{unstable} stationary patterns for $D_c < D_m$. These patterns can be constructed in the same way as stable stationary patterns for $D_c > D_m$. Their range of existence for $D_c = 0.9 D_m$ is shaded in red in (b), delimited by a dashed red line.

Figure~\ref{app-fig:geometry-bifurcations}c shows the geometrically constructed $(\bar{n},D_c)$-bifurcation diagram for $k = 0.045$, i.e.\ for a bistable nullcline (corresponding to the schematic bifurcation diagram shown in Fig.~\ref{fig:n-Dc-diagram-bistable} in the main text). The $(\bar{n},D_c)$-bifurcation diagram for a monostable nullcline is shown in Fig.~\ref{app-fig:n-Dc_finite-size} in Appendix~\ref{app-sec:continuation}, where we also show the bifurcation structure for finite domain size, $L$, obtained by numerical continuation.

\section{Numerical continuation of stationary patterns} \label{app-sec:continuation}

To calculate steady states and their bifurcation structures for systems with finite size, we use a standard numerical continuation scheme (pseudo-arclength continuation, see e.g.\ chapter 4 in Ref.~\cite{HandbookDynSys:Book}). The stationarity condition Eq.~\eqref{eq:m-stat-fbs} was discretized using finite differences, yielding a set of equations for the concentrations at the grid points. These equations, together with the flux-balance subspace Eq.~\eqref{eq:flux-balance-subspace} and the constraint of average total density Eq.~\eqref{eq:total-mass}, are used to numerically determine the stationary patterns and their bifurcations (in the Mathematica software). To continue the fold bifurcations of stationary patterns, we use a bordered matrix method \cite{HandbookDynSys:Book}.

To determine the stability of the stationary patterns, we use a finite difference discretization of the reaction--diffusion dynamics \eqref{eq:two-comp-dyn} linearized around the steady state. The resulting eigenvalue problem is solved with Mathematica. The eigenvalue with the largest real part (``dominant eigenvalue'') determines the pattern stability (see Fig.~\ref{app-fig:pattern-stability}, which is discussed in Appendix~\ref{app-sec:FBS-stability}).  

\begin{figure}[tb]
	\includegraphics{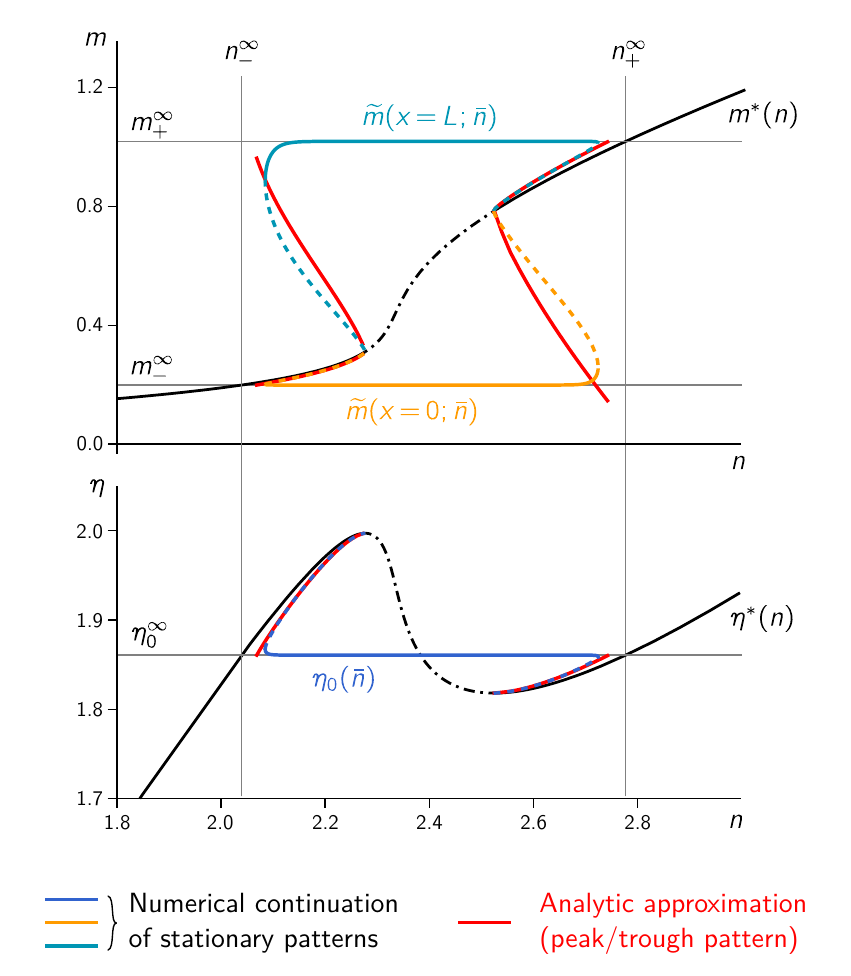}
	\caption{Numerically determined bifurcation diagram for a 2C-MCRD system with the reaction kinetics Eq.~\eqref{app-eq:wave-pinning} for the control parameter $\bar{n}$ (average total density). The figure supplements Fig.~\ref{fig:n-bifurcation-mesa-patterns} in the main text, where the bifurcation structure is shown for the pattern amplitude $|\mstat(L)-\mstat(0)|$. The homogeneous steady state is shown as black line, dash-dotted in the regime of lateral instability (note that the slope criterion, Eq.~\eqref{eq:slope-criterion}, can be written as $\partial_n \eta^* < 0$). For stationary patterns, concentrations at the domain boundaries $\mstat(0)$ and $\mstat(L)$ (yellow and teal lines in the top panel), and the FBS-position $\eta_0$ (blue line in the bottom panel) are shown. Thin gray lines indicate the plateau densities $m_\pm^\infty$ and the FBS-position, $\etaInfty$, in the large system size limit $L \to \infty$. Red lines show the heuristic approximation of peak/trough patterns Eq.~\eqref{app-eq:peak-profile-approx}, which are the unstable transition states in the multistable regimes (cf.\ Fig.~\ref{fig:n-bifurcation-mesa-patterns}). Note the almost perfect agreement of analytic approximation and numerical solutions for the FBS-position $\eta_0$.
	}
	\label{app-fig:wp-model_n-cont}
\end{figure}

\subsubsection{Bifurcation structure for $\bar{n}$}

Figure~\ref{fig:n-bifurcation-mesa-patterns} in the main text shows the $\bar{n}$-bifurcation structure of stationary patterns determined by numerical continuation for the reaction kinetics \eqref{app-eq:wave-pinning}. In Fig~\ref{fig:n-bifurcation-mesa-patterns}, the pattern amplitude is plotted against $\bar{n}$. For the same bifurcation structure, Fig.~\ref{app-fig:wp-model_n-cont}, shows additional plots of the maximum and minimum concentrations in (a) and the FBS-position, $\eta_0$, in (b). For mesa patterns $\mstat(0)$ and $\mstat(L)$ are the plateau concentrations and therefore slaved to $m_\pm^\infty = m_\pm(\etaInfty)$, while the FBS-position $\etaInfty$ is almost constant. At the boundaries of the range  where patterns exist (limited by $[n_-^\infty,n_+^\infty]$, for $L \to \infty$), the mesa patterns undergo fold bifurcations where they meet the branches of unstable peak/trough patterns (dashed lines) that emanate from the homogeneous steady state (black line, dash-dotted in the regime of lateral instability). In both plots, the prediction from the analytic approximation of (unstable) peak/trough patterns (see Appendix~\ref{app-sec:peak-patterns}) is shown as red solid lines.   

\subsubsection{Two-parameter $(\bar{n},D_c)$-bifurcation diagram}

\begin{figure}[b]
	\includegraphics{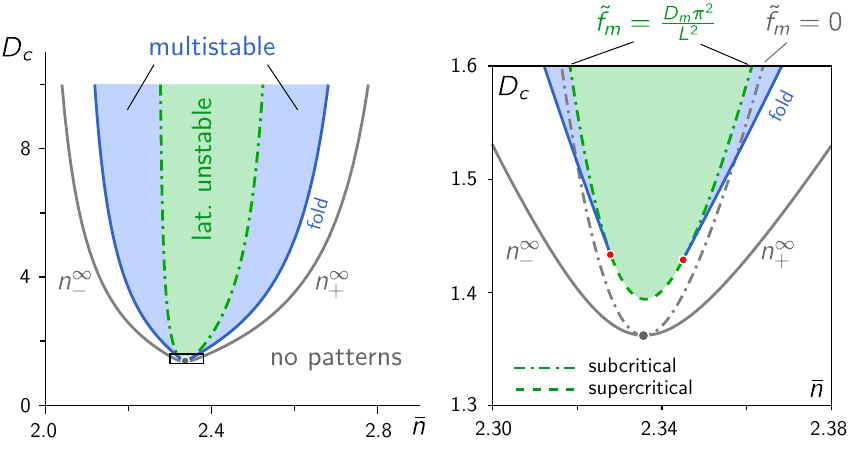}
	\caption{Right: Numerically determined $(\bar{n},D_c)$-bifurcation structure of stationary patterns in a finite sized domain; left: blow-up of the vicinity of the critical point. The continuous blue line marks the fold bifurcation of stationary patterns where stable and unstable stationary patterns meet (cf.\ Fig.~\ref{fig:n-bifurcation-mesa-patterns}b). The geometrically constructed bifurcation lines for $L\to\infty$ are shown in gray. The fold bifurcation where stable and unstable patterns merge terminates in the points where the Turing bifurcation switches from sub- to supercritical ($F_3 = 0$, cf.\ Eq.~\eqref{eq:steady-state-amplitude}). Along the line of supercritical Turing bifurcation (dashed green line), stable patterns emerge directly in a supercritical pitchfork bifurcation.
	Reaction term: Eq.~\eqref{app-eq:wave-pinning}; see Fig.~\ref{app-fig:wp-model-stat-pattern} for the nullcline shape and a typical pattern profile. Parameters: $k = 0.07$, $D_m = 1$, $L = 100$.
	}
	\label{app-fig:n-Dc_finite-size}
\end{figure}

Figure~\ref{app-fig:n-Dc_finite-size} shows the two-parameter $(\bar{n},D_c)$-bifurcation diagram for a monostable reactive nullcline  (corresponding to the schematic diagram in Fig.~\ref{fig:sub-vs-supercrit}a; the fixed parameters are the same as in Fig.~\ref{fig:n-bifurcation-mesa-patterns}). The fold-bifurcation lines (solid blue lines) of stationary patterns at finite domain size were obtained by numerical continuation. The respective bifurcation lines in the infinite system size limit, $n_\pm^\infty$ (solid gray lines), were geometrically constructed (cf.\ Appendix~\ref{app-sec:geometric-bifurcations}. The laterally unstable regime, bounded by the dash-dotted green line, was determined by linear stability analysis. On the right, a blow-up of the region around the critical point $(\bar{n}_\text{inf},D_c^{min})$. The finite sized system, 

The tip of the laterally unstable regime in a finite-sized system is shifted upwards by an amount $\sim L^{-2}$ because of the stability condition $\tilde{f}_m = \pi^2 D_m / L^2$. Close to critical point, the patterns emerge in a supercritical pitchfork bifurcation (dashed green line). The points where the onset becomes sub-critical are marked by in red disks. At these points the two lines of fold bifurcations of stationary patterns originate. The sub-critical lateral instability bifurcation is shown as green dash-dotted line.

\section{Approximation of peak/trough patterns} \label{app-sec:peak-patterns}

\begin{figure}
	\includegraphics{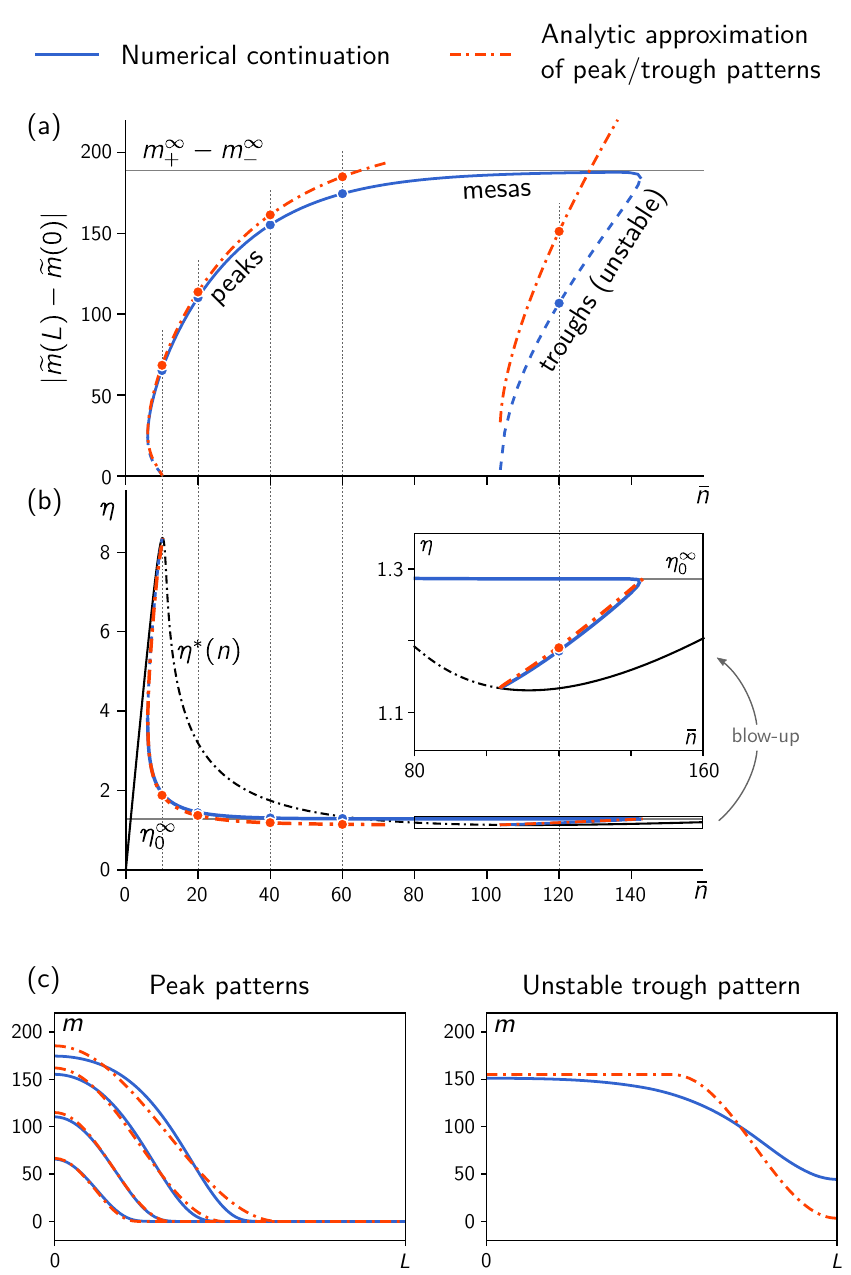}
	\caption{Numerically determined $\bar{n}$-bifurcation structure of stationary patterns for the reaction term Eq.~\eqref{app-eq:peak-model} (see Fig.~\ref{app-fig:peak-model-stat-pattern}a for the nullcline shape). 
	(a) Amplitude $|\mstat(L)-\mstat(0)|$ and FBS-position $\eta_0$ (b) of stationary patterns from numerical continuation (blue line, dashed for unstable patterns) and the analytic peak-approximation Eq.~\eqref{app-eq:peak-profile-approx} (dash-dotted red line). Peak-type patterns transition to mesa patterns as total average density is increased. The transition threshold can be estimated by the point where the peak approximation (red, dash-dotted line) for the pattern amplitude exceeds the (geometrically determined) plateau amplitude $|m^\infty_+ \,{-}\, m^\infty_-|$ (thin gray line).
	In the $(\bar{n},\eta_0)$-plot (b), the homogeneous steady states (equivalent to the reactive nullcline via $\eta^* = \ceq + \meq D_m/D_c$) are shown as solid black line (dash-dotted where laterally unstable). Inset: blowup of the $\eta$-axis in the trough pattern region.
	(c) Stationary pattern profiles from numerical continuation (solid blue lines) and from the  analytic approximation (dash-dotted red lines) for various total average densities, corresponding to the dots in (a). Left: $n = 10, 20,40,60$, stable peak patterns; right: $\bar{n} = 120$, unstable trough pattern.
	Fixed parameters: $k_\text{fb} = 0.3$, $k_\text{off} = 20$, $L = 20$, $D_m = 1$, and $D_c = 200$.
	}
	\label{app-fig:peak-approximation}
\end{figure}

In the following, we detail the construction of approximate peak/trough patterns that was briefly introduced in Sec.~\ref{sec:pattern-types}. For specificity, we present the construction for peak patterns---generalization to trough patterns is straight forward.

In our analysis of pattern types (Sec.~\ref{sec:pattern-types}), we characterized peak patterns as composed of an interface region at a system boundary connected to a plateau region (cf. Fig.~\ref{fig:profile-types}b). We have also characterized interfaces by linearization around the inflection point in Sec.~\ref{sec:interface-width} which yielded a sinusoidal interface shape with a width $\Lint(\eta_0)$ (cf.\  Eq.~\eqref{eq:Lint-approx}). We now construct a peak pattern by piecing together such an (approximate) interface at the left domain boundary and a plateau at $m_-(\eta_0)$ in the reminder of the system:
\begin{equation} \label{app-eq:peak-profile-approx}
	\mstat_\text{peak}(x) \approx \begin{cases}
 			m_0 + A \sin {\pi \left(\displaystyle \frac{x}{\Lint}  - \frac{1}{2}\right)} & x < \Lint \\
 			m_- & x > \Lint
 		\end{cases}.
\end{equation}
Within this approximation, the pattern inflection point is always at $x_0 = \Lint/2$. To match the interface to the plateau continuously at $x = \Lint$, the amplitude must be $A = m_- - m_0$. The plateau $m_-(\eta_0)$ and inflection point $m_0(\eta_0)$ are geometrically determined. To close the approximation, one has to find the FBS-position consistent with the given average total density $\bar{n}$ (to fulfill the constraint Eq.~\eqref{eq:total-mass}). From Eq.~\eqref{app-eq:peak-profile-approx}, one obtains the approximate total density average
\begin{align} 
	\bar{n}(\eta_0,L) \approx{}& \eta_0 + (1-D_m/D_c) \Big[m_-(\eta_0) + {} \notag \\
	&{}+\frac{\Lint(\eta_0)}{L} \big(m_0(\eta_0) - m_-(\eta_0)\big) \Big]. \label{app-eq:peak-eta-approx}
\end{align}
This relation can be inverted to obtain a relation $\eta_0(\bar{n},L)$ for peak patterns with a density-profile approximated by Eq.~\eqref{app-eq:peak-profile-approx}.

Peak/trough patterns are encountered in two contexts. First, stable peak patterns are typical for reaction kinetics that exhibit a strongly asymmetric nullcline shape, e.g.\ Eq.~\eqref{app-eq:peak-model}, when $D_m \ll D_c$ (see Fig.~\ref{app-fig:peak-model-stat-pattern} for a typical peak pattern). Secondly, \emph{unstable} peak/trough patterns form the unstable branches that connect the subcritical Turing bifurcation with the stable pattern branch (see Fig.~\ref{fig:n-bifurcation-mesa-patterns}). These unstable peak/trough patterns play the role of `transition' states since they lie on the separatrix that separates the basins of attraction of the homogeneous steady state and the stationary pattern in the multistable regimes. 

For both scenarios, we compared the analytic approximation Eq.~\eqref{app-eq:peak-profile-approx}, where $\eta_0$ is determined via Eq.~\eqref{app-eq:peak-eta-approx}, with numerical numerical continuation of the stationary patterns. 

The approximation of unstable peak/trough patterns for the reaction kinetics Eq.~\eqref{app-eq:wave-pinning} is shown in the bifurcation structure Fig.~\ref{app-fig:wp-model_n-cont}. 

Figure~\ref{app-fig:peak-approximation} shows the $\bar{n}$-bifurcation diagram of stationary patterns for the reaction kinetics Eq.~\eqref{app-eq:peak-model}. There is a large regime of peak patterns where the pattern amplitude keeps increasing with average total density $\bar{n}$. Peak patterns transition to mesa patterns when peak saturates in the third FBS-NC intersection point $m_+$. The amplitude of mesa patterns is almost independent of $\bar{n}$, because a change of total density merely shifts the interface position of mesa patterns (compare Fig.~\ref{fig:n-bifurcation-mesa-patterns}). Ultimately, mesa patterns transition to trough patterns which then undergo a fold bifurcation where they meet with the unstable branch of trough patterns that emerges from the homogeneous steady state. The asymmetry of the reactive nullcline (Fig. \ref{app-fig:peak-model-stat-pattern}a) is reflected by the asymmetry of the bifurcation structure. The dot-dashed red lines in Fig.~\ref{app-fig:peak-approximation} show the analytic approximation for pattern amplitude (a), the FBS-position (b) and pattern profiles (c). The approximation of peak patterns becomes less accurate as the average total density increases and ultimately breaks down at the transition to mesa patterns (around $\bar{n} \approx 60$ in Fig.~\ref{app-fig:peak-approximation}). The approximation of trough patterns is less accurate because the trough saturates more ``abruptly'' in $m_-$: Recall that the approximation underlying Eq.~\eqref{app-eq:peak-eta-approx} is a linearization of the interface region around the inflection point (cf.\ Sec.~\ref{sec:interface-width}). This approximation breaks down in regions of high nullcline curvature, indicative of high nonlinearities. Interestingly, even though the pattern profile is not well approximated for troughs (see Fig.~\ref{app-fig:peak-approximation}c), the estimate for the FBS-position $\eta_0$ is close to the true value (see inset in Fig.~\ref{app-fig:peak-approximation}c), indicating that the relevant physics (total turnover balance) is still captured. 

\section{Stimulus-induced pattern formation} \label{app-sec:stim-induced}

\begin{figure}
	\includegraphics{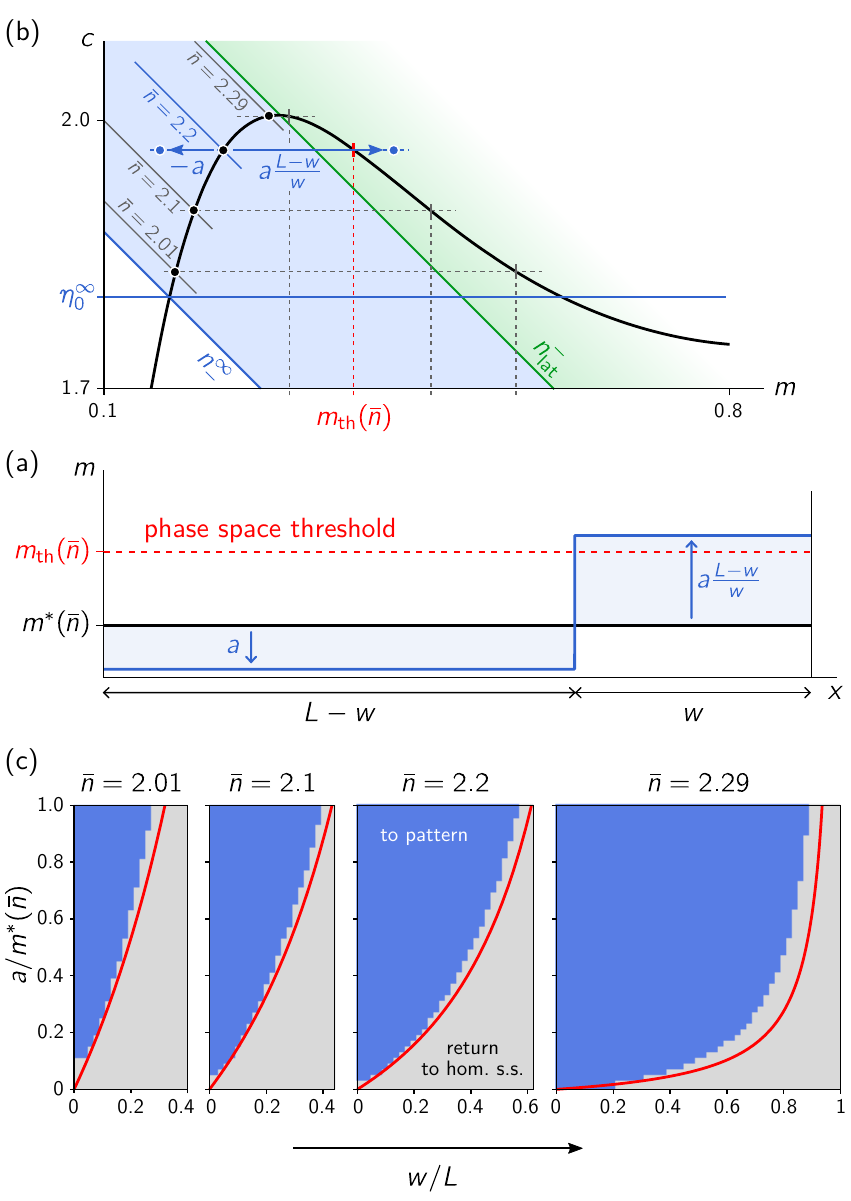}
	\caption{Test of the geometric heuristic for the perturbation threshold perturbation by numerical simulations.
	(a) We consider a prototypical type of perturbation of the homogenous steady state $\meq(\bar{n})$: a `step function' profile moving membrane density from the region $x < L-w$ into the region $x > L-w$; cf. Eq.~\eqref{eq:step-func-perturbation}. Membrane density is lowered by an amount $a$ on the left and increased by an amount $a\frac{L-w}{w}$ so that total mass is conserved.
	(b) In phase space, the threshold $m_\text{th}(\bar{n})$ for a perturbation of membrane density is determined by the intersection point of the line $c = \ceq(\bar{n})$ with the laterally unstable section of the nullcline. Colored dashed lines show this construction for various total densities $\bar{n} = 2.01, 2.1, 2.2, 2.29$ in the range of multistability $n_-^\infty < \bar{n} < n_\text{lat}^-$ (shaded in blue). The laterally unstable region is shaded in green.
	(c) The geometrically determined threshold $m_\text{th}(\bar{n})$ predicts (red line, $a_\text{th}(w;\bar{n})$) which `shapes' of perturbations, parametrized by amplitude $a$ and width $w$, trigger formation of a stationary pattern. This prediction is in good agreement with the basins of attraction of stationary pattern (shaded in blue) and homogeneous state (shaded in gray) in the parameter space of perturbation `shapes'.
	(Reaction term: Eq.~\eqref{app-eq:wave-pinning}, parameters: $k = 0.067$, $D_m = 0.1$, $D_c = 10, L = 20$.)
	}
	\label{app-fig:stim-induced}
\end{figure}

In Sec.~\ref{sec:stim-induced}, we argued that to trigger pattern formation from a laterally stable homogenous steady state, a perturbation (stimulus) must induce a \mbox{(self-)}sus\-tained laterally unstable region. Based on this intuition we provided a simple geometric heuristic for the perturbation threshold: a perturbation of the membrane concentration profile must be such that concentrations in a spatial region are pushed beyond the laterally unstable part of the nullcline in phase space. Hence, the intersection point of the line $c = \ceq(\bar{n})$ with the laterally unstable section of the nullcline, provides an estimate $m_\text{th}(\bar{n})$ for the threshold that the membrane perturbation has to exceed in a spatial region (see Fig.~\ref{app-fig:stim-induced}b, cf. Fig.~\ref{fig:stim-induced}). This criterion does not take into account the spatial shape of a perturbation, but only its characteristics in phase space. We tested how robust the estimate is against different spatial profiles using numerical simulations of the reaction--diffusion dynamics Eqs.~\eqref{eq:two-comp-dyn} with the reaction term Eq.~\eqref{app-eq:wave-pinning}. We consider prototypical perturbations with a `step-like' profile (see Fig.~\ref{app-fig:stim-induced}a):
\begin{equation} \label{eq:step-func-perturbation}
	m_\text{pert}(x) = \begin{cases}
		\meq(\bar{n}) - a & x < L-w \\
		\meq(\bar{n}) + b & x > L-w .
	\end{cases}
\end{equation}
For the perturbation to conserve the global average total density, we must set $b = a\frac{L-w}{w}$. The `step-like' profiles therefore form a two-parameter family of perturbations with the shape parameters $w$ (width of the region where density is increased) and $a$ (density removed uniformly from the rest of the system). Because the concentration may not drop below zero, only perturbations with $a <  \meq(\bar{n})$ are physically sensible. The (heuristic) threshold in phase space $m_\text{th}(\bar{n})$ (see Fig.~\ref{app-fig:stim-induced}b) is exceeded in the (high density) region $x > L-w$ when
\begin{equation} \label{eq:exc-threshold}
	a > a_\text{th}(w;\bar{n}) := \frac{w}{L-w} [m_\text{th}(\bar{n}) - m^*(\bar{n})].
\end{equation}
Note that the threshold $m_\text{th}(\bar{n})$ in phase space is a function of the average total density. We tested various total average densities $\bar{n}$ across the multistable regime $n_-^\infty < \bar{n} < \nLat^-$, and varied the `shape parameters' of the perturbation---amplitude $a$ and width $w$---throughout their respective maximal ranges: $0 < a < \meq(\bar{n})$ and $0 < w < L$. Figure~\ref{app-fig:stim-induced} shows that there is good agreement between the geometrically estimated threshold and the actual basins of attraction determined by numerical simulation.

\section{Stability of stationary patterns} \label{app-sec:FBS-stability}

In our analysis of stationary patterns, we have touched the question of stability of these patterns only peripherally in Sec.~\ref{sec:n-Dc-bistable}. Coarsening, i.e.\ the instability of multi-peak / multi-mesa patterns, in 2C-MCRD systems has been studied before both numerically \cite{Otsuji:2007a,Chiou:2018a} and semi-analytically \cite{Ishihara:2007a} for specific choices for the reaction kinetics $f(m,c)$. For specific reaction terms $f(m,c)$ that allow a mapping of the reaction--diffusion dynamics to an effective gradient dynamics, stability of patterns can be analyzed with the help of the effective free energy that is minimized by the stationary pattern \cite{Goh:2011a,Morita:2010a}. In the broader class of two-component systems without conserved total density, stability of stationary patterns has been subject to numerous mathematical studies, see e.g.\ Refs.~\cite{Nishiura:1987a,Kolokolnikov:2005a,McKay:2012a}.

Instead of the technical tools typically employed there, we choose a more heuristic approach here, building on the physical intuition we have gained throughout this work. We restrict our analysis to the case of mesa patterns with a small interface width compared to the system size ($\Lint \ll L$).

Our starting point is the insight that the stationary pattern $(\mstat(x),\cstat(x))$ is embedded in a flux-balance subspace, Eq.~\eqref{eq:flux-balance-subspace}, whose position $\etaInfty$ is determined by total turnover balance, Eq.~\eqref{eq:turnover-balance}. We hence write the stationary pattern as a pair $(\mstat(x),\etaInfty)$, where only $\mstat(x)$ depends on $x$. Next, recall that the pattern itself is scaffolded by local equilibria. In particular, the plateaus are slaved to the plateau scaffolds $m_\pm(\etaInfty)$, and the pattern inflection point is determined by $m_0(\etaInfty)$. Following a perturbation $\delta m(x,t)$ of the stationary pattern profile, the plateaus will quickly return to their stable local equilibria, the plateau scaffolds $m_\pm$. On the other hand, a perturbation of the mass-redistribution potential (``FBS-position'') $\delta \eta$, will not only shift the plateau scaffolds, but also cause an imbalance of total reactive turnover. This imbalance drives the dynamics of $\eta(x,t)$ and thus determines the stability of the pattern, as we will see in the following.

The dynamics of $\eta(x,t) = c(x,t) + m(x,t) D_m/D_c$ follows straightforwardly from the reaction--diffusion dynamics Eq.~\eqref{eq:two-comp-dyn} and read:
\begin{align}
	\partial_t \eta(x,t) 
	={}& D_c \gradx^2 \eta+ (D_m/D_c - 1) \, \partial_t m \notag \\
	={}& D_c \gradx^2 \eta+ D_m (D_m/D_c - 1) \gradx^2 m \notag \\
	&{}+ (D_m/D_c - 1)\tilde{f}(m,\eta)
\end{align}
In linearization around a stationary pattern $(\mstat(x),\etaInfty)$, we have
\begin{align} \label{app-eq:eta-dyn-lin}
	\partial_t \delta \eta(x,t) =  D_c \gradx^2 \delta \eta + D_m (D_m/D_c - 1) \gradx^2 \delta m \notag \\
	{}+ (D_m/D_c - 1)\big[\tilde{f}_m \delta m + \tilde{f}_\eta \delta \eta\big]_{(\mstat(x),\etaInfty)},
\end{align}
where the membrane perturbation $\delta m = \delta m(x,t)$ is governed by the linearization of the reaction--diffusion dynamics Eq.~\eqref{eq:m-dyn}. The intuition is that $\delta m(x,t)$ quickly relaxes to the scaffold of local equilibria. We therefore focus on the dynamics of $\delta \eta(x,t)$, which will affect the scaffold itself by shifting local equilibria, in particular the plateau scaffolds $m_\pm(\eta)$.

Reactive turnover balance is primarily determined in the interfacial region (cf. Fig.~\ref{fig:stat-pattern-and-phase-space}) around the pattern inflection point $x_0$. We therefore focus on the interface region to learn how an imbalance of total reactive turnover affects the perturbation of the mass-redistribution potential (``FBS-shift'') $\delta \eta(x,t)$. To that end, we use that the gradient of the membrane profile, $\gradx \mstat(x)$, is negligible in the plateaus and whereas it peaks at the inflection point $x_0$. We hence multiply Eq.~\eqref{app-eq:eta-dyn-lin} by $\gradx \mstat(x)$ and integrate over the whole domain $[0,L]$ to obtain
\begin{align} \label{app-eq:eta-int-linearized}
	\partial_t \delta \eta(x_0, t) 
	\approx{}& \delta \eta(x_0, t) \, \frac{D_m/D_c \,{-}\, 1}{m_+ \,{-}\, m_-} \integral{m_-}{m_+}{m} \tilde{f}_\eta(m,\etaInfty) \notag \\
	& + D_c \gradx^2 \delta \eta(x_0,t) + \mathcal{O}\big(\delta m(x,t)\big).
\end{align}
We neglect contributions $\mathcal{O}(\delta m(x,t))$ that correspond to perturbation along the direction of the FBS and quickly relax to the scaffold on the timescale $|\sigma_\text{loc}|^{-1}$, fast compared to the contribution by the first term in Eq.~\eqref{app-eq:eta-int-linearized}. Furthermore, because mass-redistribution $\eta(x,t)$ quickly homogenizes in the (small) interface region, we can neglect the second term $~\gradx^2 \delta \eta(x_0,t)$. 

\begin{figure}
	\includegraphics{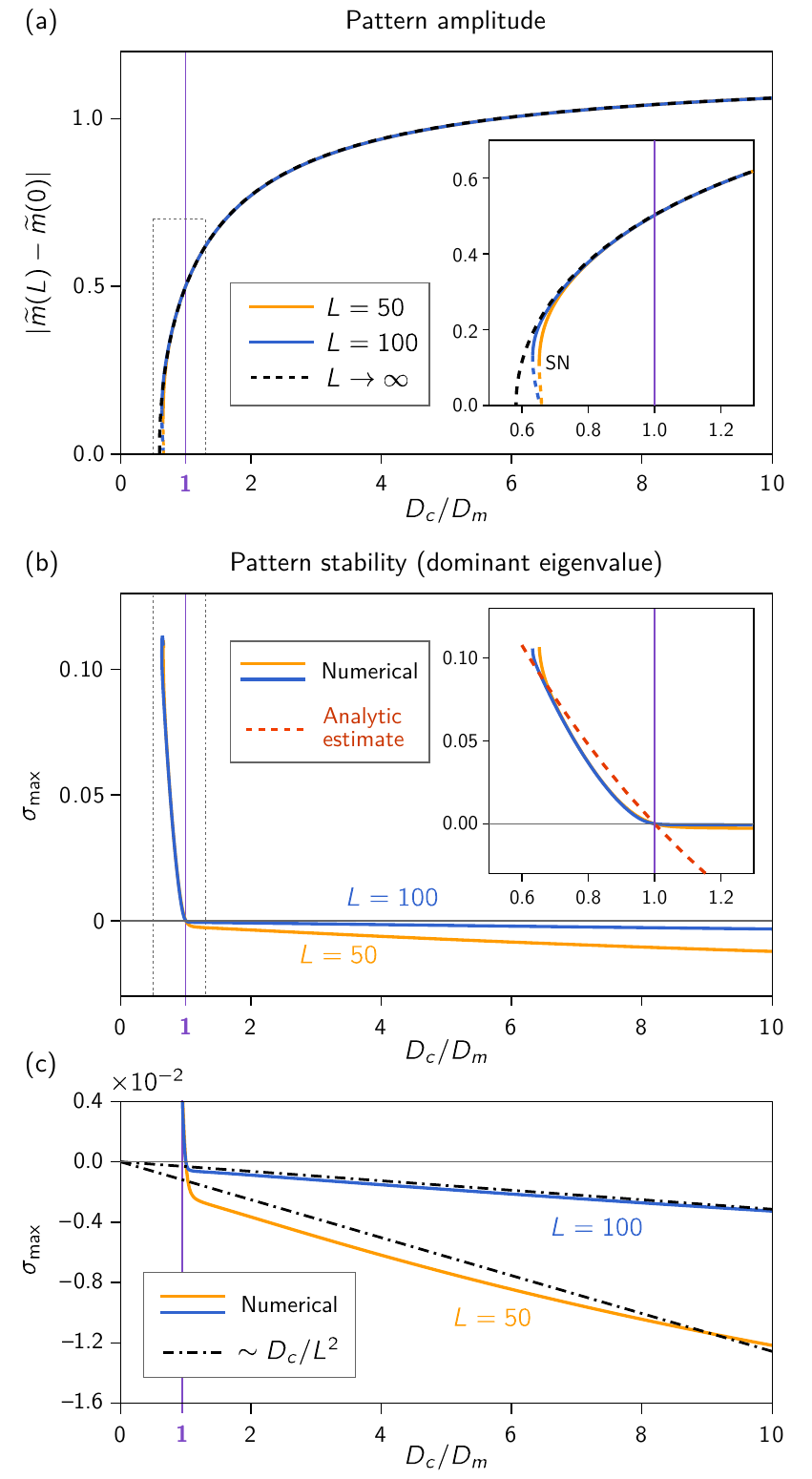}
	\caption{Stability of stationary patterns as a function of $D_c$.
	(a) Pattern amplitude for a $D_c$-sweeps at constant $\bar{n}=n_\text{stat}$ for two system sizes $L = 50$ (yellow) and $L = 100$ (blue), obtained by numerical continuation, and geometric construction (corresponding to $L\to\infty$, dashed black line). The fixed parameters are the same as in Fig.~\ref{fig:n-bifurcation-mesa-patterns}. Finite size affects the stationary pattern amplitude only in the vicinity of the saddle-node bifurcation at $\approx \, D_c^\text{min}$ (see inset: zoom on gray box). The purple line marks $D_c = D_m$. 
	(b) Numerically determined dominant eigenvalues (maximal real part shown as solid lines) indicating linear stability of the patterns. Patterns are stable for $D_c > D_m$ and unstable for $D_c < D_m$, as predicted by the geometric arguments. The dominant eigenvalue for the unstable patterns is almost independent of system size. The approximation based on this geometric intuition of turnover imbalance (first term in Eq.~\eqref{eq:eta-int-linearized}) is shown as red, dashed line in the inset. 
	(c) Blow-up of the $\sigma$-axis, for negative values. For stable patterns ($D_c > D_m$), decay of perturbations is mainly determined by the timescale $\sim D_c/L^2$ of mass-transfer from one end of the system to the other. (The dot-dashed lines show the relation $-\alpha D_c/L^2$, where the prefactor $\alpha \approx 3.1$ that depends on the system specifics has been fitted by eye.)
	}
	\label{app-fig:pattern-stability}
\end{figure}

Fig.~\ref{app-fig:pattern-stability} shows a comparison of the heuristic estimates of perturbation growth/decay rate based on Eq.~\eqref{eq:eta-int-linearized} to numerically determined dominant eigenvalues $\sigma_\text{max}(D_c)$ (linear stability analysis of stationary patterns determined by numerical continuation). The dominant eigenvalue $\sigma_\text{max}(D_c)$ crosses over from system size independent growth rate of perturbations for $D_c < D_m$ to system size dependent decay of perturbations for $D_c > D_m$. The instability is well estimated by turnover imbalance term in Eq.~\eqref{eq:eta-int-linearized} (dashed red line in the inset in (b)), while the rate at which perturbations decay in the stable regime ($D_c > D_m$) is limited by diffusive transport $\sigma_\text{diff} \sim D_c/L^2$ between the far ends of the system. Reactive timescales will become limiting in the stable regime only when $\sigma_\text{diff} \approx \sigma_\text{loc}$, that is, for fast enough cytosolic diffusion or a small system.

\section{Weakly nonlinear analysis} \label{app-sec:weakly-nonlinear}

Our goal is to find the stationary pattern in the vicinity of the onset of lateral instability (Turing bifurcation). To that end, we expand the stationary state $(\mstat(x), \eta_0)$ in harmonic functions (eigenmodes of the Laplace operator under no-flux boundary conditions):
\begin{subequations} 
\begin{align}
	\mstat(x) \approx {} & \meq + \delta m_0 + \delta m_1 \cos(\pi x/L) \notag \\
	& + \delta m_2 \cos(2\pi x/L), \label{app-eq:weakly-nonlinear-ansatz} \\
	\eta_0 \approx {}& \eta^* + \delta \eta_0,	
\end{align}	
\end{subequations}
where $\eta^* = D_m/D_c \meq+\ceq$ is the FBS position of the homogenous steady state. Mass conservation necessitates $\delta m_0 + \delta \eta_0 - D_m/D_c \delta m_0 = 0$, hence 
$$
\delta \eta_0 = (D_m/D_c - 1) \, \delta m_0
$$
We plug the ansatz Eq.~\eqref{app-eq:weakly-nonlinear-ansatz} into the stationarity condition, Eq.~\eqref{eq:m-stat-fbs}, Taylor expand $\tilde{f}$, and project onto the zeroth harmonic (this is most conveniently done in CAS software, e.g Wolfram Mathematica)
\begin{align}
	0 ={}&  \big(\tilde{f}_m + (1-D_m/D_c) \tilde{f}_\eta\big) \delta m_0 + \frac{1}{4} \tilde{f}_{mm}\delta m_1^2 \notag \\
	&{}+\mathcal{O}(\delta m_0^2, \delta m_1^2 \delta m_2, \delta m_0 \delta m_1^2, \delta m_0 \delta m_2^2),
\end{align}
and and onto the second harmonic
\begin{align}
	0 ={}& \big(\tilde{f}_m - 4 D_m \pi^2/L^2\big) \delta m_2 + \frac{1}{4} \tilde{f}_{mm}\delta m_1^2 \notag \\
	&{}+ \mathcal{O}(\delta m_2^2, \delta m_0 \delta m_1^2, \delta m_0^2 \delta m_2, \delta m_1^2 \delta m_2).
\end{align}
Solving for $\delta m_0$ and $\delta m_2$ we get
\begin{subequations}
\begin{align}
	\delta m_0 &=\frac{1}{4} \frac{\tilde{f}_{mm}}{\tilde{f}_m + (1-D_m/D_c) \tilde{f}_\eta} \delta m_1^2 + \mathcal{O}(\delta m_1^4),\\
	\delta m_2 &=\frac{1}{4} \frac{\tilde{f}_{mm}}{\tilde{f}_m -  4 D_m \pi^2/L^2} \delta m_1^2 + \mathcal{O}(\delta m_1^4).
\end{align}	
\end{subequations}
These equations describe how asymmetry of the nullcline shape (and thereby reactive turnover) influences the pattern profile. For $\tilde{f}_{mm} = 0$ the turnovers on either side of the inflection point grow symmetrically (with opposite sign) as the amplitude $\delta m_1$ of the pattern increases. This symmetry of the turnovers only occurs at the inflection point of the nullcline where its shape is point-symmetric. Away from the inflection point of the nullcline, its shape is no longer symmetric around the steady state $(\meq,\ceq)$, and, respectively, the turnovers grow asymmetrically in the two halves of the system. This asymmetry creates an imbalance of the total turnover that is compensated by two effects: (i) the flux-balance subspace shifts (i.e.\ $\eta_0$ will deviate from $\eta^*$), and (ii) the pattern itself becomes asymmetric as the second harmonic will have a non-zero amplitude $\delta m_2$. Together, these two effects compensate the asymmetry of the turnovers, such that total turnover balance is reached.

The amplitude $\delta m_1$ of the stationary pattern is obtained by projecting \eqref{eq:m-stat-fbs} onto the first harmonic $\cos(\pi x/L)$\begin{equation} \label{app-eq:steady-state-amplitude}
	0 = F_1 \, \delta m_1 + F_3 \, \delta m_1^3 + \mathcal{O}(\delta m_1^5),
\end{equation}
where
\begin{equation}
	F_1 = \tilde{f}_m - D_m \pi^2/L^2,
\end{equation}
and
\begin{align} 
	F_3 ={}& \frac{\tilde{f}_{mmm}}{8} -\frac{\tilde{f}_{mm}}{8} \frac{\tilde{f}_{mm}}{\tilde{f}_m -  4 D_m \pi^2/L^2} \notag \\
		&-\frac{\tilde{f}_{mm}}{4}  \frac{\tilde{f}_{mm} - (1-D_m/D_c)\tilde{f}_{m\eta}}{\tilde{f}_m - (1-D_m/D_c)\tilde{f}_\eta}. \label{eq:F3-raw}
\end{align}
Since there is no second order term in Eq.~\eqref{app-eq:steady-state-amplitude}, patterns always originate in a pitchfork bifurcation. At the bifurcation point, the first order coefficient vanishes ($F_1 = 0$). The system is laterally unstable if $F_1$ is positive (cf. Eq.~\eqref{eq:qmax}). Hence, only if $F_3 < 0$, the third order coefficient can saturate the pattern amplitude (supercritical bifurcation). For $F_3 > 0$ the bifurcation is subcritical.

The third order coefficient can be simplified further: A simple calculation shows that
\begin{equation}
	\tilde{f}_m - (1-D_m/D_c)\tilde{f}_\eta = f_m - f_c = \sigma_\text{loc},
\end{equation}
and therefore
\begin{align}
	\tilde{f}_{mm} - (1-D_m/D_c)\tilde{f}_{m\eta} &= \partial_m \sigma_\text{loc}(m, \eta - m \, D_m/D_c) \notag \\
	&= \tilde{\partial}_{m} \sigma_\text{loc}(m,\eta)
\end{align}
where $\tilde{\partial}_{m} = \partial_m - (D_m/D_c) \partial_c$ is the derivative along the flux-balance subspace.
With that, the second summand in the brackets in $F_3$ (cf. Eq.~\eqref{eq:F3-raw}) can be rewritten and we obtain
\begin{equation}
	F_3 = \frac{\tilde{f}_{mmm}}{8} - \frac{\tilde{f}_{mm}}{2} \left(
		\frac{\tilde{f}_{mm}/2}{\tilde{f}_m -  4 D_m \pi^2/L^2} +
		\frac{\tilde{\partial}_{m} \sigma_\text{loc}}{\sigma_\text{loc}}
	 \right).
\end{equation}
We further rewrite the denominator of the first summand in the brackets as $F_1 - 3 D_m \pi^2/L^2$, and use that $F_1$ vanishes at the bifurcation point (i.e.\ is small in its vicinity). We thus have  
\begin{equation} \label{eq:F3-simplified}
	F_3 = \frac{\tilde{f}_{mmm}}{8} + \frac{\tilde{f}_{mm}^2}{24} \frac{L^2}{\pi^2 D_m} -
		\frac{\tilde{f}_{mm}}{4} \frac{\tilde{\partial}_{m} \sigma_\text{loc}}{\sigma_\text{loc}} + \mathcal{O}(F_1).
\end{equation}
Note that in weakly nonlinear approximation Eq.~\eqref{eq:steady-state-amplitude} in the main text, we have pulled the $\mathcal{O}(F_1)$ out from $F_3$ to simplify notation. The role and physical interpretation of the three terms in $F_3$, as written in the form Eq.~\eqref{eq:F3-simplified}, are discussed in the main text in Sec.~\ref{sec:sub-supercrit}.

\section{Nullcline curvature approximation}
\label{app-sec:curvature-approx}

In the following, we show that the nullcline curvature $\kappa$ can be approximated by
\begin{equation} \label{eq:curvature-approx}
	\kappa \approx - \frac{f_c^2}{(f_m^2+f_c^2)^{3/2}} \, \tilde{f}_{mm},
\end{equation}
in the vicinity of the Turing bifurcation (onset of lateral instability; recall the slope criterion Eq.~\eqref{eq:slope-criterion}).
Start by rewriting the second derivative $\tilde{f}_{mm}$ in terms of derivatives of $f$:
\begin{align}
	\tilde{f}_{mm} &= \partial_m^2 f\!\left(m,\eta - \frac{D_m}{D_c} m\right) \notag \\
	&= f_{mm}-2 \frac{D_m}{D_c}  f_{mc} + \left(\frac{D_m}{D_c}\right)^2 f_{cc}
\end{align}
In the vicinity of the Turing bifurcation, we have $-D_m/D_c \approx -f_m/f_c$ (from the slope criterion for lateral instability, Eq.~\eqref{eq:slope-criterion}), so we obtain
\begin{equation} \label{eq:fmm-approx}
	\tilde{f}_{mm} \approx f_c^{-2}\big[f_c^2 f_{mm} -2 f_m f_c f_{mc} + f_m^2 f_{cc}\big].
\end{equation}
Comparing to the formula for the curvature $\kappa$ of an implicitly determined curve $f(m,c) = 0$
\begin{equation} \label{eq:curvature-formula}
	\kappa = - \frac{f_c^2 f_{mm} -2 f_m f_c f_{mc} + f_m^2 f_{cc}}{(f_m^2+f_c^2)^{3/2}},
\end{equation}
one sees that the numerator of the curvature formula is identical to the term in the square brackets in Eq.~\eqref{eq:fmm-approx}. Thus, by combining Eqs.~\eqref{eq:fmm-approx} and~\eqref{eq:curvature-formula}, we obtain the approximation Eq.~\eqref{eq:curvature-approx}.

\vspace{2ex}
\section{Topological equivalence of 2C-MCRD systems shear banding in complex fluids} \label{app-sec:shear-banding}

Analogies between the shear banding in complex fluids, phase separation near thermal equilibrium, and reaction--diffusion systems have been drawn before on mathematical grounds, i.e.\ using similarity of the equations used describe these phenomena \cite{Radulescu:1999a,Fardin:2012a}.
The phase space analysis of MCRD pattern formation presented in the present work establishes a connection to shear banding on the basis of phase space geometry. More specifically, these two phenomena can be regarded as \emph{topologically equivalent}, i.e.\ they can be understood in terms of equivalent geometric objects in phase space. We shall briefly outline this connection in the following.

Complex fluids can exhibit a non-monotonic \emph{constitutive} relationship $\Sigma(\dot{\gamma})$ between the total stress $\Sigma$ and the (homogeneous) strain rate $\dot{\gamma}$~\cite{Olmsted:2008a,Divoux:2016a}. When the total stress decreases upon an increase in strain rate, $\partial_{\dot{\gamma}}\Sigma < 0$, a mechanical instability results, which leads to a separation of a sheared fluid into ``shear bands'' with different viscosities and strain rates, which coexist at a common total stress.
The ``common total stress'' construction on the \emph{constitutive} curve $\Sigma(\dot{\gamma})$ employed to analyze this phenomenon (in a one-dimensional system) is analogous to our flux-balance construction on the reactive nullcline for 2C-MCRD systems (see~Sec.~\ref{sec:scaffolding-introduction}), by means of a mapping $(n,\eta) \leftrightarrow (\dot{\gamma},\Sigma)$ between the phase space variables.
The average strain rate $\bar{\dot{\gamma}} = L^{-1} \int_0^L \mathrm{d} x \, \dot{\gamma}$ is analogous to the average total density $\bar{n}$. 
The constitutive $\Sigma(\dot{\gamma})$ curve is analogous to the reactive nullcline.
The steady state conditions are spatially uniform total stress and flux-balance (spatially uniform mass-redistribution potential; see~\ref{sec:flux-balance-subspace}) respectively.
The selection of the common total stress generally depends on the details of the model \cite{Dhont:2008a}, in particular on stress diffusion \cite{Fardin:2015a}. For simple models, it can be pictured similarly to a Maxwell-construction (cf.\ total turnover balance in a 2C-MCRD system illustrated in Fig.~\ref{fig:stat-pattern-and-phase-space}).
Furthermore, momentum propagation due to stress gradients in complex fluids is equivalent to mass redistribution due to concentration gradients in MCRD systems.
Accordingly, the low Reynolds number limit is analogous to the $D_c \to \infty$ limit in the 2C-MCRD system.

Taking these analogies together, we conclude that these physically distinct phenomena are topologically equivalent and can be studied with similar phase-space geometric tools.
Such a connection might benefit both fields as more involved scenarios are investigated, for instance coupling to additional degrees of freedom: 
For models of complex fluids, additional spatial dimensions \cite{Fielding:2010a}, and coupling to internal structure of the fluid \cite{Fielding:2007a, Turcio:2018a}, can lead to a variety of intricate spatiotemporal patterns; 
for mass-conserving reaction--diffusion systems, additional components or additional conserved species can equally lead to a broad range of phenomena \cite{Murray:2017a, Halatek:2018a, Alonso:2010a}.
Studying such systems in terms of local equilibria theory offers an exciting new perspective for future research.


\begin{thebibliography}{186}
\bibitem [{Note:Trough()}]{Note:Trough}%
  \BibitemOpen
  \bibinfo {note} {The complement to a peak is a `trough' pattern, composed of
  an interface connected to a high density plateau.}\BibitemShut {Stop}%
\bibitem [{Note:RollingBallPeak()}]{Note:RollingBallPeak}%
  \BibitemOpen
  \bibinfo {note} {Given $\eta_0$, and the boundary condition $m(x \to -
  \infty) \to m_-(\eta_0)$ one can determine the entire peak profile using the
  rolling-ball analogy: The profile corresponds to the Newtonian trajectory
  $m(x)$ of a mass point in a potential $V(m) = -\int \mathrm{d}m \, f$, where
  $m$ is the ``position'' of the mass point and $x$ is ``time''.}\BibitemShut
  {Stop}%
\bibitem [{Note:SaddleNode()}]{Note:SaddleNode}%
  \BibitemOpen
  \bibinfo {note} {The slope $\snc$ of a continuous reactive nullcline smoothly
  approaches the slope $-1$ at the saddle-node bifurcations of the reactive
  equilibria. Hence, as $D_c$ is decreased, the laterally unstable regions
  shrink, but disappear fully only at $D_c = D_m$. In the vicinity of the
  saddle-node bifurcations for sufficiently small $D_c \gtrsim D_m$, no
  stationary patterns exist, but there are still small regions of lateral
  instability.}\BibitemShut {Stop}%
\bibitem [{Note:GammaInterpolation()}]{Note:GammaInterpolation}%
  \BibitemOpen
  \bibinfo {note} {Alternatively, one may introduce $\gamma \in [0,1]$ as an
  interpolation parameter to interpolate between the original system ($\gamma =
  1$) and the mass-conserving subsystem ($\gamma = 0$). One can then use
  analytic or numeric methods to continue the solution of the mass-conserving
  subsystem to the original system and thus systematically study the effect of
  the source terms.}\BibitemShut {Stop}%
\end{thebibliography}

%

\end{document}